\documentclass[11pt,a4paper,onecolumn,usenatbib,referee]{mnras}
\usepackage{graphicx}
\usepackage{color}
\usepackage{amssymb}
\usepackage{amsmath}
\usepackage{mathabx}
\usepackage{stmaryrd}
\usepackage{multirow}
\usepackage{multicol}
\usepackage{amscd}
\usepackage{mfpic}
\usepackage{verbatim}
\usepackage{hyperref}
\usepackage{pstricks,pst-node,pst-text,pst-3d}
\usepackage{ftnxtra}
\usepackage{fnpos} 

\definecolor{cream}{rgb}{.97, .95, .88}
\definecolor{darkcream}{rgb}{1., .88, .5}
\definecolor{lightpink}{rgb}{0.98, 0.88, 0.87}
\definecolor{lightwhite}{rgb}{1., 0.98, 0.95}
\definecolor{lightsalmon}{rgb}{1., 0.95, 0.90}
\definecolor{lightviolet}{rgb}{0.9, 0.8, 0.9}
\definecolor{lightgray}{rgb}{.96, .96, .96}  
\definecolor{lgray}{rgb}{.75, .75, .75}
\definecolor{LemonChiffon}{rgb}{0.95, 1., 0.7}
\definecolor{lightolivegreen}{rgb}{0.84, 0.89, 0.25}
\definecolor{lightgreen}{rgb}{.664, 1., .52}
\definecolor{llgreen}{rgb}{.900, .983, .960}
\definecolor{tristle}{rgb}{0.87, 0.67, 0.77} 
\definecolor{pink}{rgb}{0.95, 0.45, 0.75}
\definecolor{magenta}{rgb}{1., 0, 1.}
\definecolor{violet}{rgb}{0.9, 0.20, 0.85}
\definecolor{darkolivegreen}{rgb}{0.55, 0.65, 0.35}
\definecolor{maroon}{rgb}{0.7, 0.26, 0.56}
\definecolor{lightmaroon}{rgb}{0.85, 0.38, 0.58}
\definecolor{darkmaroon}{rgb}{0.604, 0.169, 0.451}
\definecolor{ddarkmaroon}{rgb}{0.2, 0.03125, 0.150}
\definecolor{mediumorchid}{rgb}{0.8, 0.33, 0.83}
\definecolor{mediumorchidd}{rgb}{1., 0.33, 0.63}
\definecolor{darkgreen}{rgb}{0.1, 0.6, 0.13}
\definecolor{lightyellow}{rgb}{1., 1., 0.82}
\definecolor{turquoise}{rgb}{0.042, 0.586, 0.512}
\definecolor{turquoisel}{rgb}{0.66, 0.94, 0.83}
\definecolor{darkturquoise}{rgb}{0.21, 0.55, 0.50}
\definecolor{coral}{rgb}{1., 0.6, 0.21}
\definecolor{lightorange}{rgb}{1., 0.88, 0.75}
\definecolor{orangered}{rgb}{1., 0.5, 0.}
\definecolor{orange}{rgb}{1., 0.65, 0.1}
\definecolor{orangel}{rgb}{1., .85, .3}
\definecolor{darkorange}{rgb}{0.875, 0.4, 0.204}
\definecolor{ddarkorange}{rgb}{.675, .218, .05}
\definecolor{bluesky}{rgb}{0.48, 0.53, 1.}
\definecolor{gold}{rgb}{1., 0.85, 0.25}
\definecolor{goldd}{rgb}{0.95, 0.75, 0.05}
\definecolor{darkviolet}{rgb}{0.54, 0.04, 0.84}
\definecolor{ddarkviolet}{rgb}{.382, .063, .657}
\definecolor{lightblue}{rgb}{0.30, 0.86, 0.89}
\definecolor{LightBlue}{rgb}{0.68, 0.85, 0.9}
\definecolor{lblue}{rgb}{0.78, 0.90, 0.95}
\definecolor{darkblue}{rgb}{.105, .308, .707}
\definecolor{lightmaroon}{rgb}{0.85, 0.38, 0.58}
\definecolor{darkmaroon}{rgb}{0.604, 0.169, 0.451}
\definecolor{darkpink}{rgb}{0.879, 0.020, 0.766}
\definecolor{ddarkpink}{rgb}{0.738, 0.195, 0.406}
\definecolor{grey}{rgb}{0.717, 0.717, 0.717}
\definecolor{lightgrey}{rgb}{0.800, 0.800, 0.800}
\definecolor{brown}{rgb}{0.740, 0.323, 0.182}
\definecolor{redbrown}{rgb}{.575, .158, .05}
\definecolor{darkbrown}{rgb}{0.34, 0.25, 0.05}
\definecolor{orangebrown}{rgb}{0.433, 0.262, 0.06}
\definecolor{pinkl}{rgb}{1., 0.788, 0.918}
\definecolor{salmon}{rgb}{1., 0.66, 0.5}
\definecolor{lightbrown}{rgb}{0.703, 0.508, 0.121}

\def\etal{{\it et al.}}

\def\Name#1#2 {{#2} {#1, }}

\def\Journal#1#2#3#4{{#3}, {#1}, {\bf #2}, #4} 
\def\cir#1{{\GCN} #1}
\def\rep#1{{\GCR} #1}
\def\AA{\em A.\& A.}

\def\AIA{\em Advance in Astro.}
\def\AIP{\em AIP Conf.Proc.}

\def\APJ{\em ApJ.}
\def\APJL{\em ApJ.Lett.}
\def\APJS{\em ApJ.Suppl.}

\def\ARA{\em Annu.Rev.A\&A}

\def\APS{\em APS }
\def\ASP{\em ASP Conf.Ser.}

\def\CJA{\em Chin. J. Astron. Astrophys.}

\def\CQG{\em Class.Quant.Grav.}

\def\GCN{\em GCN Circ.}
\def\GCR{\em GCN Rep.}

\def\JCA{\em J. Cosmol. Astrop. Phys.}

\def\MRA{\em MNRAS}

\def\NAT{\em Nature}
\def\NATA{\em Nature Astro.}

\def\PRD{{\em Phys. Rev.} D}
\def\PRL{\em Phys. Rev. Lett.}

\def\POS{\em Proc. of Science}

\def\RPP{\em Rept. Prog. Phys.}
\def\SCI{\em Science}

\def\SSC{\em Space Sci.}
\def\SSR{\em Space Sci. Rev.}

\def\be{\begin{equation}}
\def\ee{\end{equation}}
\def\bea{\begin{eqnarray}}
\def\eea{\end{eqnarray}}
\def\bes{\begin{equation*}}
\def\ees{\end{equation*}}
\def\beas{\begin{eqnarray*}}
\def\eeas{\end{eqnarray*}}

\title{Prompt gamma-ray emission of GRB 170817A associated to GW 170817: A consistent picture}

\author[H. Ziaeepour]{Houri~Ziaeepour$^{1,2}$\thanks{Email: houriziaeepour@gmail.com} \\
$^1$Institut UTINAM, CNRS UMR 6213, Observatoire de Besan\c{c}on, Universit\'e de Franche Compt\'e, 41 bis ave. de l'Observatoire, \\BP 1615, 25010 Besan\c{c}on, France \\
$^2$Mullard Space Science Laboratory, Holmbury St Mary, Dorking, Surrey RH5 6NT, UK}

\date{Accepted XXX. Received YYY; in original form ZZZ}
\pubyear{2018}

\begin{document}


\label{firstpage}
\pagerange{\pageref{firstpage}--\pageref{lastpage}}
\maketitle

\begin{abstract}
{The short GRB 170817A associated to the first detection of gravitation waves from a Binary Neutron 
Star (BNS) merger was in many ways unusual. Possible explanations are emission from a cocoon or 
cocoon break out, off-axis view of a structured or uniform jet, and on-axis ultra-relativistic jet 
with reduced density and Lorentz factor. Here we use a phenomenological model of shock evolution and 
synchrotron/self-Compton emission to simulate the prompt emission of GRB 170817A and to test above 
proposals. We find that synchrotron emission from a mildly relativistic cocoon with a Lorentz factor 
of 2-3, as considered in the literature, generates a too soft, too long, and too bright prompt 
emission. Off-axis view of an structured jet with a Lorentz factor of about 10 can reproduce 
observations, but needs a very efficient transfer of kinetic energy to electrons in internal shocks, 
which is disfavored by particle in cell simulations. We also comment on cocoon breakout as a mechanism 
for generation of the prompt gamma-ray. A relativistic jet with a Lorentz factor of about 100 and a 
density lower than typical short GRBs seems to be the most plausible model and we conclude that 
GRB 170817A was intrinsically faint. Based on this result and findings of relativistic 
magnetohydrodynamics simulations of BNS merger in the literature we discuss physical and astronomical 
conditions, which may lead to such faint short GRBs. We identify small mass difference of progenitor 
neutron stars, their old age and reduced magnetic field, and anti-alignment of spin-orbit angular 
momentum induced by environmental gravitational disturbances during the lifetime of the BNS as causes 
for the faintness of GRB 170817A. We predict that BNS mergers at lower redshifts generate on average 
fainter GRBs.}
\end{abstract}


\section{Introduction} \label{sec:intro}
The discovery of the Gravitational Wave (GW) event GW 170817~\citep{gw170817ligo} and accompanying 
electromagnetic transient 
GRB 170817A~\citep{gw170817fermi,gw170817integral,gw170817multimess,gw170817fermimulti}, and its 
afterglow in X-ray~\citep{gw170817xray,gw170817cxc,gw170817swiftnustar} and other energy bands~\citep{gw170817optdes,gw170817optdlt,gw170817optsss1,gw170817optsss,gw170817rprocess,gw170817optkilonova,gw170817cocoon,gw170817optkilonovath,gw170817ir,gw170817earlyradio,gw170817earlyradio0,gw170817earlyradio1,gw170817hubble} are 
revolutionizing astronomy and 
fundamental physics. Association of GW 170817 to merger of a binary neutron star, based on the masses 
of the progenitors and the length of GW event, is the first direct evidence for formation of short 
GRBs by collision and merging of ultra-compact astronomical objects. Although observation of 
supernova-like behaviour of late time afterglow of long GRBs has confirmed the hypothesis of their 
formation during core collapse of massive stars, a direct evidence for the origin of short GRBs 
had to wait the historic detection of GW/GRB 170817A. 

Despite excitements about its observation, GRB 170817A is very far from being a typical 
representative of hundreds of short GRB events detected during the past 3 decades or so by high 
energy space observatories such as BATSE~\citep{batse}, Neil Gehrels Swift Observatory~\citep{swift}, 
Fermi~\citep{fermi}, Integral~\citep{integral}, Konus Wind~\citep{konuswind}, etc. It is much softer 
than most short GRBs, a few orders of magnitude fainter than short bursts with known redshift, and 
falls on the boundary of short-long GRB separation. The unusual characteristics of GRB 170817A are 
evidently noticed and widely discussed in the articles published immediately after the announcement 
of GW/GRB 170817 detection. 

The simplest explanation is an off-axis view~\citep{structuredjet} of a uniform (top hat) or 
structured ultra-relativistic jet similar to those of other short GRBs~\citep{gw170817rprocess}. 
Alternatively, the burst might have been formed by a mildly relativistic magnetized 
cocoon~\citep{gw170817cocoon,gw170817cocoon0} at its breakout~\citep{grbcocoon}. However, it seems 
an extra-ordinary coincidence if we have detected an off-axis GRB or one generated by the cocoon 
breakout in the first detection of gravitational waves from a BNS merger. 
Although~\citep{grboffaxis,gw170817latexraystructjet,grboffaxprobab} argue that due to the small 
opening angle of relativistic jets, electromagnetic counterparts of GW events from binary mergers 
must be dominated by relatively soft emission of a jet viewed off-axis or a cocoon or sheath 
surrounding the jet, X-ray light curves for simulated afterglows with non-zero viewing 
angle~\citep{grboffaxis,grbbbagsimul,grbbbagsimul0} deviate significantly from Swift-XRT observations 
of more than 100 short GRBs followed up by this instrument so far, including 
kilonova/GRB 130603B~\citep{grb130603bxrt} (see also sections \ref{sec:faintness} and 
\ref{sec:implic} for further discussion). 

Evidence for (semi-)thermal emission from a cocoon~\citep{grbcocoon} is also very rare and mostly in 
low energies. Therefore, either most short GRBs belong to a completely different population, or the 
dynamics of their progenitors is such that the probability of a close to on-axis view is large. 
In conclusion, although with a statistical sample of one event it is not possible to rule out a 
rare coincidence of GW with off-axis or cocoon emission, we should consider other possibilities.

In this work we first briefly review observed properties of GRB 170817A in Sec. \ref{sec:grbcomp} 
and compare them with those of other short GRBs. This opens our discussion and arguments in 
Sec \ref{sec:faintness} about the small probability that faint soft short GRBs such as GRB 170817A 
be off-axis view of an otherwise normal GRB. We raise other possibilities as reasons behind 
faintness and softness of some short GRBs, including GRB 170817A, which have their root in the 
physics of formation and acceleration of jets, and production of GRB emission. In 
Sec. \ref{sec:prompt} we use a phenomenological model for formation and evolution of GRB emission by 
internal shocks~\citep{hourigrb,hourigrbmag} to simulate the prompt emission of GRB 170817A and 
compare simulation parameters with those of GRB 130603B - the only other short GRB with evidence of 
an accompanying kilonova~\citep{grb130603bkilonova,grb130603bkilonova0,kilonovarev}. The aim of this 
exercise is to have a quantitative estimation of physical properties of GRB producing processes 
and their progenitor stars, notably jet density, Lorentz factor, and Poynting energy, which can be 
compared with findings of BNS merger simulations. Results of our simulations are discussed and 
interpreted in Sec. \ref{sec:interp}. In Sec. \ref{sec:progdy} we use conclusions of 3D General 
Relativistic Magneto-Hydro-Dynamics (GRMHD) simulations from literature to investigated which 
configuration and properties of progenitors may lead to a thin jet with a relatively low Lorentz 
factor, as estimated for GRB 170817A. Implications of our findings are discussed in 
Sec. \ref{sec:implic}. We provide an overall qualitative picture of GW/GRB 170817 event and its 
difference with intrinsically brighter GRBs. Outlines and prospectives are presented in 
Sec. \ref{sec:outline}.

\section{GRB and other electromagnetic emissions} \label{sec:grbcomp}
\begin{figure}
\begin{center}
\begin{tabular}{p{7cm}p{7cm}}
a) & b) \\
\hspace {-1cm}\includegraphics[width=8.5cm]{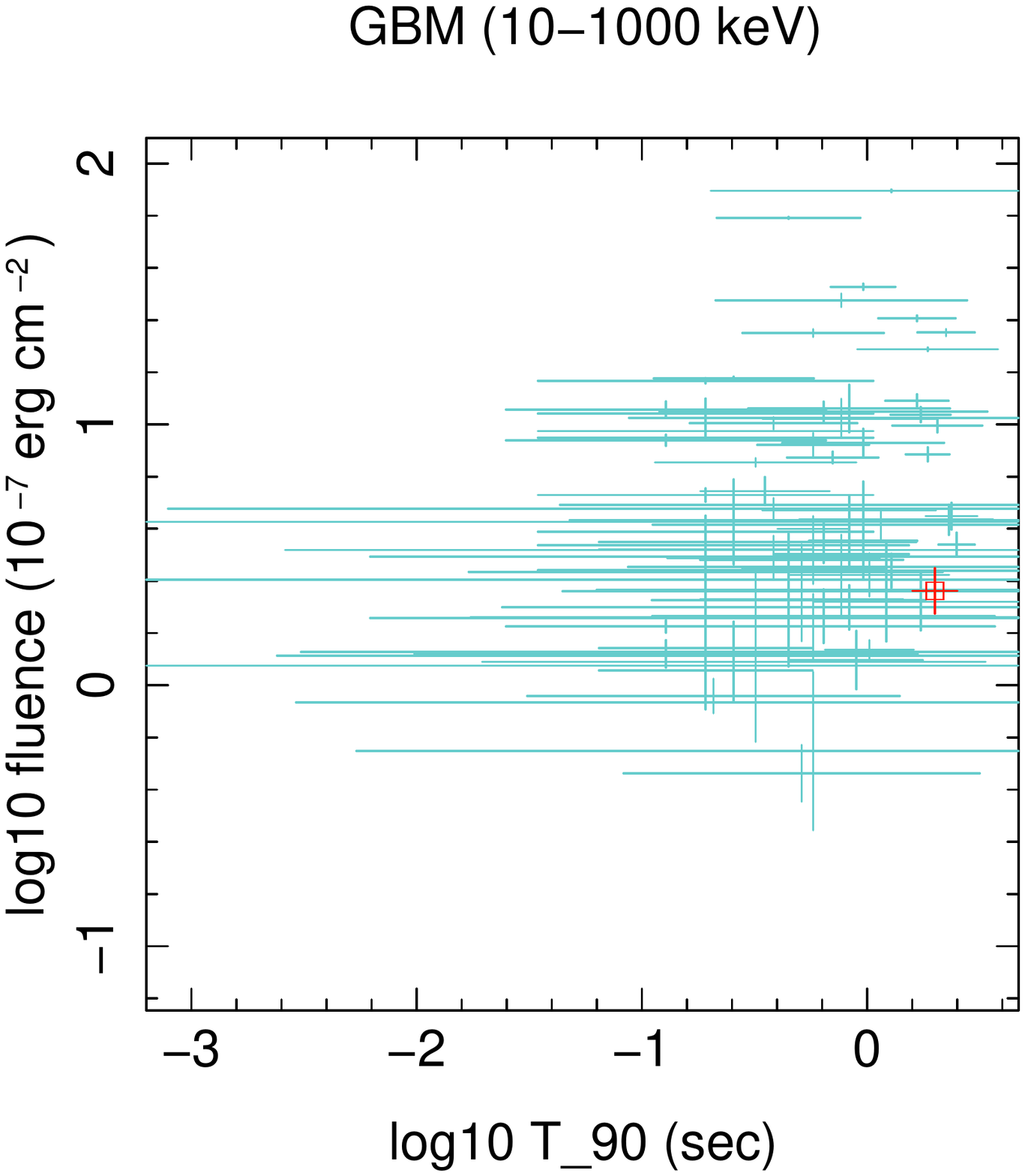} & 
\hspace {-2cm}\includegraphics[width=6cm,angle=90]{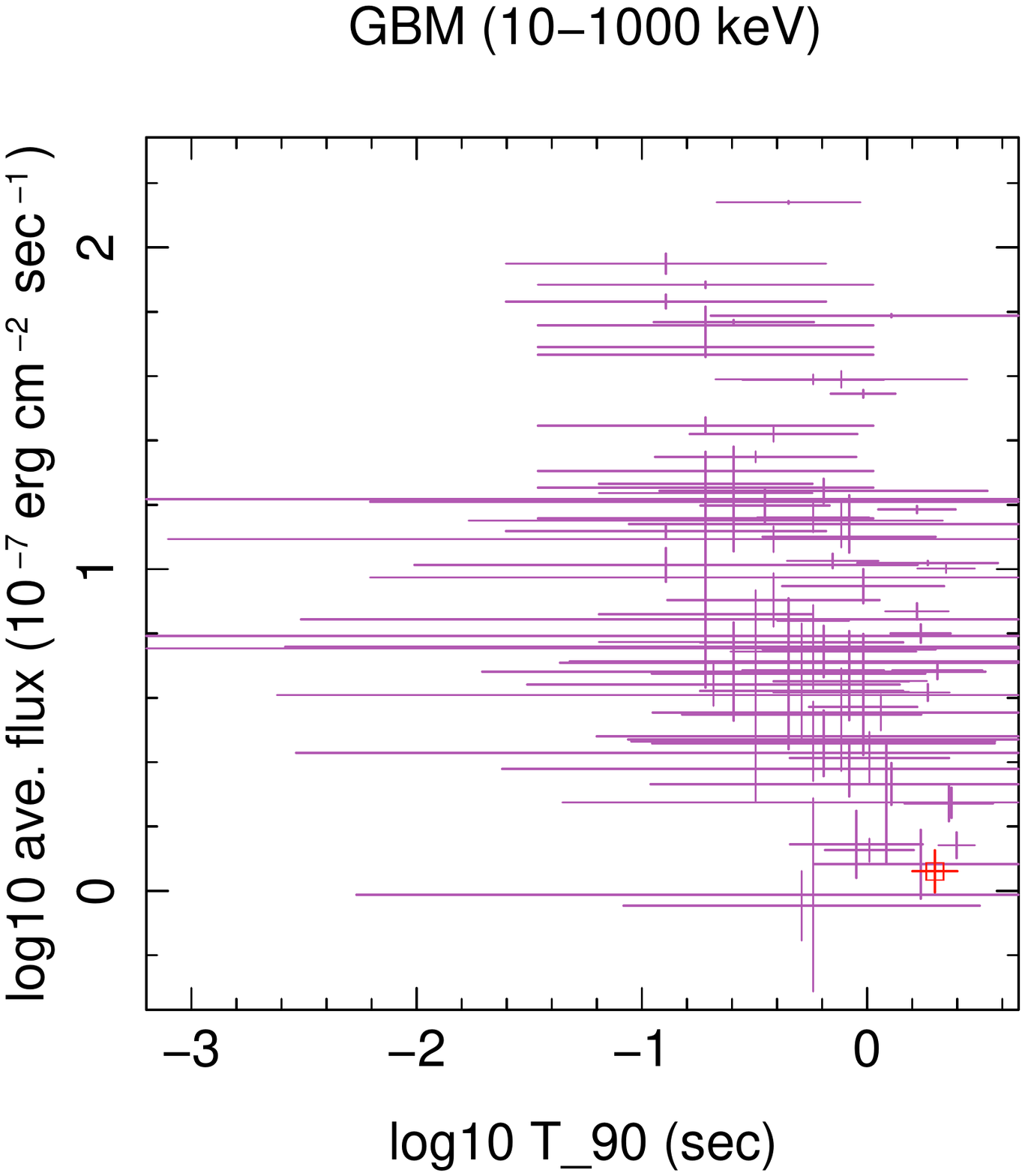} \\
c) & d) \\
\includegraphics[width=7cm,angle=90]{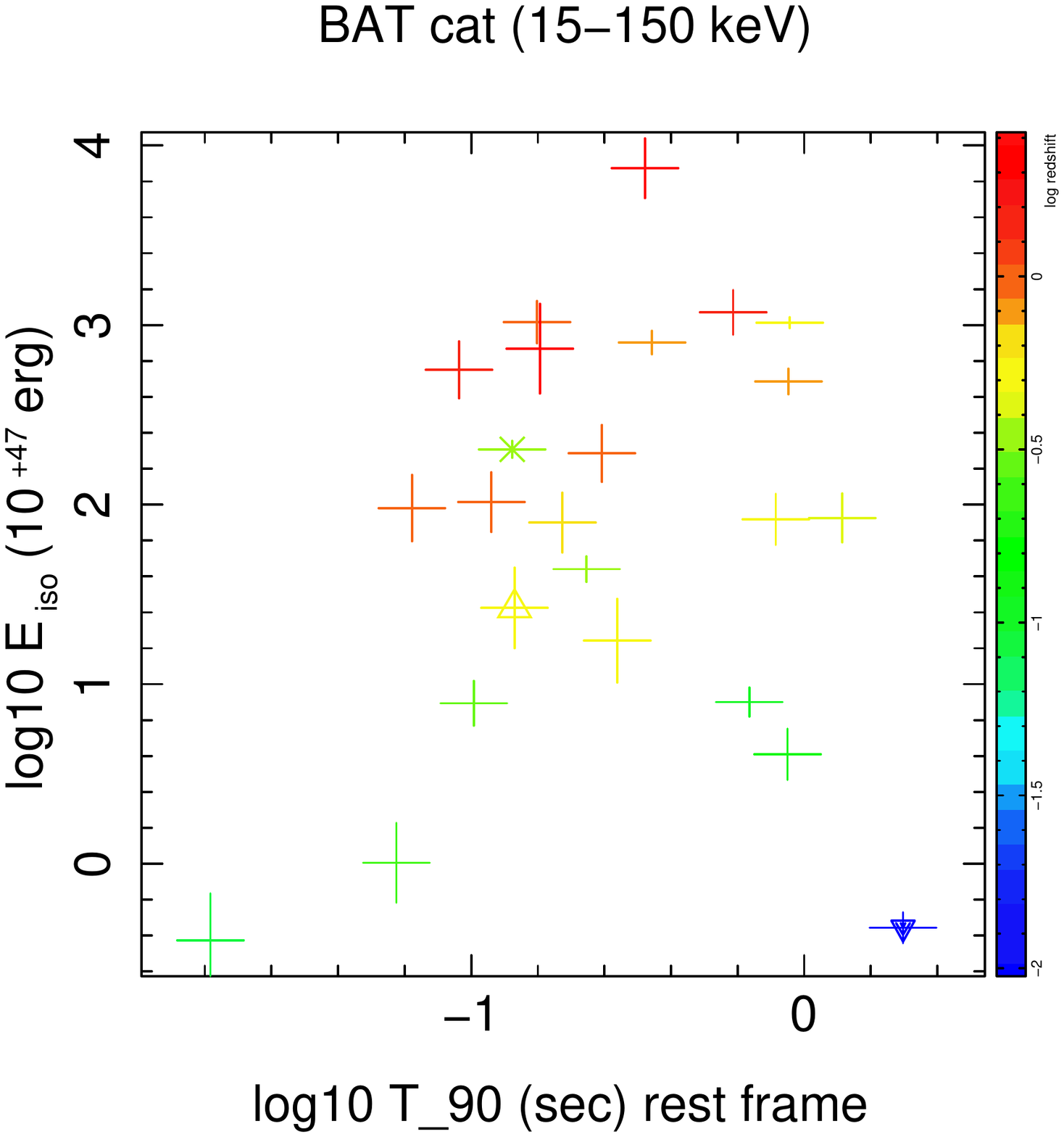} & 
\includegraphics[width=7cm,angle=90]{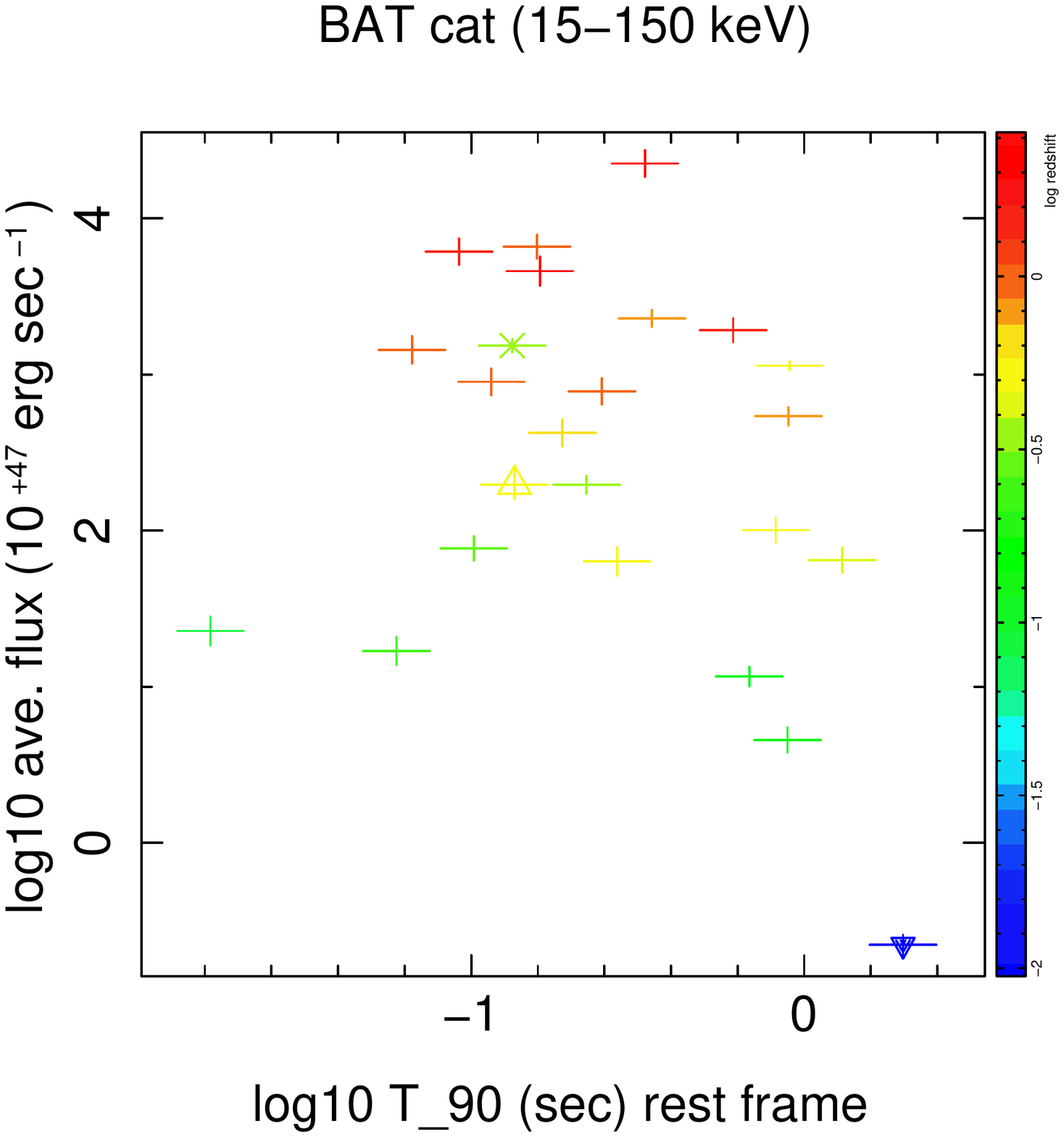}
\end{tabular}
\end{center}
\caption{a) Fluence as a function of $T_{90}$ for short GRB's observed by the Fermi-GBM detector from 
Ref.~\citep{gbmcat2012}; b) Average flux determined by dividing fluence with $T_{90}$ for the same 
data set as in plot a). In these plots GRB 170817A is single out with a square symbol. 
c) $E_{iso}$ of short GRB's with known redshift in the Swift-BAT $15-150$~keV energy band; 
d) Average flux of the same data as in c). In c) and d) redshift is color coded. The data used 
in these plots are taken from the Swift GRB on-line database 
\href{https://swift.gsfc.nasa.gov/archive/grb\_table/}{https://swift.gsfc.nasa.gov/archive/grb\_table/} 
using as selection criteria $T_{90} \leqslant 2$ sec. As GRB 170817A was not in the FoV of the 
Swift-BAT, in c) and d) we have used fluence measured in the Fermi-GBM 10~keV-2~MeV band. Thus, 
$E_{iso}$ and average flux of GRB 170817A shown in these plots are upper limits and shown with an 
inverse triangle as the symbol of upper limit. Star symbol presents kilonova/GRB 130603B and up-right 
triangle shows GRB 160624A at $z=0.483$, the only GRB with known redshift since 01 September 2015, 
considered as the beginning of the Advanced LIGO operation, which its GW could be apriori observed 
if it was at a lower redshift. \label{fig:allsgrb}}
\end{figure}

GRB 170817A was detected by the GBM detector of the Fermi satellite~\citep{gw170817fermi} about 
1.7 sec after the end of gravitational wave event GW 170817. It lasted for about $2$ sec, had an 
integrated fluence in 10~keV to 1~MeV band of 
$(2.8 \pm 0.2) \times 10^{-7}$ erg cm$^{-2}$~\citep{gw170817fermi} 
($(1.4 \pm 0.4 \pm 0.6) \times 10^{-7}$ erg~cm$^{-2}$ in Integral-IBIS 75~keV to 2~MeV 
band~\citep{gw170817integral}), a photon count rate of $3.7 \pm 0.9$ photon~sec$^{-1}$~cm$^{-2}$ 
for 64 msec binning, and its peak energy was $E_{peak} = 229 \pm 78$ keV~\citep{gw170817fermi}. In 
comparison with other short GRBs these characteristics correspond to properties of a mildly faint 
short GRB, see Fig. \ref{fig:allsgrb}-a,b. The follow up of this event by a plethora of ground and 
space based telescopes~\citep{gw170817multimess,gw170817fermimulti} allowed to find the 
optical/IR/radio counterparts AT 2017 gfo, its host galaxy NGC 4993, and thereby its redshift 
$z = 0.0095$ and its distance of $\sim 40$~Mpc\footnote{In this work we use vanilla $\Lambda$CDM 
cosmology with $H_0 = 70$ km sec$^{-1}$~Mpc$^{-1}$, $\Omega_m = 0.3$ and $\Omega_\Lambda = 0.7$.}, making 
GRB 170817A the closest GRB with known distance so far, see e.g.~\citep{sgrbrev} for a review of 
properties of short GRBs and their hosts. Using the distance to the host galaxy, GRB 170817A had an 
isotropic luminosity $E_{iso}\sim 5 \times 10^{46}$ erg in 10 keV to 1 MeV band, which makes it the 
most intrinsically faint short burst with known redshift, see Fig. \ref{fig:allsgrb}-c,d. Moreover, 
the peak energy of the burst is close to lowest peak energy of short bursts observed by Fermi-GBM 
(see Fig. 31 in~\citep{grbgbmspect}). 

Unfortunately at the time of prompt emission GRB 170817A was not in the field of view of Swift-BAT 
and no early follow-up observation is available, except for an upper limit of $> 4$ sigma on any 
excess from background in the time interval $(T+2.944~,~T+100)$ sec, where $T$ is the prompt 
gamma-ray trigger time, in 10~keV to 10~MeV band - assuming $E_{peak} = 128$~keV and a CPL spectrum for 
this time interval - from Konus-Wind satellite~\citep{gw170817konus}. Indeed even in the time interval 
of detection by Fermi-GBM and Integral-IBIS, the burst was too faint for Konus-Wind and only an upper 
limit is reported~\citep{gw170817konus}.

\subsection{X-ray counterpart} \label{sec:xray}
The earliest observation of GW/GRB 170817A in X-ray was at about $T+0.6$~days 
$ = T+51840$~sec~\citep{gw170817swiftnustar}. Nonetheless, from preliminary observations by 
the Swift-XRT in the sky area calculated from gravitational wave observations by the Advanced 
LIGO-Virgo, a flux limit of $\sim 10^{-12}$~erg~sec$^{-1}$~cm$^{-2}$ in 0.3-10~keV band can be put on 
the X-ray afterglow of GRB 170817A around $T+0.2$~days. There is also an upper flux limit of 
$3.5 \times 10^{-15}$ erg~sec$^{-1}$~cm$^{-2}$ at $\sim T+2$ days for any X-ray 
afterglow~\citep{gw170817cxc2day}.

Although the X-ray afterglow of some short GRBs have been brighter than these limits, others, e.g. 
GRB 070724A~\citep{grb070724a,grb070724aopt,grb070724arpoc}, GRB 111020A~\citep{grb111020a}, 
GRB 130912A~\citep{grb130912a,grb130912aopt}, and GRB 160821B~\citep{grb160821b,grb160821bir} had 
smaller fluxes at $\sim 0.3$~days after trigger. GRB 111020A is an interesting case because its host 
galaxy is most probably at redshift 0.02~\citep{grb111020aredshift}. Therefore, its progenitor stars 
might have had properties similar to those of the progenitors of GRB 170817A. It had a total 
gamma-ray fluence about 50\% less than GRB 170817A, but an average flux about 3 time larger than the 
latter. It was observed by the Swift-XRT from $\sim T+100$~sec up to $\sim T+3 \times 10^5$~sec. 
However, its X-ray flux at $T+0.2$ days and $T+0.6$ days was smaller than upper limits reported 
by Swift and NuStar at these epochs for X-ray afterglow of GRB 170817A. Therefore, in contrast to 
suggestions in the literature, in absence of early observations we cannot conclude that GRB 170817A 
was unusually faint in X-ray at early times.

Another evidence for the presence of an early X-ray afterglow is the Swift-UVOT observations at 
$\sim T+0.6$ days. They show a bright UV afterglow at this epoch. Giving the faintness and softness 
of the prompt $\gamma$-ray emission, the early X-ray afterglow of this GRB had to be equally soft 
and quickly decaying. This is consistent with the observation of a relatively bright early UV 
afterglow, which classifies this event as a blue kilonova at early 
times~\citep{gw170817bluekilonova,gw170817bluekilonovapol,gw170817bluekilonovamod}, see also 
Sec. \ref{sec:implic} for more details. 

Evidence of a X-ray counterpart was ultimately observed by Chandra 
observatory~\citep{gw170817cxc,gw170817xray} at $T+9$ days - only a lower limit flux of 
$2.7 \times 10^{-15}$ erg~sec$^{-1}$~cm$^{-2}$ in 0.3-8~keV - and a measurement of slightly brighter 
flux of $3.5 \times 10^{-15}$ erg~sec$^{-1}$~cm$^{-2}$ at $\sim T+16$ days. Similar late time 
brightening in X-ray and optical is observed in some short~\citep{grb130603bxray,grb091109b} and 
long~\citep{grb060712latex,grb060807latex} GRBs. They can be due to: MHD instabilities leading to 
increase in magnetic energy dissipation~\citep{grbjetsimulinstab,grbjetsimulinstab0}; external shock 
generated by the collision of a mildly relativistic thermal cocoon - ejected along with a relativistic 
GRB making jet - with the ISM or circumburst material~\citep{nakarpoynting,cocoon}; or late outflows 
from an accretion disk~\citep{latexrayexcess}. Therefore, it is not certain that late X-ray 
counterpart of GRBs is directly related to the relativistic jet, which is believed to generate the 
prompt gamma-ray, see also the commentary~\citep{latexcomment} about this issue.

For comparison, at $T+9$ days and $T+16$ days the X-ray flux of GRB 111020A in 0.3-10~keV 
band was $\sim 8 \times 10^{-15}$ erg~sec$^{-1}$~cm$^{-2}$ and 
$\sim 5 \times 10^{-15}$ erg~sec$^{-1}$~cm$^{-2}$, respectively. Considering the lower prompt gamma-ray 
flux of GRB 170817A, the above fluxes are proportionally similar to the Chandra observations of 
the late afterglow of GRB 170817A. However, Chandra upper limit at $\sim T+2$ days is much 
smaller than the flux of GRB 111020A at the same epoch. Nonetheless, it is consistent with much 
steeper decay slope observed in other short GRBs, notably kilonova/GRB 130603B. 

In conclusion, taking into account the faintness of the prompt gamma-ray emission of GRB 170817A, 
its X-ray afterglow did not probably seem too unusual if it had been observed a few hundreds 
of seconds after the Fermi-GBM trigger.

\section {Plausibility of an off-axis observation} \label{sec:faintness}
One of the explanation for the faintness of GRB 170817A is its off-axis view. Here we argue that based 
on statistical arguments and properties of other GRBs, it seems unlikely that the weakness of this 
GRB can be fully explained by geometry and viewing angle.

A far observer receives radiation of a relativistic emitter only from a cone around the source velocity 
direction with half-opening angle of $\theta = \sin^{-1} (1/\Gamma) \approx 1/\Gamma$ for 
$\Gamma \gg 1$, where $\Gamma$ is Lorentz factor of emitter in the rest frame of 
observer~\citep{emissionbook}. Lorentz factor of GRB jets are estimated to be 
$\Gamma \sim \mathcal{O}(100)$, see e.g.~\citep{grblorentz,grblorentz0,grblorentz1}. Even in GRB 
models in which the prompt emission is assumed to be produced by a magnetically dominated Poynting 
flow~\citep{poytingflow,critisism,nakarpoynting}, the Lorentz factor must be $\sim \mathcal{O}(10)$. 
Therefore, as long as the opening angle of the jet $\theta_j > [\sin^{-1} (1/\Gamma) \lesssim 6^\circ$ 
for $\Gamma \gtrsim \mathcal{O}(10)]$, the viewing can be considered as on-axis, unless the jet is 
strongly structured and its Lorentz factor at high latitudes is much smaller than on the jet axis.

According to numerical simulations of BNS merger~\citep{nsmergerrprocsimul,nsmergerrprocsimulgw,nsmergerrprocsimulhres,nsmergerrprocsimul0,nsmergerrprocsimul1,nsmergerrprocsimulout} 
the ejecta leading to a relativistic jet is poleward and has a half-opening angle of 
$\lesssim 30^\circ$. This is much larger than minimum $\theta_j$ discussed above. However, it is 
expected that the ultra-relativistic component of ejecta have a smaller opening 
angle~\citep{grbjetsimul1,grbjetsimul}. Therefore, apriori the probability of off-axis view of short 
GRBs is much larger than on-axis. This implies unrealistically large number of compact object mergers, 
necessity for larger emission efficiency, and larger number of short bursts similar to GRB 170817, 
which is in contradiction with observations.

The solution for this conundrum, suggested long time ago, is the precession of compact objects orbits, 
specially during 
merger~\citep{nstarprecessionj,nstarprecession,nstarprecession0,nstarprecession1,nstarbhmergsimul2}. 
In addition, precession of the progenitors of GRBs may explain some of substructures in their light 
curves~\citep{nstarprecessionj,nstarprecession1,hourigrbmag}. Precession frequencies as fast as 
$\mathcal{O}(100)$~Hz are expected during BNS merger~\citep{nstarprecessiont}. Moreover, GRMHD 
simulations of jet formation~\citep{grbjetsimul1,grbjetsimul} show that the maximum Lorentz factor 
is attained in the middle or close to the outer part of jet funnel rather than on its rotation axis. 
Therefore, even in absence of precession in the central object, its rotation alone is enough for 
inducing a precession in the relativistic jet. 

In presence of a precession with maximum angle $\theta_p$, the sky surface covered by the jet is 
$4\pi |\cos (\theta_p+\theta_j) - \cos (\max[0, |\theta_p - \theta_j|])|$ rather than 
$4\pi (1 - \cos \theta_j)$ for a non-precessing jet. This relation is completely geometric and 
independent of the Lorentz factor of the jet, in contrast to opening angle, which may intrinsically 
depend on the Lorentz factor\footnote{Opening and precession angles $\theta_j$ and $\theta_p$, 
respectively, are defined for an observer at the center of progenitor. We remind that a far observer 
cannot measure the opening angle if $\theta_j > \sin^{-1}(1/\Gamma)$ and the jet axis does not 
precess.}. Without precession the probability that the line of sight fall outside the jet funnel is 
$P_{out} = \cos \theta_j > 0.9$ for $\theta_j < 25^\circ$, where as in presence of precession by a 
comparable amount, i.e. $\theta_p \sim 25^\circ$, the probability will be reduced to $P_{out} = 0.36$. 
Thus, in absence of precession, if GW/GRB 170817 were an off-axis event with our line of sight passing 
close to outer boundary of a structured jet or a cocoon with a half-opening angle $\lesssim 25^\circ$, 
statistically speaking LIGO had to have observed $\sim 9$ similar events without a GRB counterpart, 
because in 9 out of 10 events our line of sight would not intercept emission cone and we would not 
receive high energy synchrotron photons. 

When the line of sight is outside the jet funnel, the observer only receive scattered photons, which 
their flux and energy would be much smaller than the primary synchrotron emission. Notably, Compton 
scattering of photons by electrons increases the flux at energies well below the synchrotron peak, 
see simulation of~\citep{hourigrbmag} and Sec. \ref{sec:prompt}. The peak energy of GRB 170817A is 
only $\sim 0.2-0.3$ dex lower than other short GRBs, such as kilonova/GRB 130603B, and it is unlikely 
to be completely due to scattering of primary photons. Indeed some of analyses of late afterglow 
observations rule out off-axis model, see Sec. \ref{sec:evid} for more details.

Obviously, based on the above statistical argument alone and observation of just one GW and GRB from 
a BNS merger it is not possible to rule out an off-axis prompt emission from GRB 170817A. Nonetheless, 
it encourages us to consider the possibility that orders of magnitude faintness of this burst have an 
intrinsic origin.

\section{Prompt emission model} \label{sec:prompt}
To understand properties of a relativistic jet or a cocoon, which might have generated such an 
unusual GRB, we use the phenomenological model and corresponding simulation code described 
in~\citep{hourigrb,hourigrbmag}. In this model the GRB prompt emission is produced by 
synchrotron/self-Compton processes in a dynamically active region at the head front of shocks between 
density shells inside a relativistic cylindrical jet. In addition to the magnetic field generated 
by Fermi processes in the active region, the model and corresponding simulation code can include an 
external magnetic field precessing with respect to the jet axis. The origin of such a 
field is irrelevant for the model. It can be a precessing Poynting flow or the magnetic field of 
a precessing central object, which after releasing the ejecta precesses with respect to the latter. 

An essential aspect of this model, which distinguishes it from other phenomenological GRB 
formulations, is the evolution of parameters with time. In addition, simulation of each burst 
consists of a few time intervals - {\it regimes}. Each regime corresponds to an evolution rule 
(model) for phenomenological quantities such as fraction of energy transferred to fields and its 
variation; variation of the thickness of synchrotron/self-Compton emitting {\it active} region; 
etc. Division of simulated bursts to these intervals allows to change parameters and 
phenomenological evolution rules which are kept constant during one time interval. Continuity 
conditions implemented in the simulation code guarantees the continuity of physical quantities between 
this time intervals, and adjustment of ensemble of parameters and intervals leads to light curves 
and spectra which well reproduce properties of real GRBs. In addition, implementation of some 
of discoveries of Particle In Cell (PIC) simulations, such as the small thickness of layer 
containing high energy electrons responsible for inverse Compton scattering of 
photons~\citep{fermiaccspec,fermiaccspec0,fermiaccspec1,picsimul}, lead to more realistic 
simulations. 

Table \ref{tab:paramdef} summarizes parameters of this model. Despite their long list, simulations of 
typical long and short GRBs in~\citep{hourigrbmag} show that the range of values which lead to 
realistic bursts are fairly restricted. In this section we use this model to simulate the prompt 
emission of GRB 170817A and compare its properties with those of other short bursts, in particular 
GRB 130603B, which thanks to its brightness is extensively observed~\citep{grb130603b} and classified 
as a kilonova~\citep{grb130603bkilonova,grb130603bkilonova0}. 

\begin{table}
\begin{center}
\caption{Parameters of the phenomenological prompt model \label{tab:paramdef}}
\vspace{0.5cm}
\begin{tabular}{| p{2.5cm} | p{12cm} |}
\hline
Model (mod.) & Model for evolution of active region with distance from central engine; See 
Appendix \ref{app:modes} and~\citep{hourigrb,hourigrbmag} for more details. \\
$r_0$ (cm) & Initial distance of shock front from central engine. \\
$\Delta r_0$ & Initial (or final, depending on the model) thickness of active region. \\
$p$ & Slope of power-law spectrum for accelerated electrons; See eq. (3.8) of~\citep{hourigrbmag}. \\
$p_1,~p_2$ & Slopes of double power-law spectrum for accelerated electrons; See eq. (3.14) 
of~\citep{hourigrbmag}. \\
$\gamma_{cut}$ & Cut-off Lorentz factor in power-law with exponential cutoff spectrum for 
accelerated electrons; See eq. (3.11) of~\citep{hourigrbmag}. \\
$\gamma'_0$ & Initial Lorentz factor of fast shell with respect to slow shell. \\
$\tau$ & Index in the model defined in eq. (3.28) of~\citep{hourigrbmag}. \\
$\delta$ & Index in the model defined in eq. (3.29) of~\citep{hourigrbmag}. \\
$Y_e$ & Electron yield defined as the ratio of electron (or proton) number density to baryon number 
density. \\
$\epsilon_e$ & Fraction of the kinetic energy of falling baryons of fast shell 
transferred to leptons in the slow shell (defined in the slow shell frame). \\
$\alpha_e$ & Power index of $\epsilon_e$ as a function of $r$. \\
$\epsilon_B$ & Fraction of baryons kinetic energy transferred to induced magnetic field in 
the active region. \\
$\alpha_B$ & Power index of $\epsilon_B$ as a function of $r$. \\
$N'$ & Baryon number density of slow shell. \\
$\kappa$ & Power-law index for N' dependence on $r'$. \\
$n'_c$ & Column density of fast shell at $r'_0$. \\
$\Gamma$ & Lorentz factor of slow shell with respect to far observer. \\
$|B|$ & Magnetic flux at $r_0$. \\
$f$ & Precession frequency of external field with respect to the jet.\\ 
$\alpha_x$ & Power-law index of external magnetic field as a function of $r$. \\
$\phi$ & Initial phase of precession, see ~\citep{hourigrbmag} for full description. \\
\hline
\end{tabular}
\end{center}
{\small
\begin{description}
\item{$\star$} The phenomenological model discussed in~\citep{hourigrb} and its 
simulation~\citep{hourigrbmag} depends only on the combination $Y_e\epsilon_e$. For this reason 
only the value of this combination is given for simulations.
\item{$\star$} The model neglects variation of physical properties along the jet or active region. 
They only depend on the average distance from center $r$, that is $r-r_0 \propto t-t_0$.
\item{$\star$} Quantities with prime are defined with respect to rest frame of slow shell, and 
without prime with respect to central object, which is assumed to be at rest with respect to 
a far observer. Power indices do not follow this rule.
\end{description}
}
\end{table}

\subsection{Parameter selection} \label{sec:paramselect}
Because of large number of parameters in the phenomenological model, in order to find best fits 
to the data we restricted our search in the parameter space to most important characteristics, 
namely: $r_0$, $p$, $\gamma_{cut}$, $ \gamma'_0$, $\Gamma$, $Y_e\epsilon_e$, $\epsilon_B$, $N'$, 
$n'_c$ and $|B|$. Other parameters are fixed to values suitable for simulation of short GRBs 
with more typical characteristics, see~\citep{hourigrbmag} for some examples. 

Beginning with a choice for $\Gamma$, which determines kinematic of the ejecta, we changed the 
value of other parameters such that an acceptable fit to the data be found. We divide 
simulations according to the initial Lorentz factor of slow shell (bulk) $\Gamma$ to 3 categories: 
\begin{description}
\item {\bf On-axis ultra-relativistic jet} with $\Gamma \sim \mathcal{O}(100)$~\citep{grblorentz,grblorentz0,grblorentz1}, see also simulations in~\citep{hourigrbmag}; 
\item {\bf Off-axis structured relativistic jet} with $\Gamma \sim \mathcal{O}(10)$~\citep{gw170817xray,gw170817rprocess};
\item {\bf Mildly relativistic cocoon} with $\Gamma \sim \mathcal{O}(1)$~\citep{grbcocoon,{nakarpoynting},gw170817cocoon,gw170817cocoon0,{grboffaxprobab}}. 
\end{description}
We remind that here {\it cocoon} means a mildly relativistic, mildly collimated outflow with a 
Lorentz factor of $\sim 2-3$. It should not be confused with cocoon breakout model. We should also 
emphasize that as the exploration of parameter space was not systematic, the value of parameters in 
models with best fits to the data should be considered as approximation rather than exact. Another 
important issue, specially when considering the best models, is the fact that parameters of the 
phenomenological model studied in~\citep{hourigrb} are not completely independent. For instance, 
based on physical principles it is expected that fraction of kinetic energy transferred to induced 
electric and magnetic fields depend on the strength of the shock, which is determined by the density 
difference of colliding shells and their relative Lorentz factor. But there is no simple formulation 
for these dependencies, and they could not be considered in the phenomenological model. We leave 
further discussion of this issue to the next section, where we assess plausibility of selected 
simulations.

\subsection{Simulation of GRB 170817A} \label{sec:gwsimul}
\begin{figure}
\begin{center}
\begin{tabular}{p{8.6cm}p{4cm}}
\includegraphics[width=12cm]{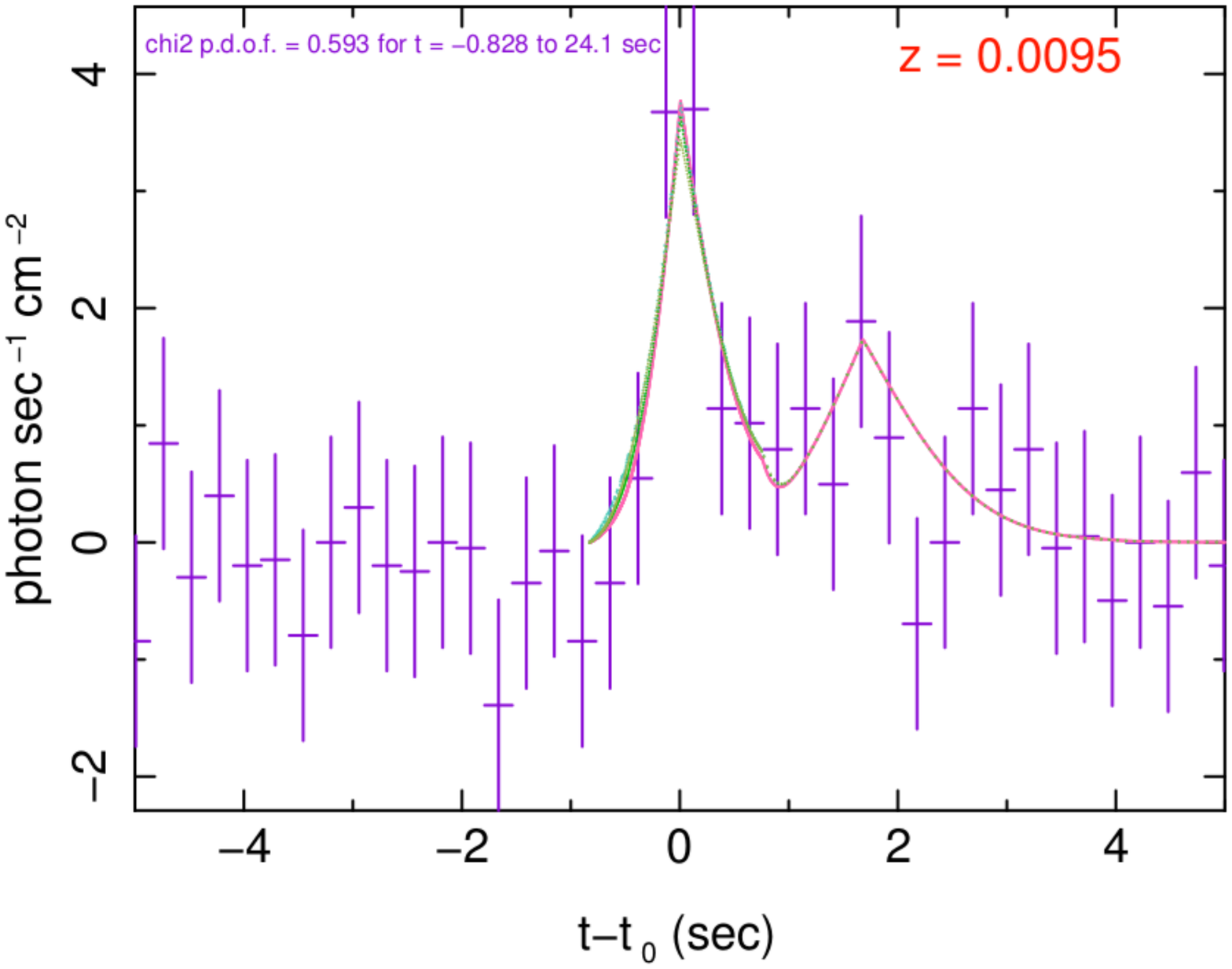} & \vspace{-6cm}\includegraphics[width=4cm]{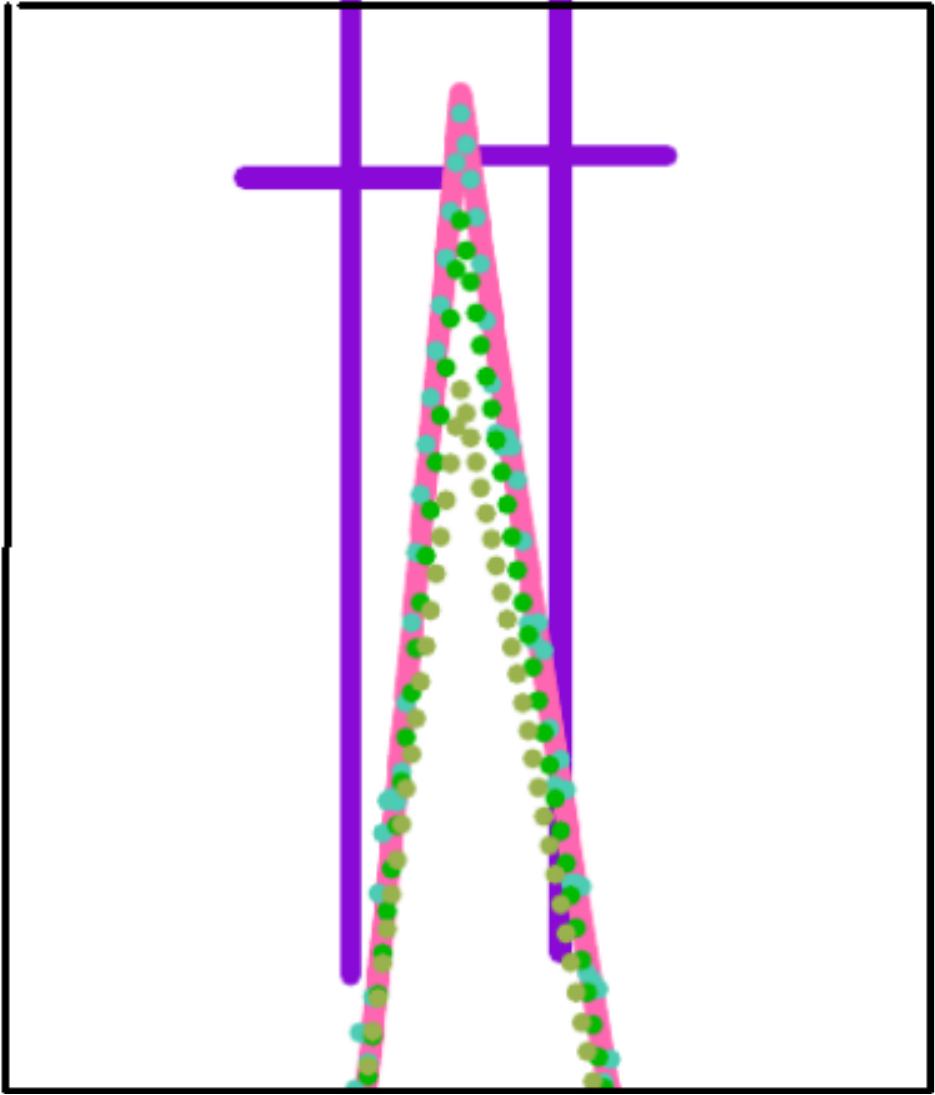}
\end{tabular}
\end{center}
\caption{Light curves of the 4 best simulations in 10 keV - 1 MeV. The data is from observations 
of Fermi-GBM~\citep{gw170817fermi}. This plot shows that these simulations have very similar light 
curves. The inset is a zoom on the first peak and shows the slight difference of the amplitude of 
the first peak in these models. The value of $\chi^2$ is for the full line corresponding to 
model No. 2 in Table \ref{tab:param} for the first peak and model No. 3 for the second peak. 
Other curves (doted lines) correspond to model No. 1 with and without an external magnetic field 
(blue and dark green curves, respectively), and an off-axis model with all parameters the same as 
model No. 2, except column density of ejecta which is $n'_c = 5 \times 10^{25}$ cm$^{-2}$ 
(light green). The value of $\chi^2$ per degree of freedom of the first two simulations are about 
$0.02$ larger than model No. 2 and that of the last model is $\sim 0.03$ larger. \label{fig:totlc}}
\end{figure}

\begin{figure}
\begin{center}
\begin{tabular}{p{6cm}p{6cm}p{6cm}}
a) & b) & c) \\
\includegraphics[width=8cm]{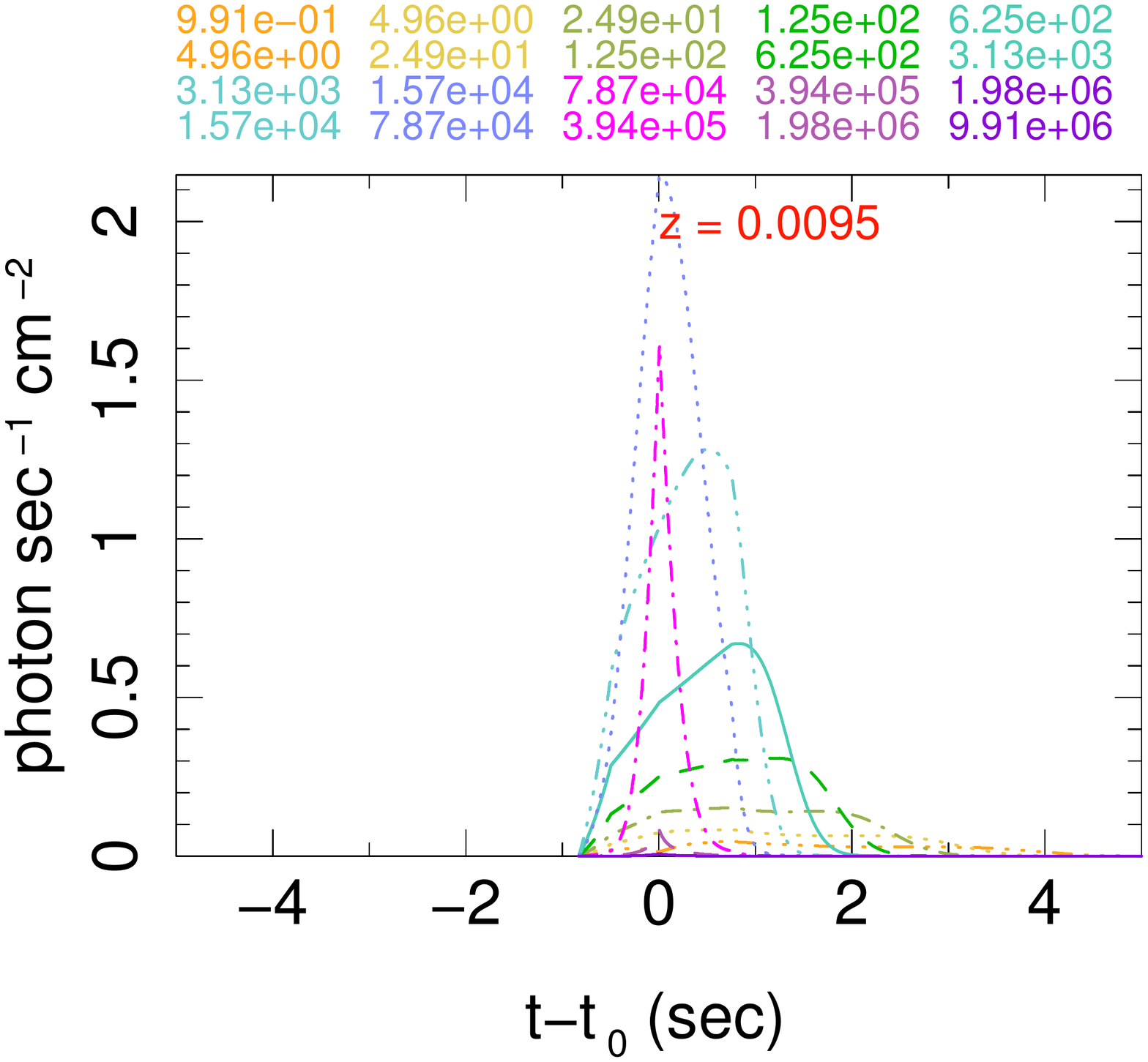} & 
\hspace{-0.5cm} \includegraphics[width=8cm]{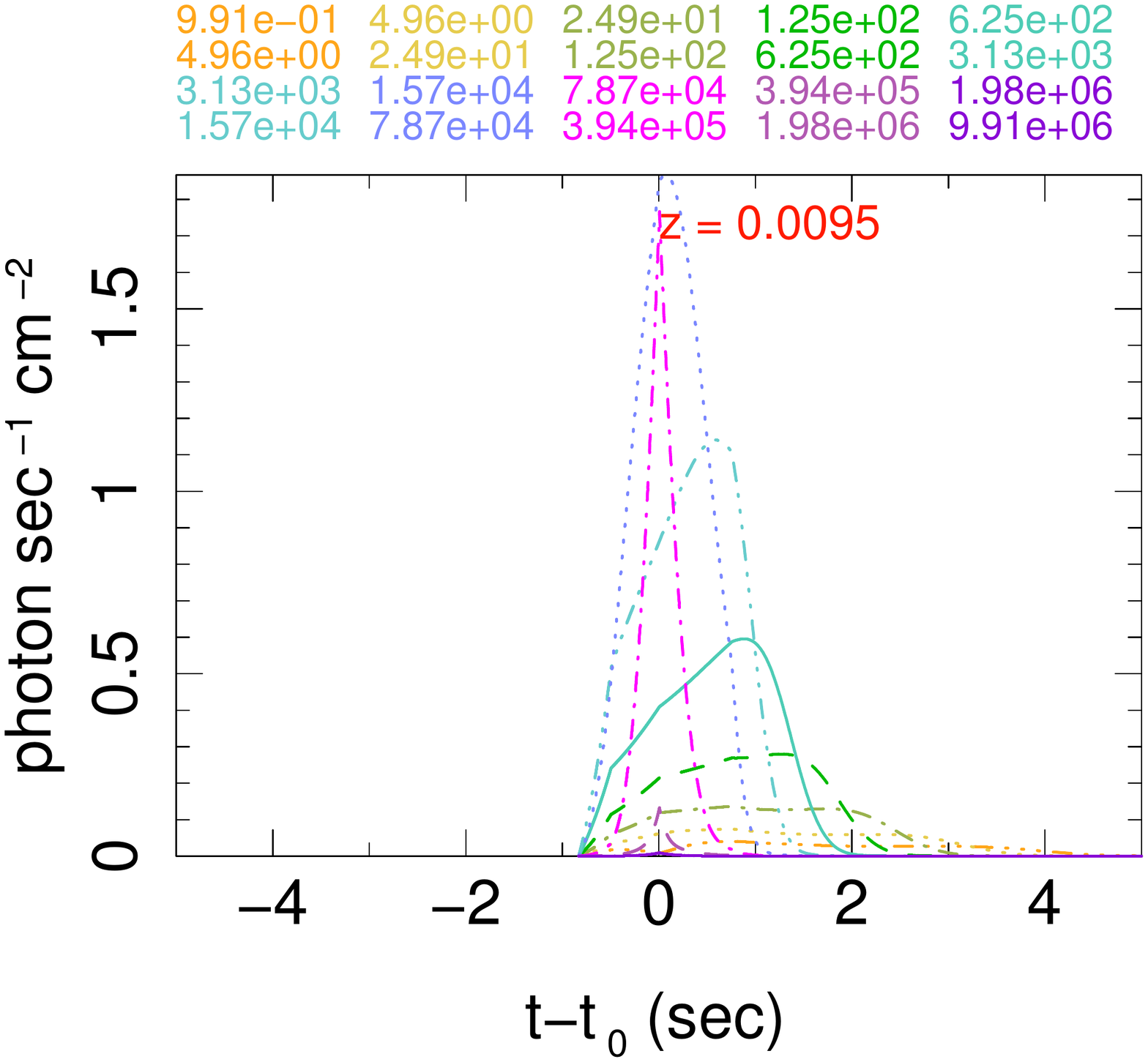} &
\hspace{-1cm} \includegraphics[width=8cm]{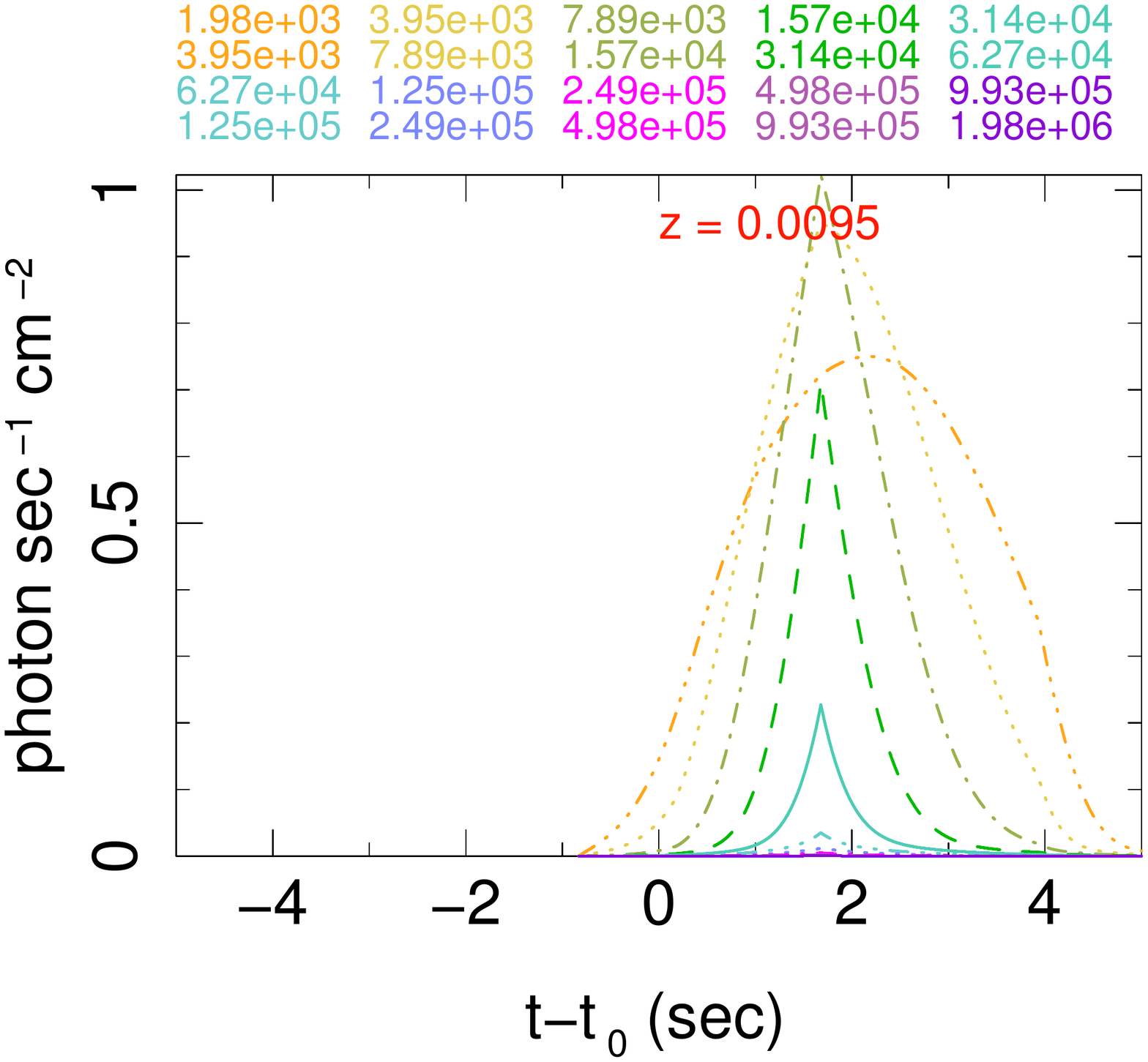}
\end{tabular}
\end{center}
\caption{Light curves of simulated models in energy bands covered by Fermi-GBM and Integral 
SPI-ACS instruments: a) Simulation No. 2; b) Simulation No. 1 without external magnetic field; 
c) Second peak, that is simulation No. 3. All simulation numbers refer to Table \ref{tab:param}. 
Minimum and maximum of each energy band in eV is written in the corresponding color on the top 
of each plot. Notice that the second peak is simulated in lower energy bands than the first 
peak. The lag between highest energy bands is roughly zero and consistent with observation of 
short GRBs.\label{fig:lcbands}}
\end{figure}

Fig. \ref{fig:totlc} shows light curves of the 4 best simulated bursts according to their 
chi-square fit in 10 keV-1 MeV band along with the Fermi-GBM data. The two peaks in the 
observed light curve are simulated separately and adjusted in time such that the sum of two peaks 
minimize $\chi^2$-fit to the data. Fig.\ref{fig:lcbands} shows light curves in narrower bands 
for each peak. Table \ref{tab:param} shows the value of parameters for these simulations. 
Table \ref{tab:notgood} shows the value of parameters we have explored to find best fits to the 
data. They correspond to simulations which do not fit the data well. The last column of this table 
describes their deficiencies and Fig. \ref{fig:notgood} shows light curves and spectra of a sample 
of them. We use these results here and in Sec. \ref{sec:interp} to assess how variation of 
parameters affects properties of simulated GRBs, and to which extent parameters are degenerate. 

\begin{figure}
\begin{center}
\begin{tabular}{p{8cm}p{8cm}}
a) & b) \\
\vspace{-1cm} \includegraphics[width=8cm]{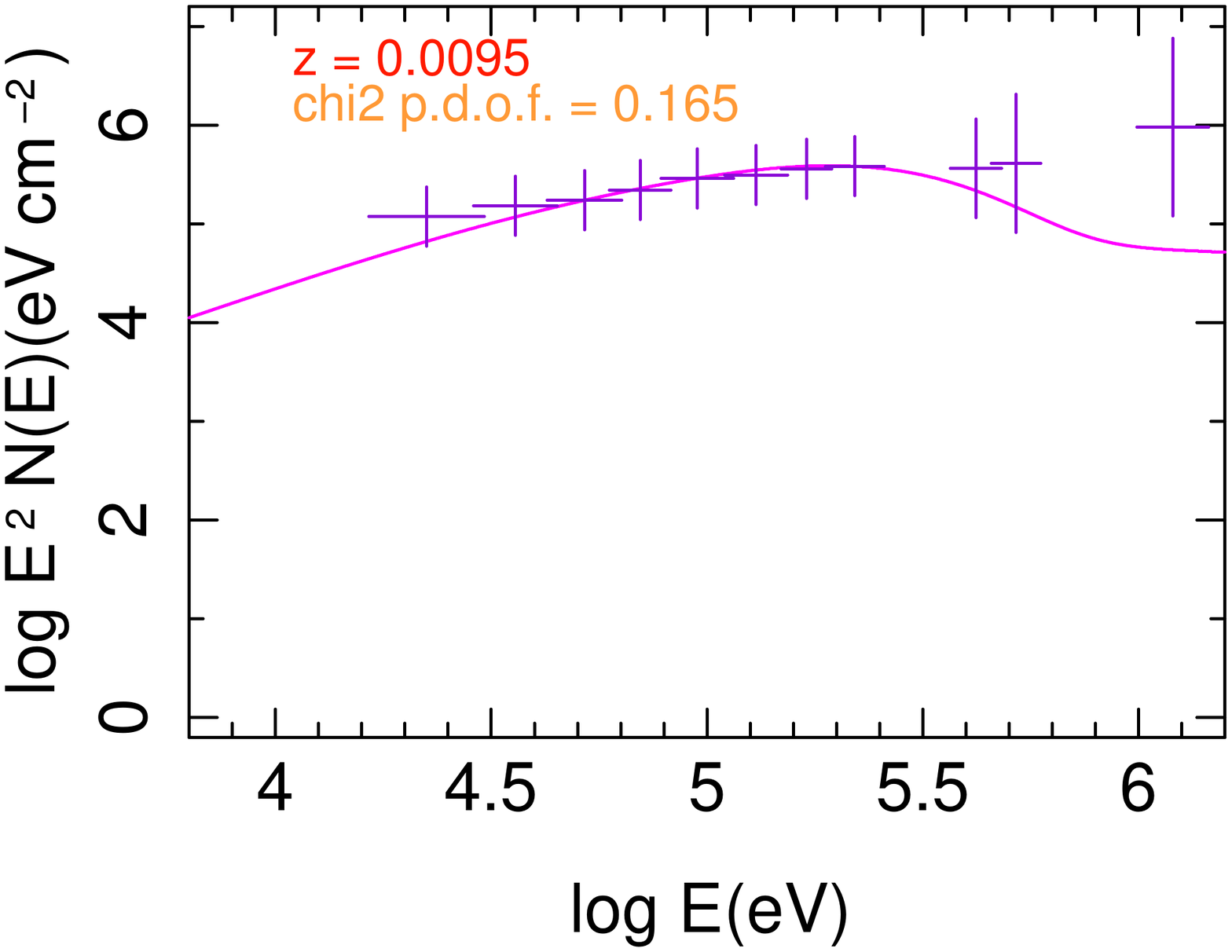} & 
\vspace{-1cm} \includegraphics[width=8cm]{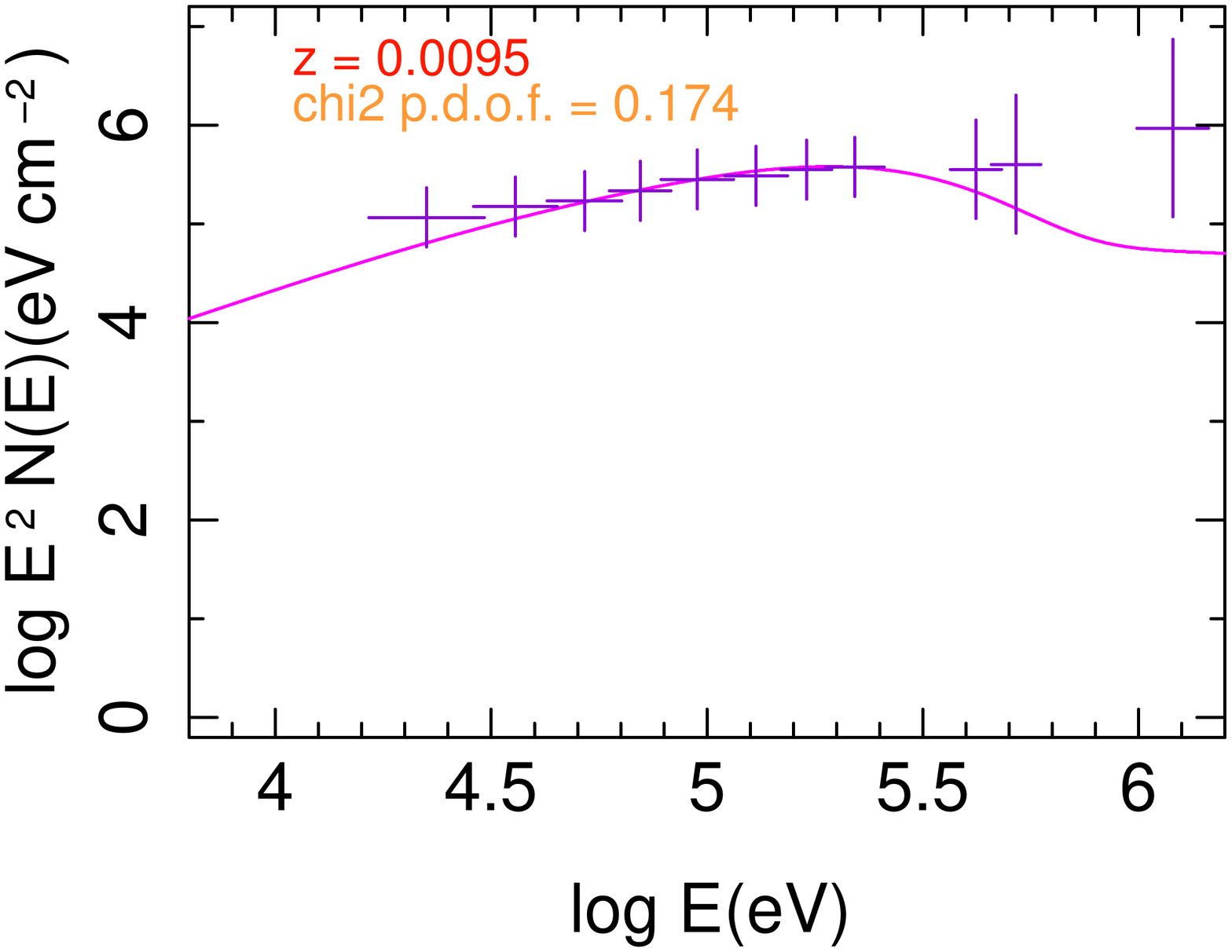} \\
c) & d) \\
\vspace{-1cm} \includegraphics[width=8cm]{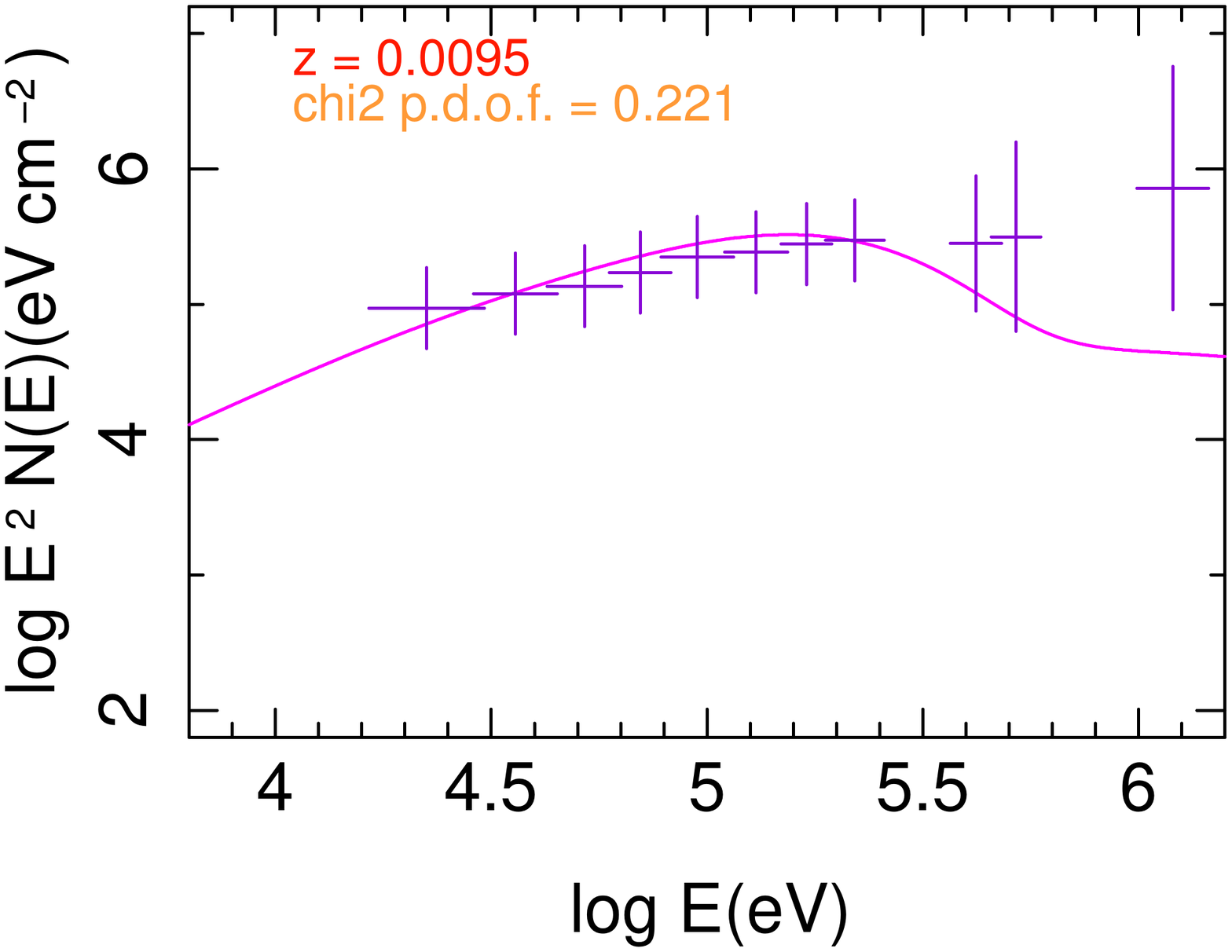} &
\vspace{-1cm} \includegraphics[width=8cm]{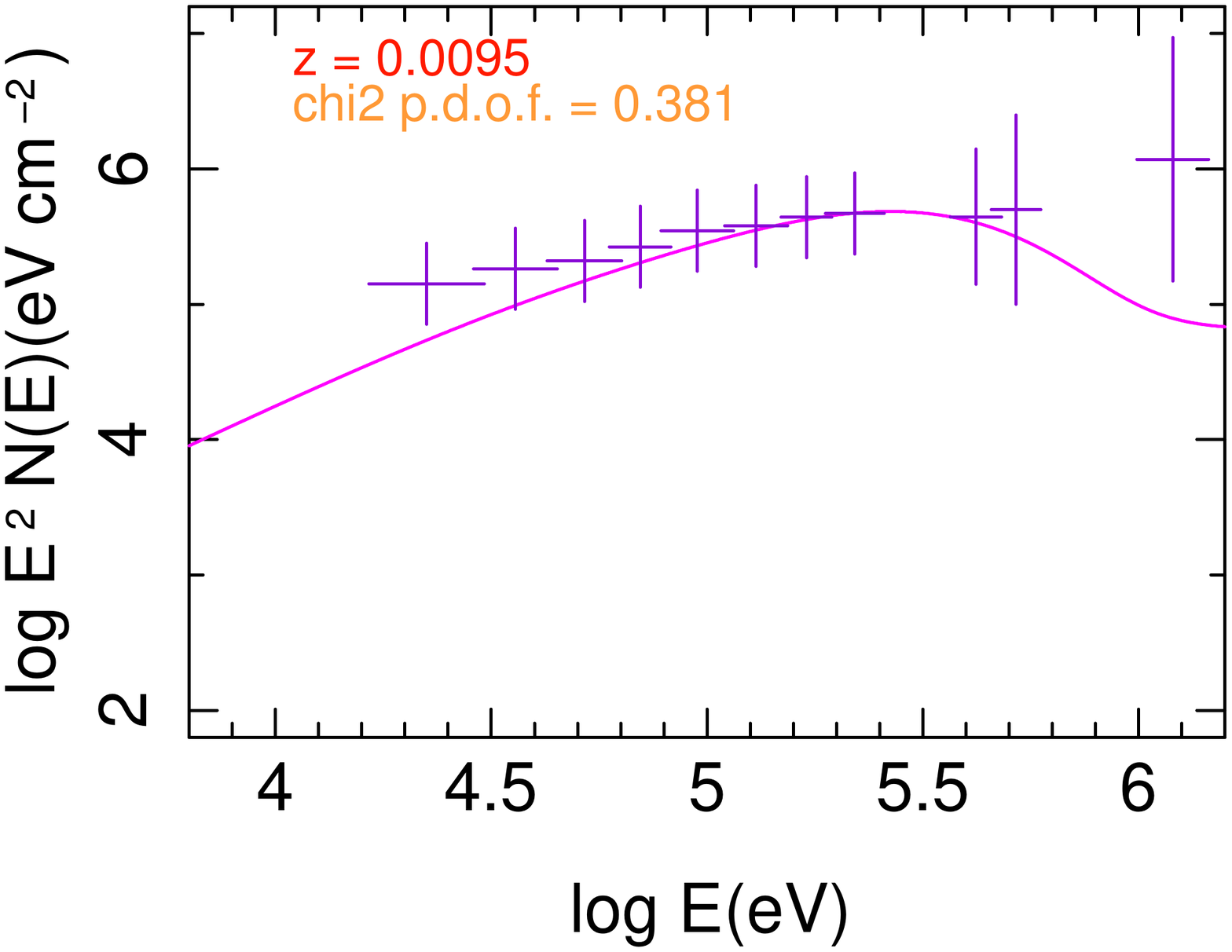}
\end{tabular}
\end{center}
\caption{Spectra of simulated models fitted to Fermi-GBM data: a) Model No. 1; b) Model No. 1 
without external magnetic field; c) Model No. 2; d) A model with the same parameters as models No. 2 
except for $n'_c = 5 \times 10^{25}$ cm$^{-2}$. As the published spectral data in~\citep{gw170817fermi} 
is in count rate, after changing it to energy flux we used peak energy from~\citep{gw170817fermi} to 
normalize data such that at $E = E_{peak} = 215 \pm 54$~keV observed and simulated spectra have the 
same amplitude. For this reason, spectra of simulated models have much smaller $\chi^2$ than their 
corresponding light curves. The data point with highest energy and uncertainty is considered to be an 
outlier and is not included in the calculation of $\chi^2$ because it may affect fitting and lead to 
large deviation from true model, see e.g.~\citep{statoutliers}. \label{fig:spect}}
\end{figure}

\begin{figure}
\begin{center}
\begin{tabular}{p{6cm}p{6cm}p{6cm}}
\vspace{-1cm}\includegraphics[width=6cm]{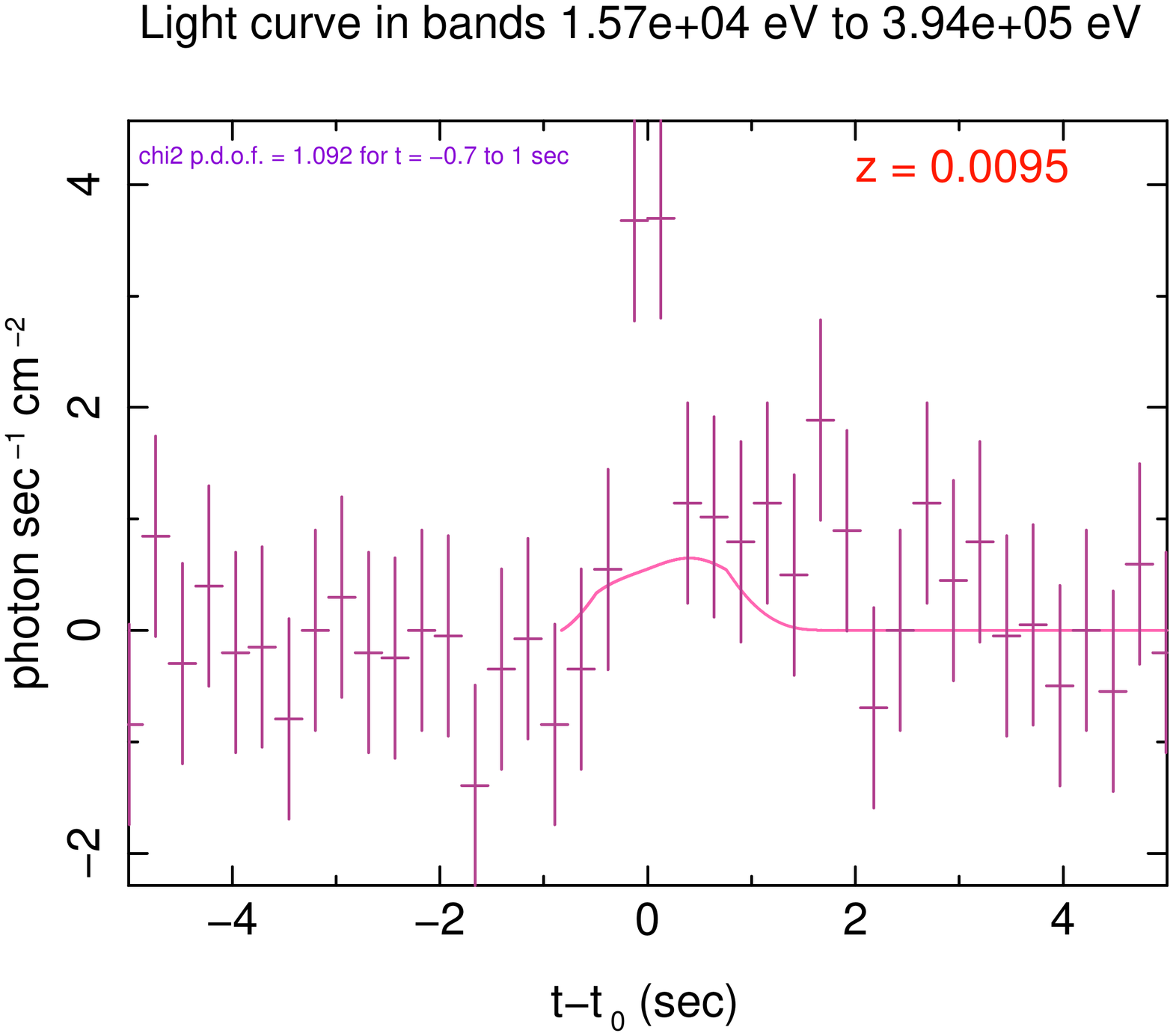} & 
\vspace{-1cm}\includegraphics[width=6cm]{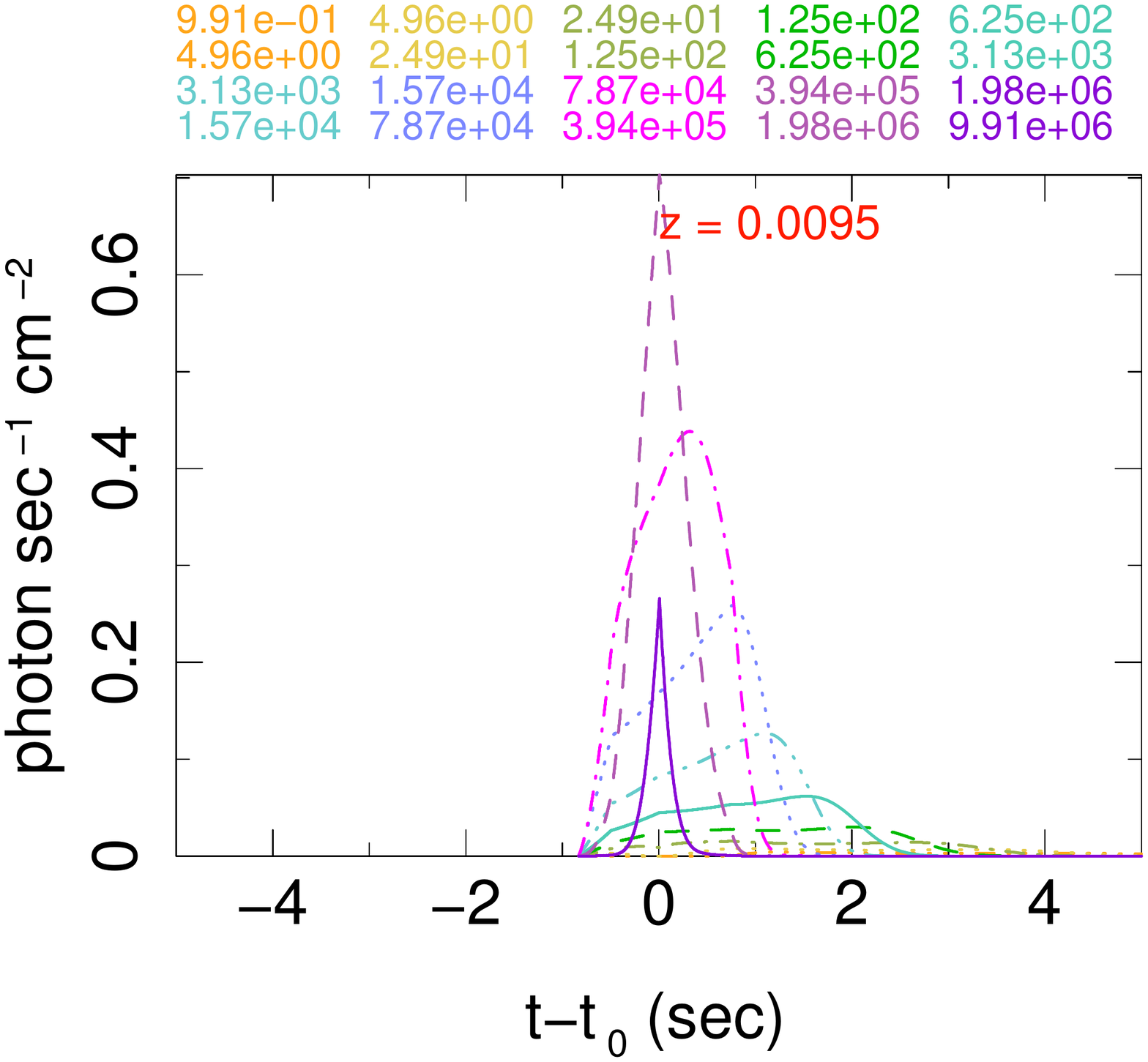} & 
\vspace{-1cm}\includegraphics[width=6cm]{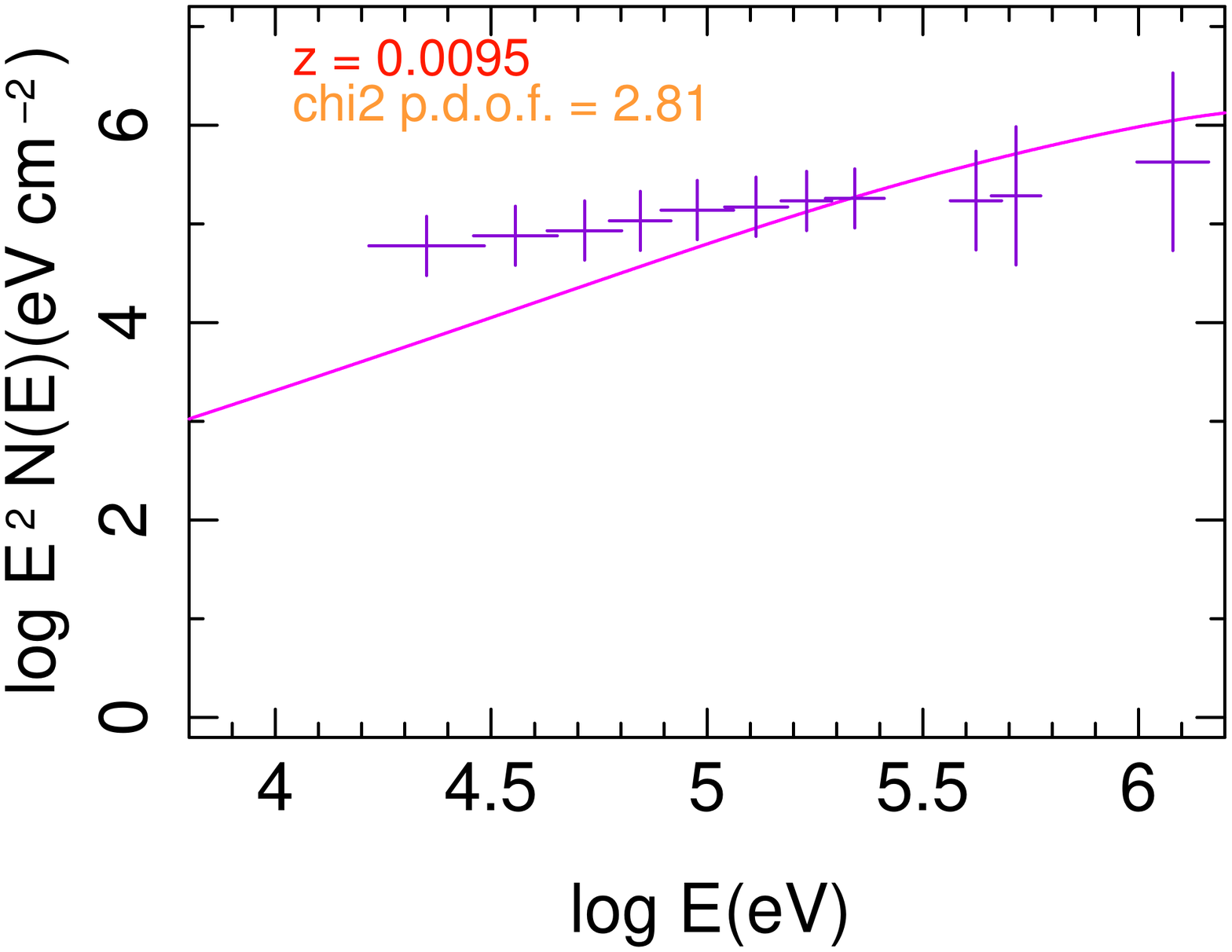} \\
\vspace{-0.7cm}\includegraphics[width=6cm]{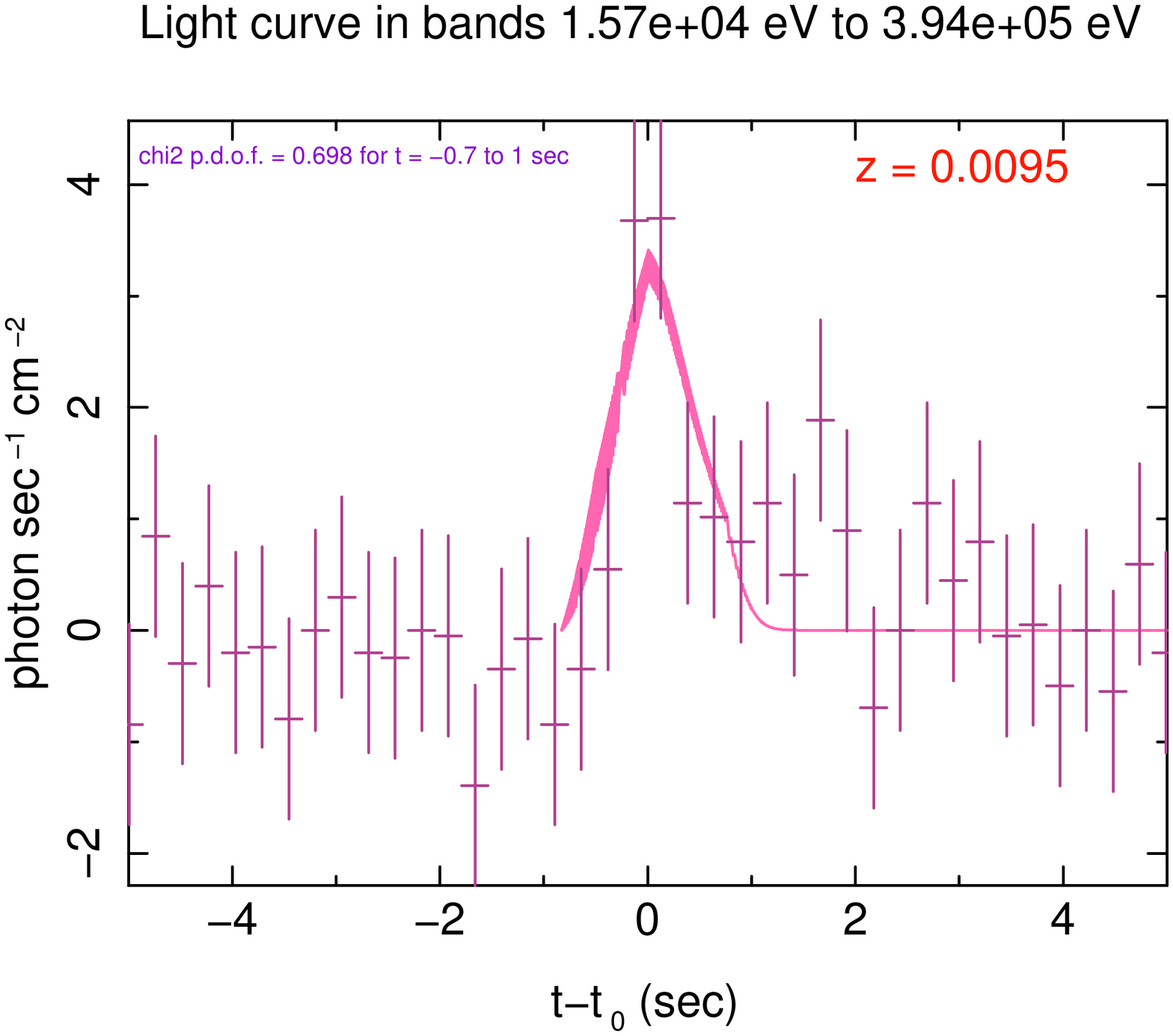} &
\vspace{-0.7cm}\includegraphics[width=6cm]{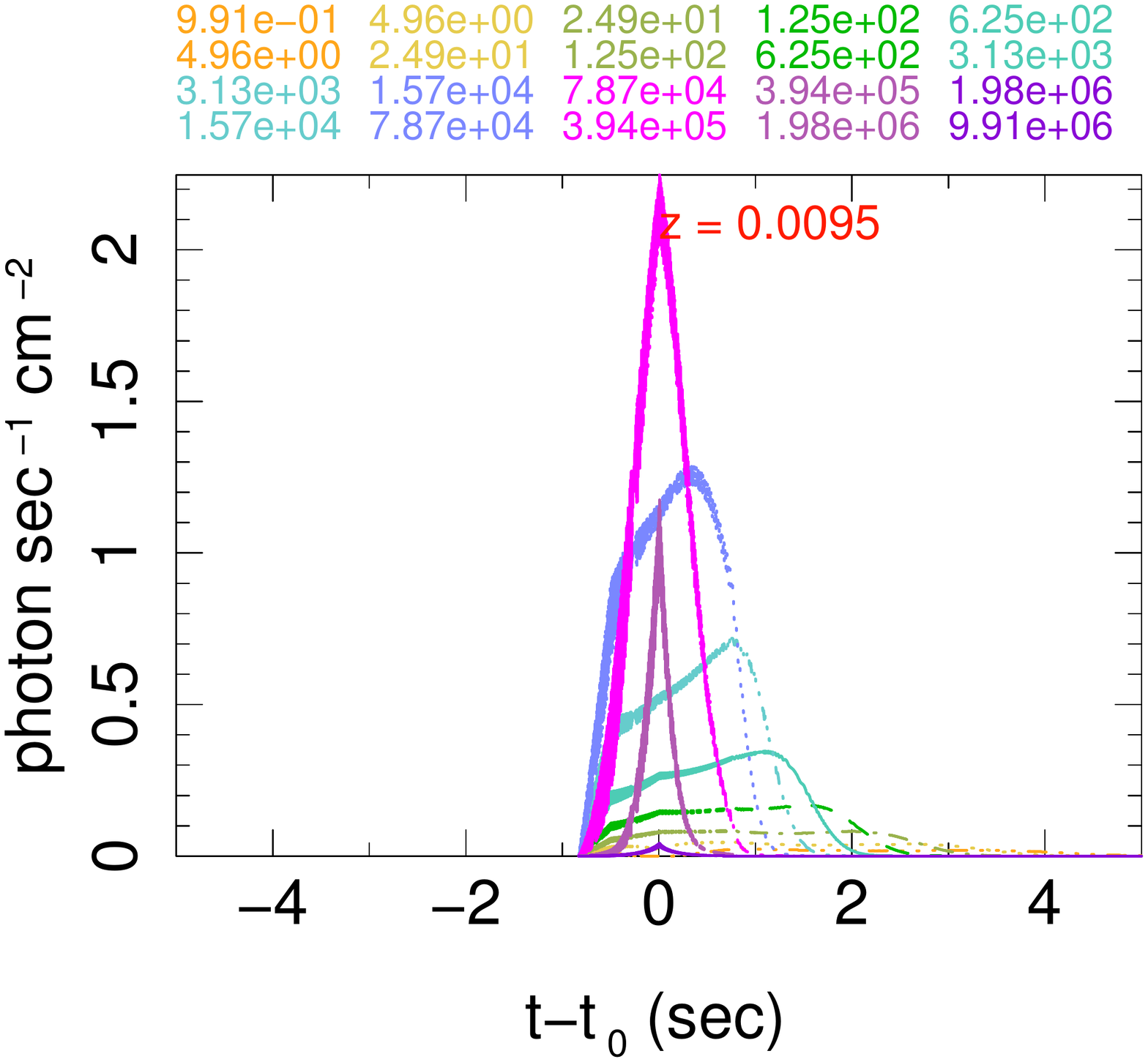} &
\vspace{-0.7cm}\includegraphics[width=6cm]{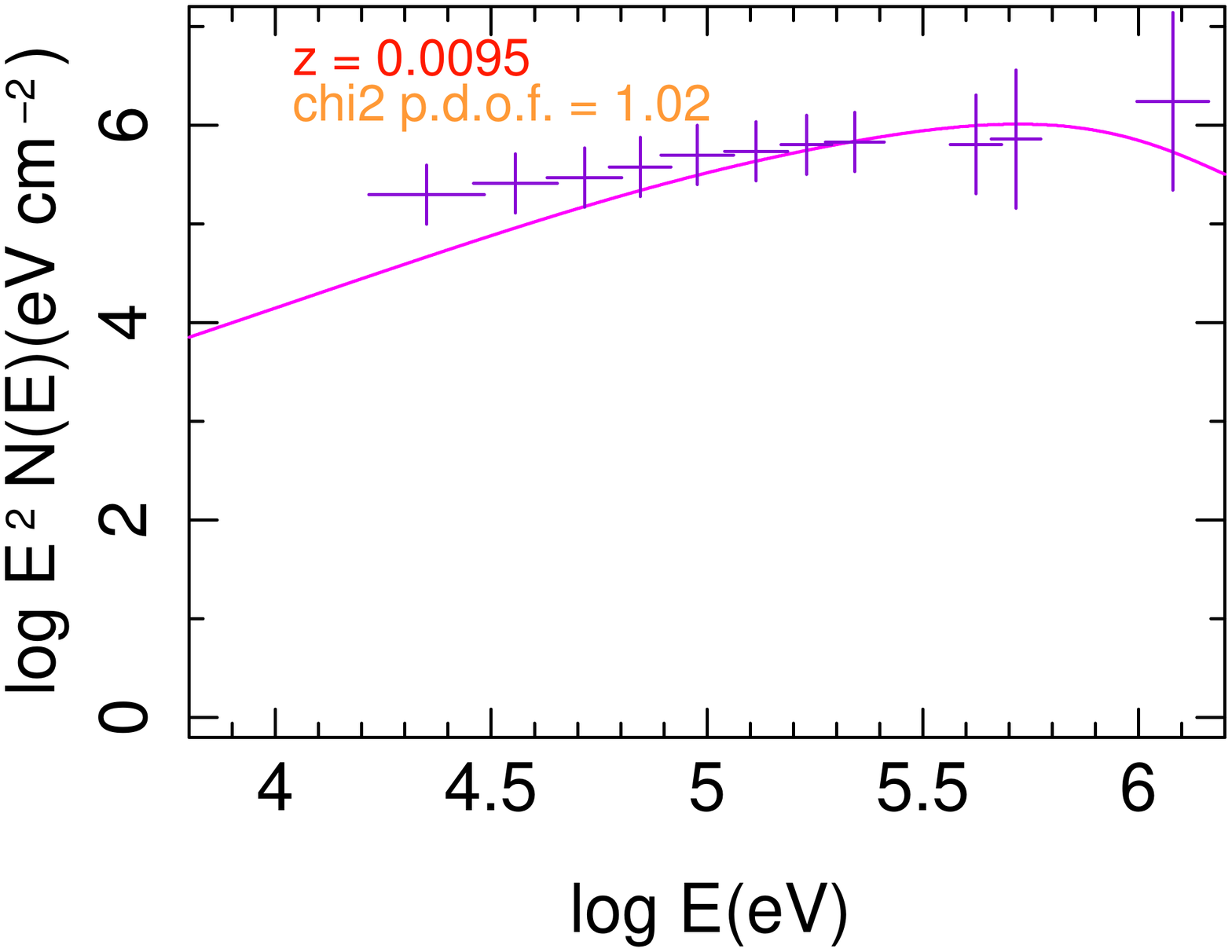} \\
\vspace{-0.7cm}\includegraphics[width=6cm]{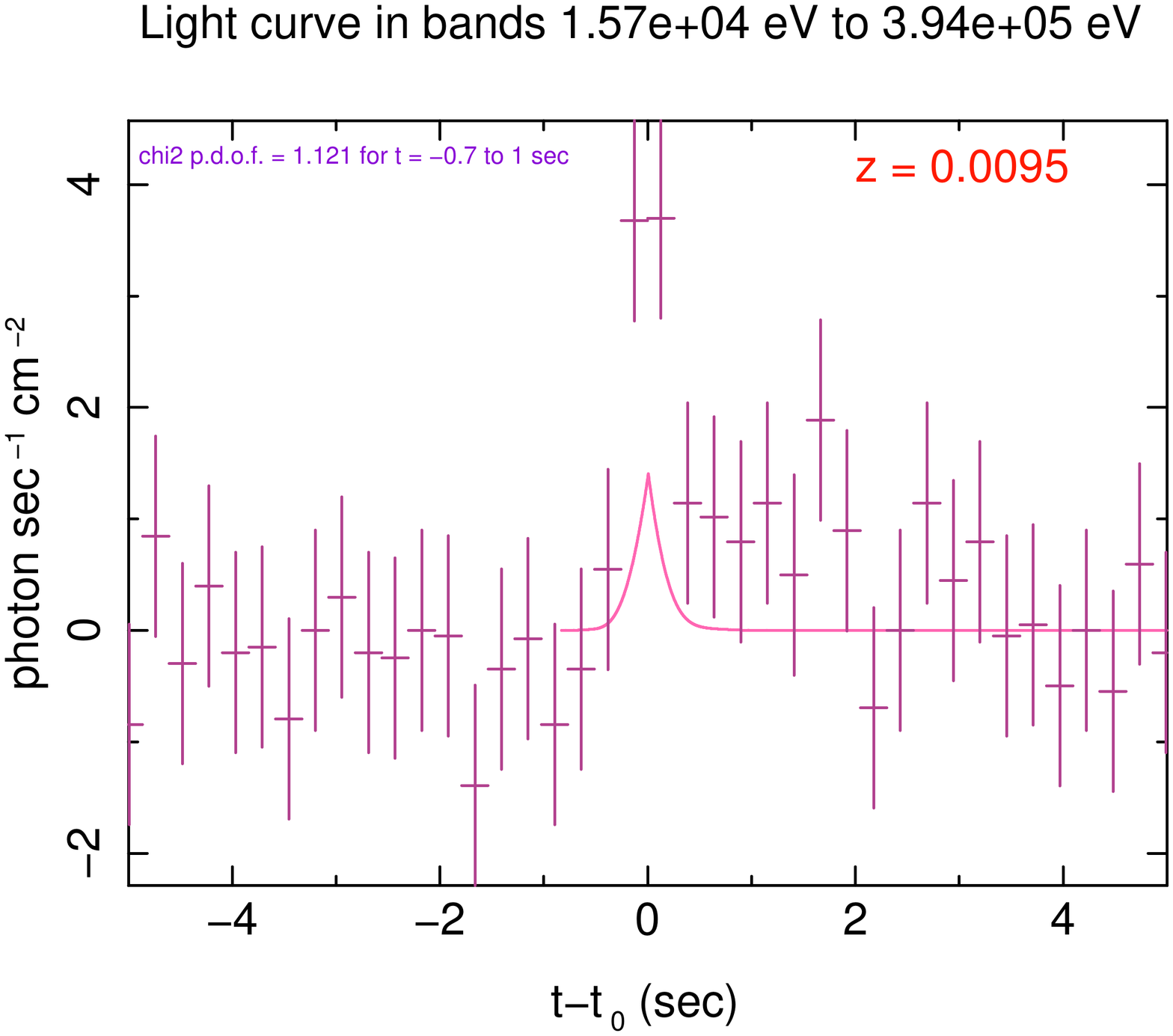} &
\vspace{-0.7cm}\includegraphics[width=6cm]{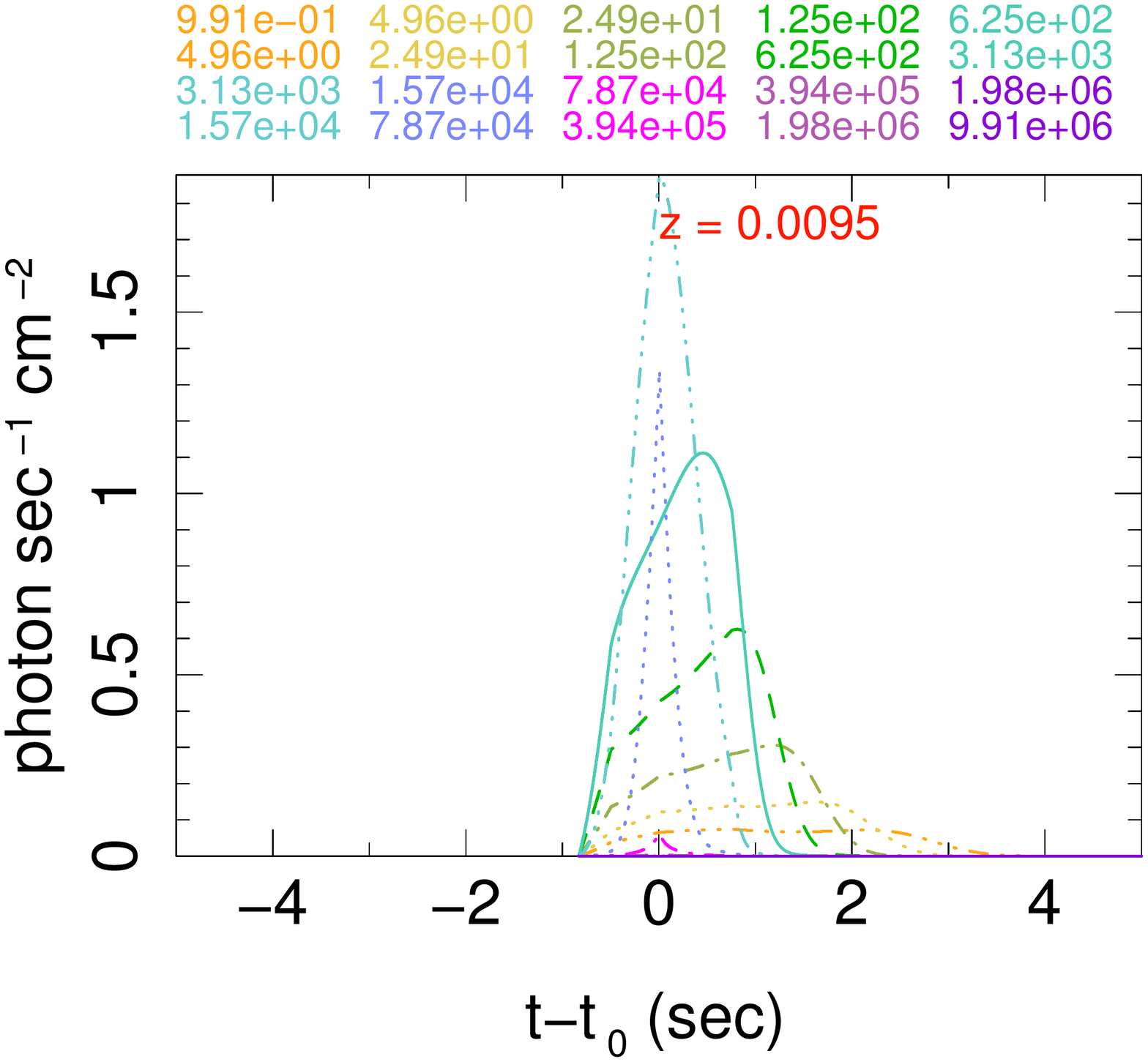} &
\vspace{-0.7cm}\includegraphics[width=6cm]{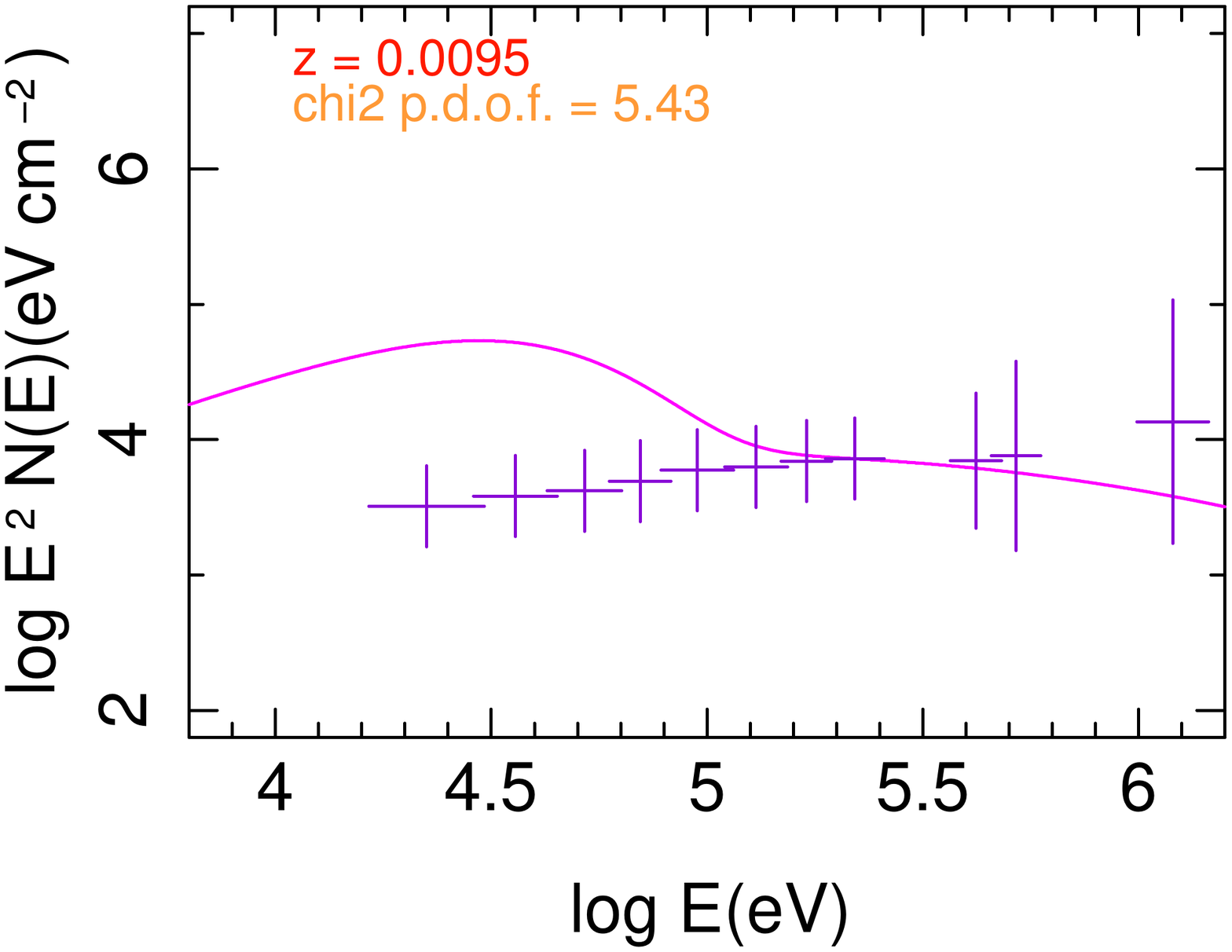} \\
\vspace{-0.7cm}\includegraphics[width=6cm]{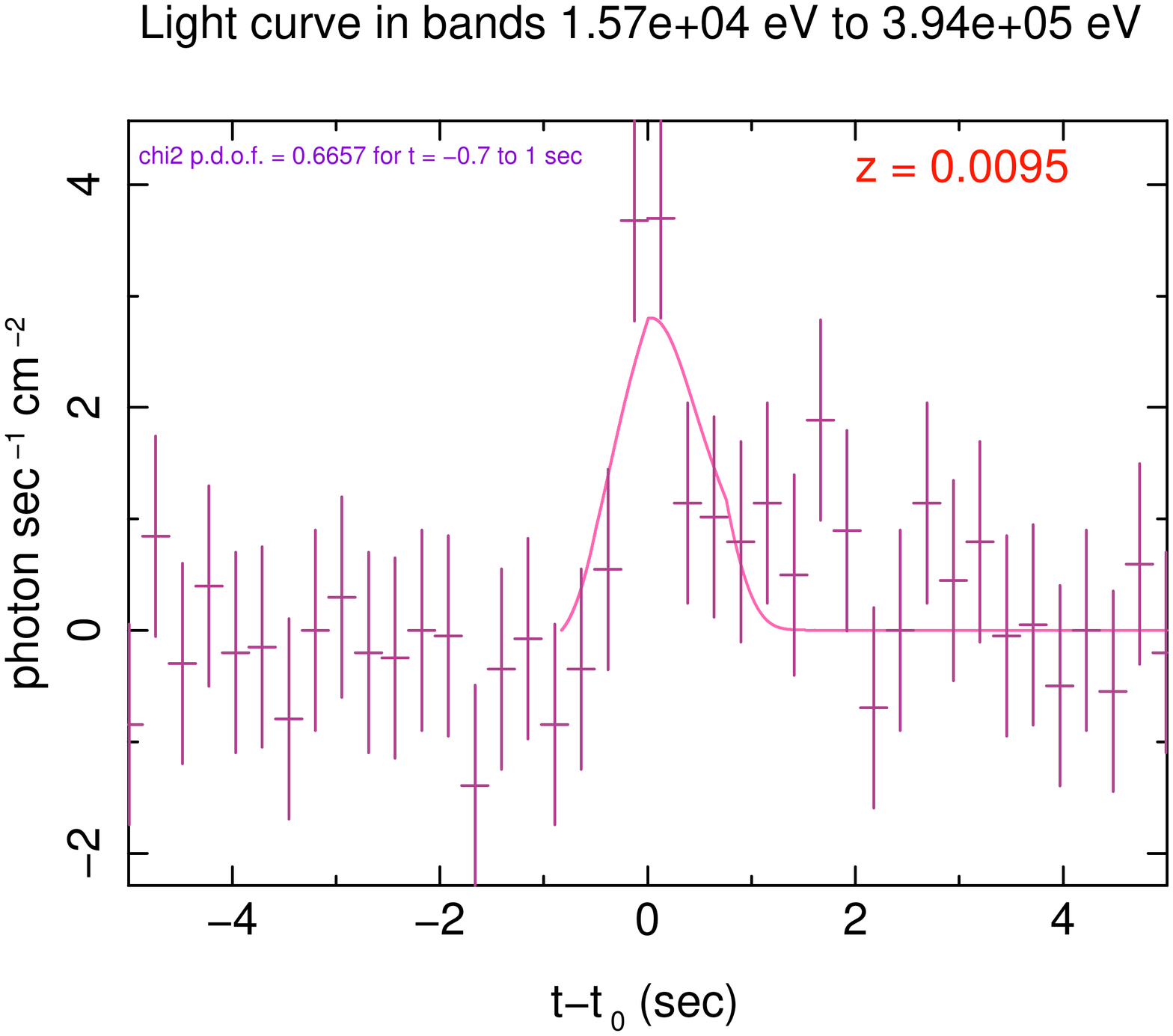} &
\vspace{-0.7cm}\includegraphics[width=6cm]{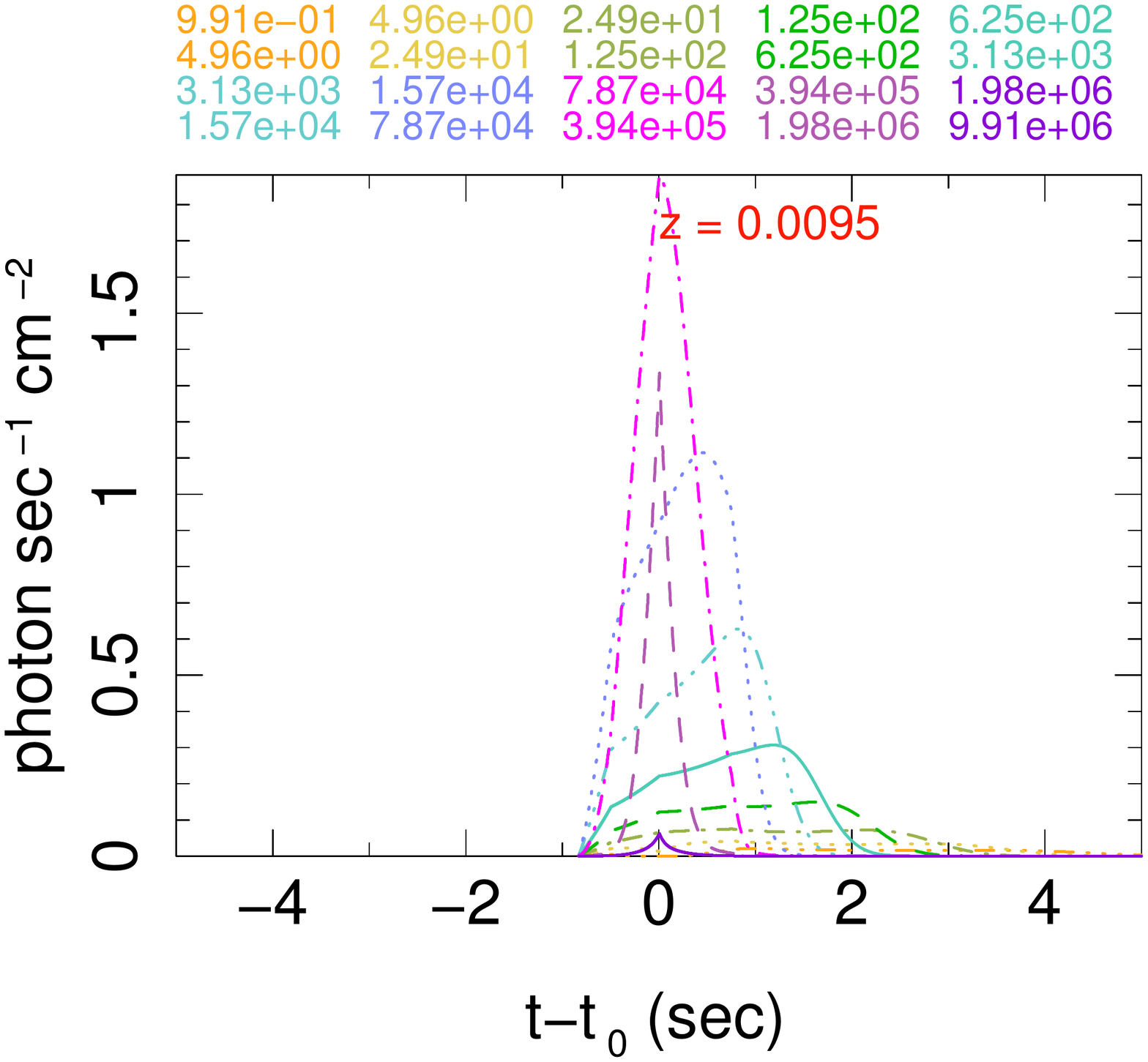} &
\vspace{-0.7cm}\includegraphics[width=6cm]{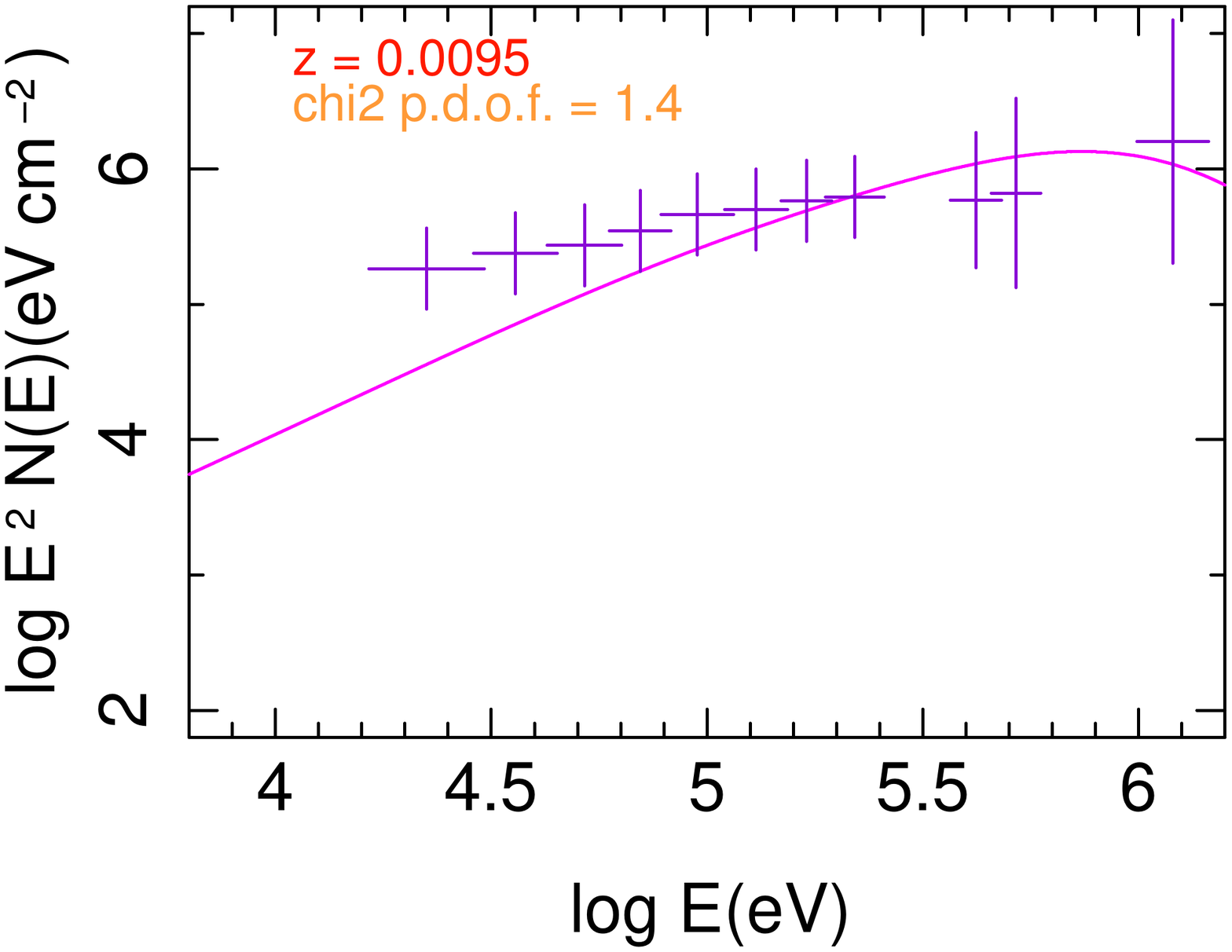} \\
\vspace{-0.7cm}\includegraphics[width=6cm]{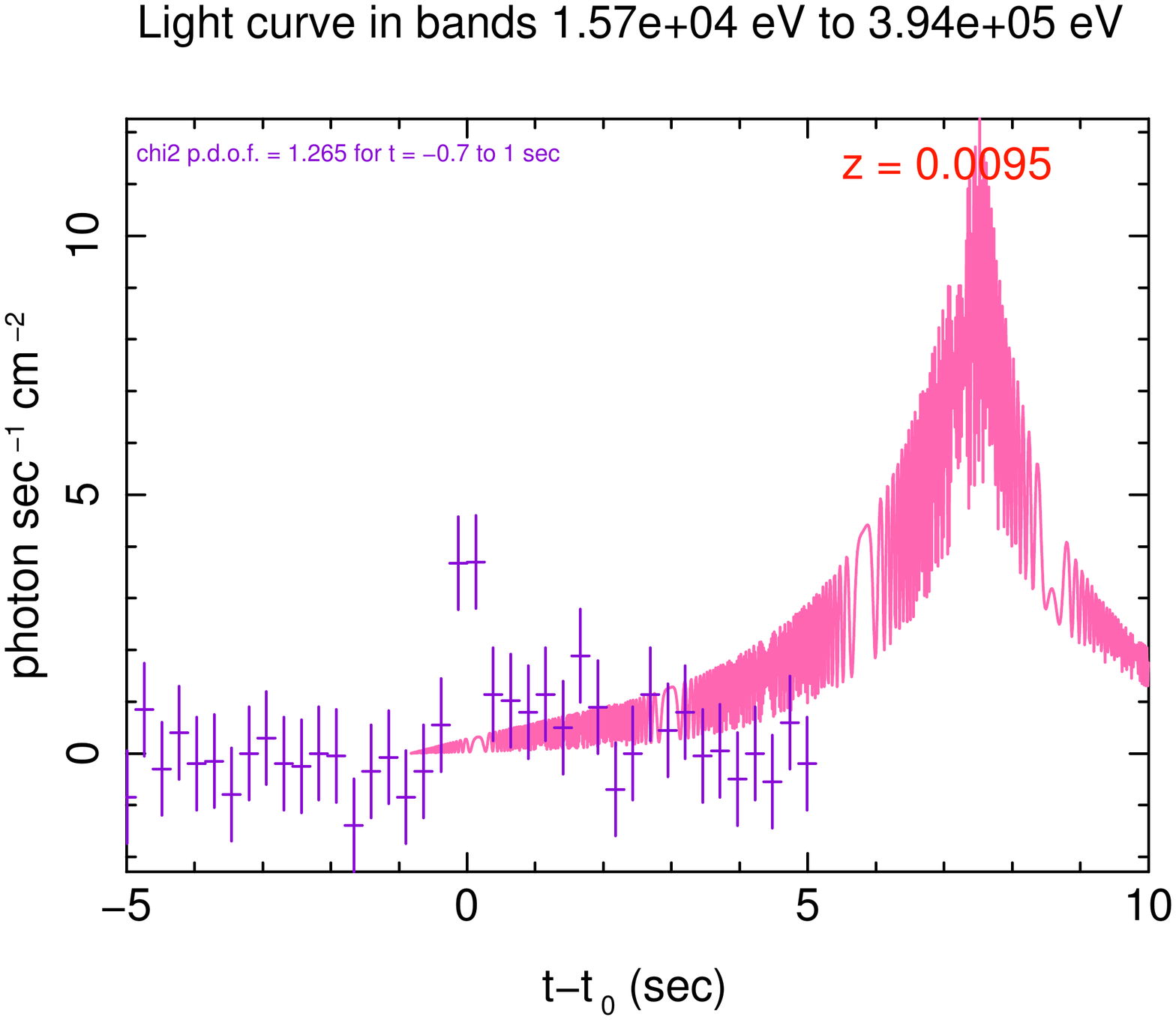} &
\vspace{-0.7cm}\includegraphics[width=6cm]{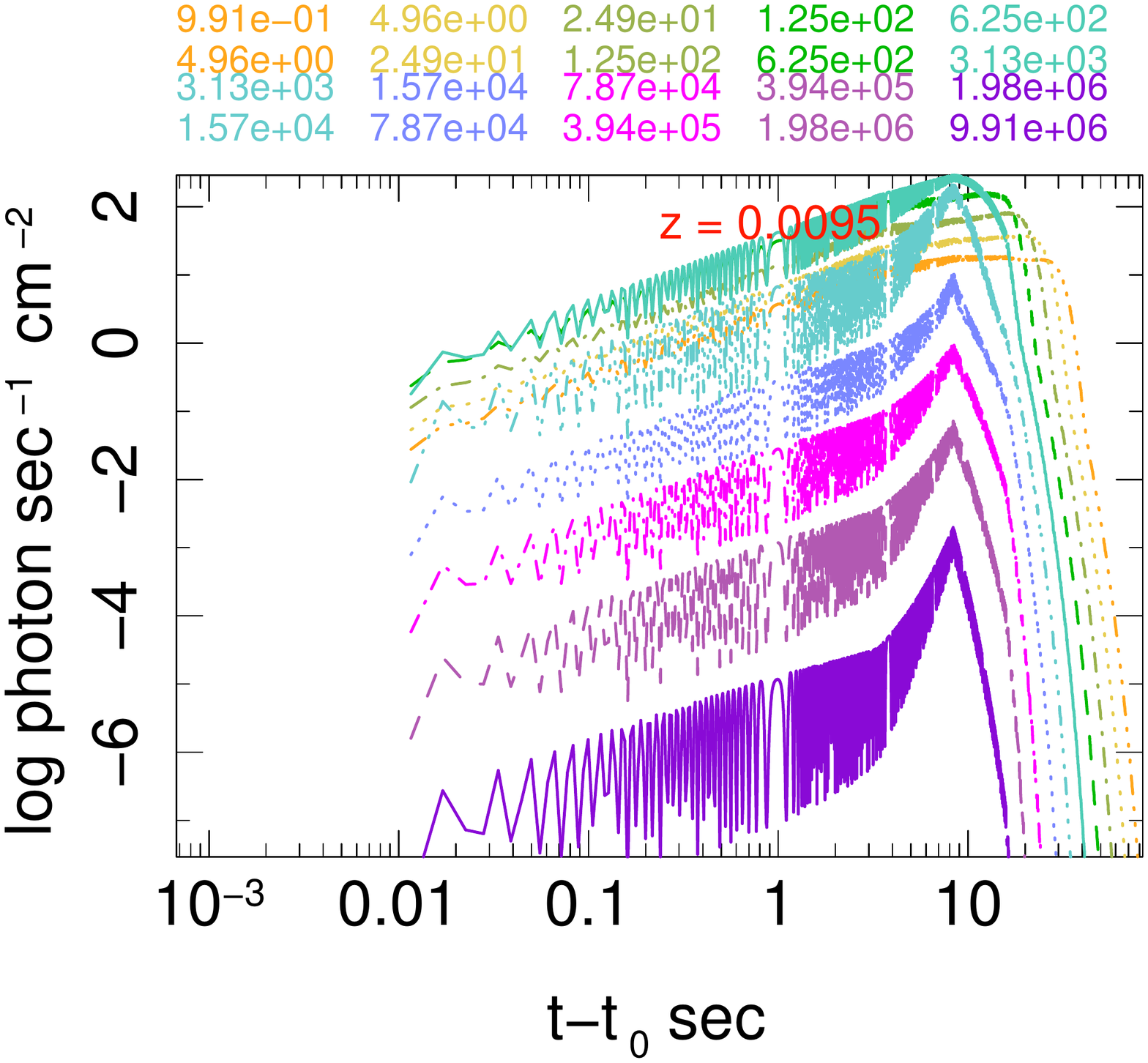} &
\vspace{-0.7cm}\includegraphics[width=6cm]{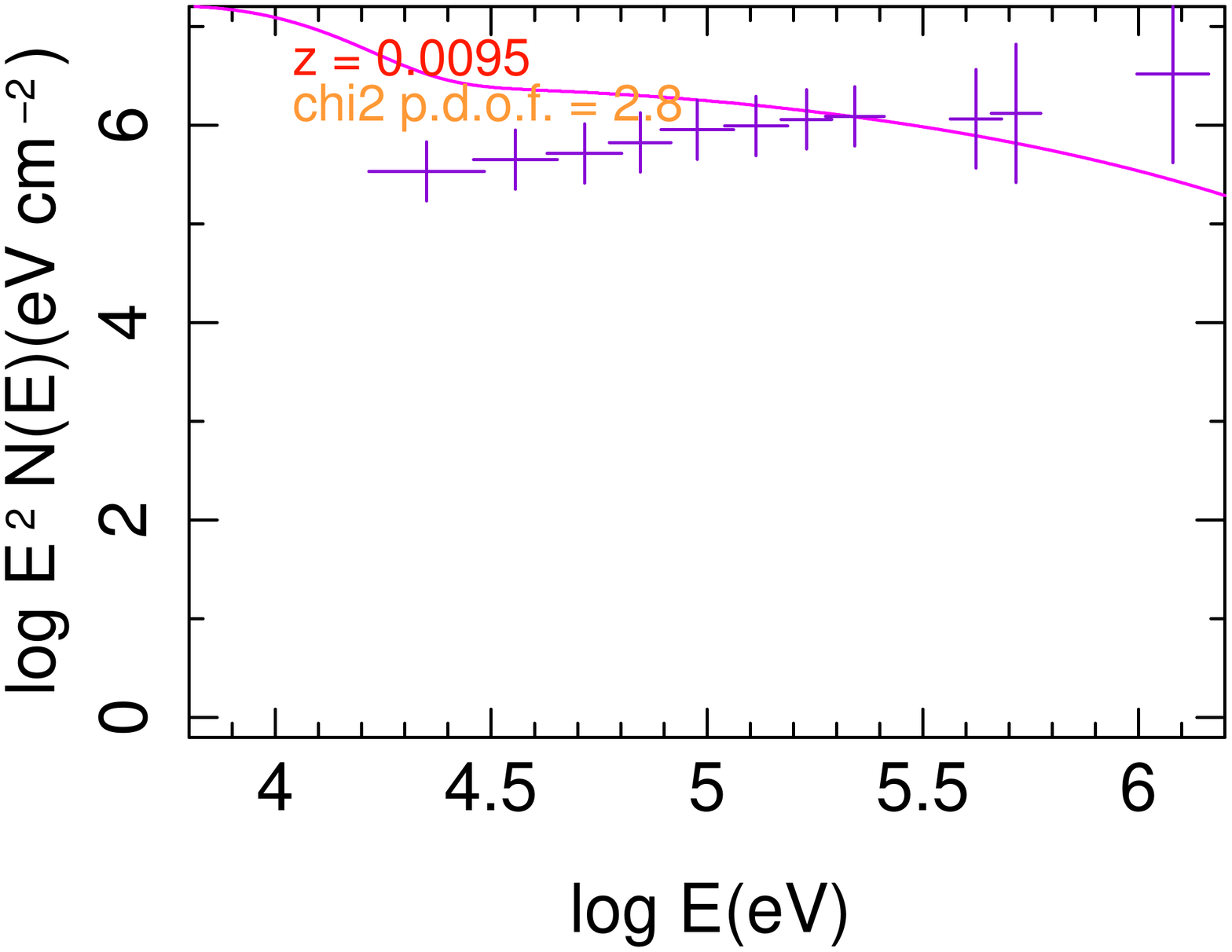}\\
\vspace{-0.7cm}\includegraphics[width=6cm]{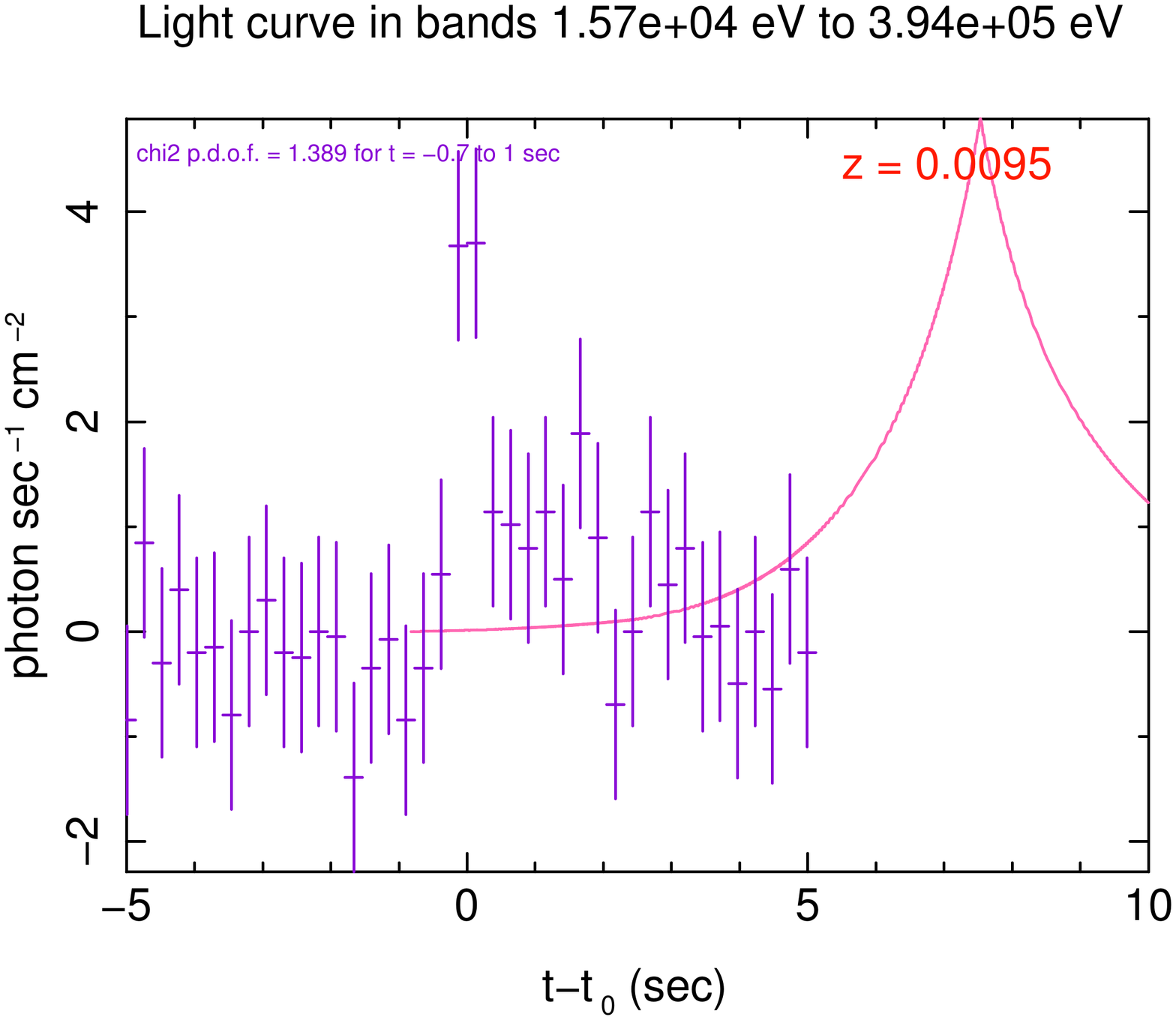} &
\vspace{-0.7cm}\includegraphics[width=6cm]{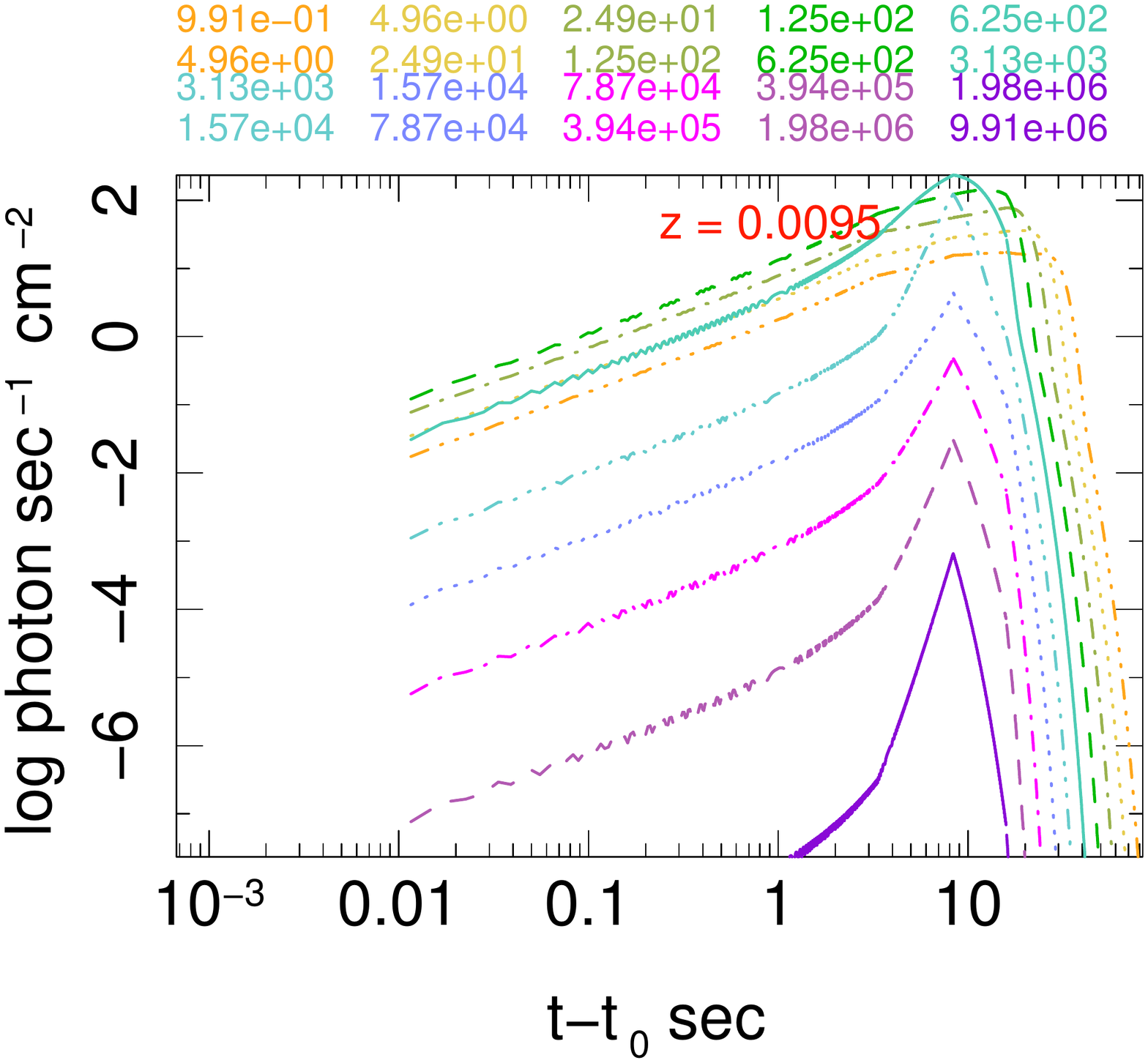} &
\vspace{-0.7cm}\includegraphics[width=6cm]{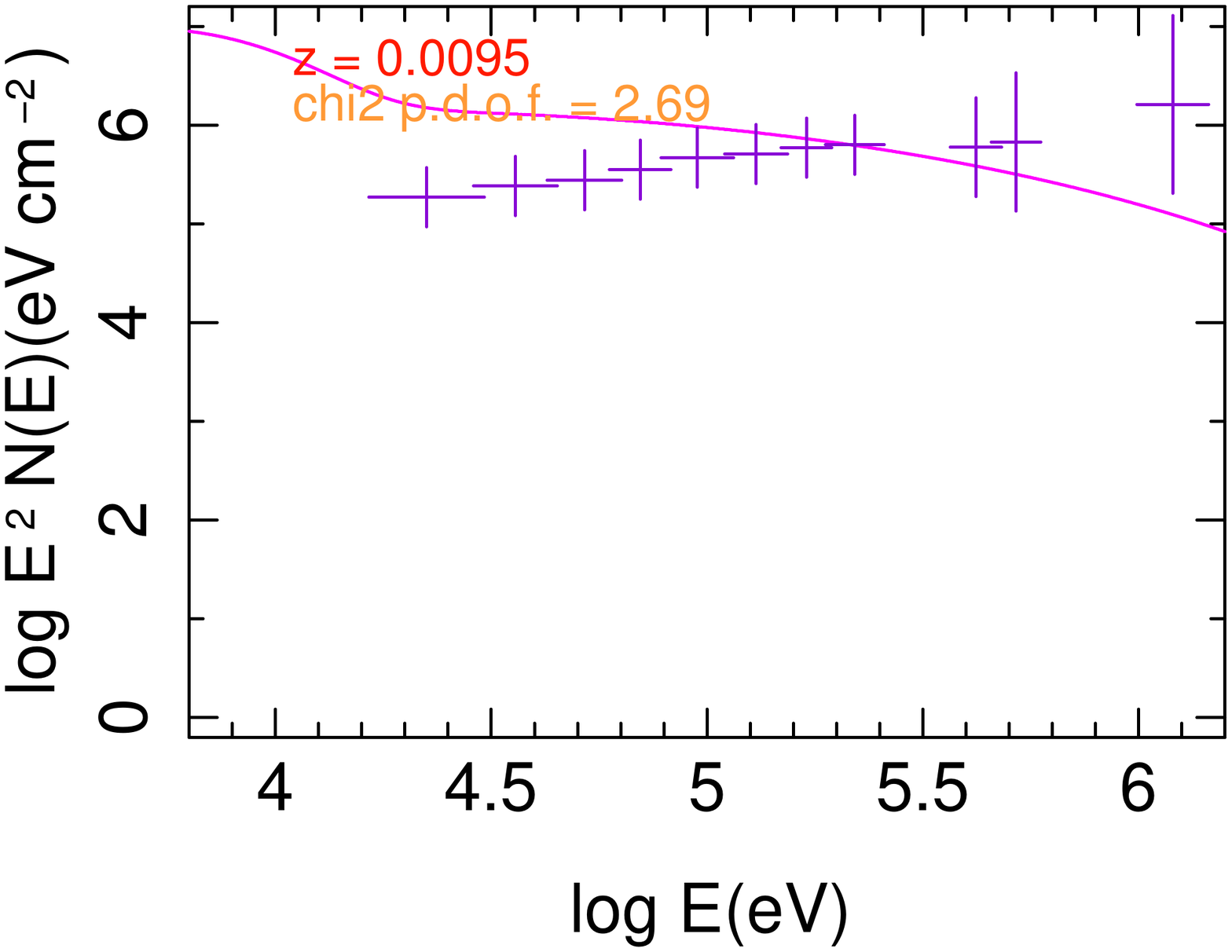}
\end{tabular}
\end{center}
\caption{Broad band (left) and narrow band (middle) light curves, and spectra (right) of a sample 
of simulations which are not good fit to the first peak of GRB 170817A. From top to bottom they 
correspond to simulations: 11, 12, 16, 17, 19, and 20 in Table \ref{tab:notgood}. Rapid variation is 
visible in the light curves when the precessing magnetic field is strong and precession is fast. 
They cannot be distinguished from shot noise if precession period is shorter than time resolution 
or binning of data. For the sake of clarity the narrow band light curves of the last two models are 
shown in logarithmic scale. We remind that GBM spectral data is normalized such that it has the same 
amplitude as simulations at $E = E_{peak} = 215 \pm 54$~keV \label{fig:notgood}}
\end{figure}

Similarity of light curves of simulations with best fits, despite large differences in some of their 
parameters, shows the degeneracy of parameters of the phenomenological model. Nonetheless, fitting the 
spectrum of the first peak to data provides further selection criteria. We did not fit the simulated 
model to the spectrum of the second peak because the data in~\citep{gw170817fermi} includes only 2 
measured data points at lowest energies for this peak and other data points are observational 
limits. Fig. \ref{fig:spect} shows spectra of 4 simulations which their light curves are shown 
in Fig. \ref{fig:totlc}. From this figure it is evident that spectrum in Fig. \ref{fig:spect}-d 
is a weaker fit to data than other models shown in this figure. It has a cumulative probability of 
random coincidence\footnote{Here the cumulative probability is defined as $P(X < \chi^2_{data})$, where 
$X$ is a random variable with chi-square distribution and $N-1$ degrees of freedom; $N$ is the number 
of data points; and $\chi^2_{data}$ is the value of chi-square fit of data to model.} of 
$P \approx 0.12$ for 10 degrees of freedom, where other 3 models have 
$P \approx \mathcal{O}(1) \times 10^{-3}$. Despite differences in goodness of fit, all these simulation 
are very similar to each others and to the data, and it is not possible to choose one of them as the 
best fit to the GBM data. However, the comparison of spectra in \ref{fig:spect}-a and 
\ref{fig:spect}-b, which their only difference is an external magnetic field in the former, may be 
interpreted as the necessity of a mild magnetic field in addition to the field induced by Fermi 
processes in the shock front.

\newcounter{x}
\setcounter{x}{\thetable}

\begin{table}
\begin{center}
\caption{Parameter set of simulated models. \label{tab:param}}
\end{center}
{\scriptsize
\begin{center}
\begin{tabular}{|p{5mm}|p{2cm}|p{5mm}p{8mm}p{10mm}p{10mm}p{8mm}p{10mm}p{5mm}p{10mm}p{5mm}p{5mm}p{5mm}p{5mm}|}
\hline
No. & GRB/peak & mod. & $\Gamma$ & $r_0$ (cm) & $\frac{\Delta r_0}{r_0}$ & $(\frac{r}{r_0})_{max}$ & $p(p_1)$ & $p_2$ & $\gamma_{cut}$ & $\kappa$ & $ \gamma'_0$ & $ \tau$ & $ \delta$  \\
\hline 
\multirow{4}{5mm}{1} & \multirow{4}{2cm}{GW/GRB 170817: first peak, rel.jet} & 1 & 100 & $2\times10^{10}$ & $5\times10^{-5}$ & 1.5 & 2.5 & - & 10 & 0 & 1.5 & - & 1 \\
 & & 0 & - & - & - & 1.5 & - & - & 10 & 0 & - & 0 & - \\
 & & 2 & - & - & - & 1.5 & - & - & 10 & 0 & - & - & 3 \\
 & & 2 & - & - & - & 4 & - & - & 10 & 0 & - & - & 5 \\
\hline
\multirow{4}{5mm}{2} & \multirow{4}{2cm}{GW/GRB 170817: first peak, off-axis} & 1 & 10 & $2\times10^{10}$ & $5\times10^{-5}$ & 1.5 & 2.5 & - & 10 & 0 & 1.5 & - & 1 \\
 & & 0 & - & - & - & 1.5 & - & - & 10 & 0 & - & 0 & - \\
 & & 2 & - & - & - & 1.5 & - & - & 10 & 0 & - & - & 3 \\
 & & 2 & - & - & - & 4 & - & - & 10 & 0 & - & - & 5 \\      
\hline
\multirow{4}{5mm}{3} & \multirow{4}{2cm}{GW/GRB 170817: second peak} & 1 & 30 & $6\times10^{10}$ & $5\times10^{-5}$ & 1.5 & 2.5 & - & 10 & 0 & 1.5 & - & 1 \\
 & & 0 & - & - & - & 1.5 & - & - & 10 & 0 & - & 0 & - \\
 & & 2 & - & - & - & 1.5 & - & - & 10 & 0 & - & - & 3 \\
 & & 2 & - & - & - & 4 & - & - & 10 & 0 & - & - & 5 \\
\hline
\multirow{3}{5mm}{4} & \multirow{3}{2cm}{GRB 130603B} & 3 & 500 & $8 \times 10^9$ & $5\times10^{-3}$ & 15 & 2.1 & 3 & 3 & 0 & 1.5 & - & 2 \\
 & & 2 & - & - & - & 15 & - & 3 & 3 & 0 & - & - & 3 \\
 & & 2 & - & - & - & 15 & - & 3 & 3 & 0 & - & - & 4 \\
\hline 
\end{tabular}
\end{center}
}
\end{table}

\setcounter{table}{\thex}
\begin{table}
\begin{center}
\caption{{\bf(continued)} Parameter set of the simulated models}
\end{center}
{\scriptsize
\begin{center}
\begin{tabular}{|p{5mm}|p{2cm}|p{10mm}p{5mm}p{5mm}p{5mm}p{15mm}p{15mm}p{12mm}p{10mm}p{10mm}p{10mm}|}
\hline
No. & GRB/peak & $\epsilon_B$ & $\alpha_B$ & $\epsilon_e Y_e$ & $\alpha_e$ & $N'$ (cm$^{-3}$) & $n'_c$ (cm$^{-2}$) & $|B|$ (kG) & $f$ (Hz) & $\alpha_x$ & $\phi$(rad.)\\
\hline
\multirow{4}{5mm}{1} & \multirow{4}{2cm}{GW/GRB 170817: first peak, rel.jet} & $10^{-4}$ & -1 & 0.01 & -1 & $2 \times 10^{14}$ & $10^{25}$ & 0.8 & 500 & - & - \\
 & & - & -2 & - & -2 & - & - & - & - & 1 & - \\
 & & - &  2 & - &  2 & - & - & - & - & 2 & - \\
 & & - &  4 & - &  4 & - & - & - & - & 3 & - \\
\hline
\multirow{4}{5mm}{2} & \multirow{4}{2cm}{GW/GRB 170817: first peak, off-axis} & $10^{-4}$ & -1 & 0.03 & -1 & $2 \times 10^{14}$ & $5 \times 10^{24}$ & 0.5 & 500 & 1 & - \\
 & & - & -2 & - & -2 & - & - & - & - & 1 & - \\
 & & - &  2 & - &  2 & - & - & - & - & 2 & - \\
 & & - &  4 & - &  4 & - & - & - & - & 3 & - \\
\hline
\multirow{4}{5mm}{3} & \multirow{4}{2cm}{GW/GRB 170817: second peak} & $10^{-4}$ & -1 & 0.01 & -1 & $2 \times 10^{13}$ & $5 \times 10^{23}$ & 0 & - & - & - \\
 & & - & -2 & - & -2 & - & - & - & - & - & - \\
 & & - &  2 & - &  2 & - & - & - & - & - & - \\
 & & - &  4 & - &  4 & - & - & - & - & - & - \\
\hline
\multirow{3}{5mm}{4} & \multirow{3}{2cm}{GRB 130603B} & $10^{-4}$ & -2 & 0.02 & -2. & $10^{15}$ & $2 \times 10^{26}$ & 26 & 500 & 1 & 0 \\
 & & - & 2 & - & 2 & - & - & - & - & 1 & 0. \\
 & & - & 3 & - & 2 & - & - & - & - & 1 & 0. \\
\hline
\end{tabular}
\end{center}
}
\begin{description}
\item {$\star$} Each data line corresponds to one simulated regime, during which quantities listed 
here remain constant or evolve dynamically according to fixed rules. A full simulation of a burst 
usually includes multiple regimes (at least two). 
\item {$\star$} Horizontal black lines separate time intervals (regimes) of independent simulations 
identified by the number shown in the first column.
\item {$\star$} A dash as value for a parameter presents one of the following cases: it is 
irrelevant for the model; it is evolved from its initial value according to an evolution equations 
described in~\citep{hourigrb,hourigrbmag}; or it is kept constant during all regimes. 
\end{description}
\end{table}

\begin{figure}
\begin{center}
\begin{tabular}{p{5cm}p{5cm}p{5cm}p{5cm}}
\hspace{-1cm}\includegraphics[width=5.5cm,angle=90]{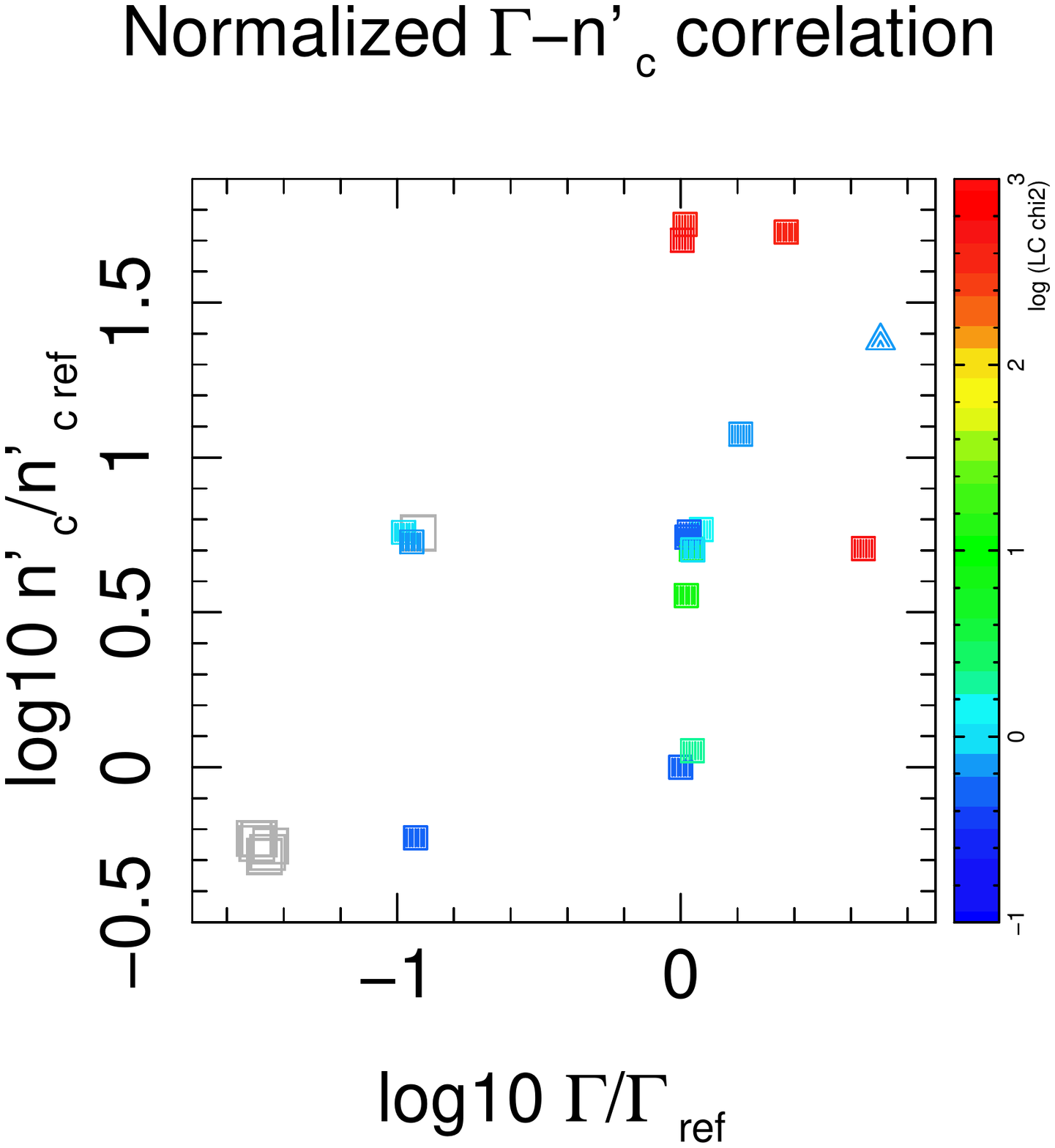} & 
\hspace{-1.75cm}\includegraphics[width=5.5cm,angle=90]{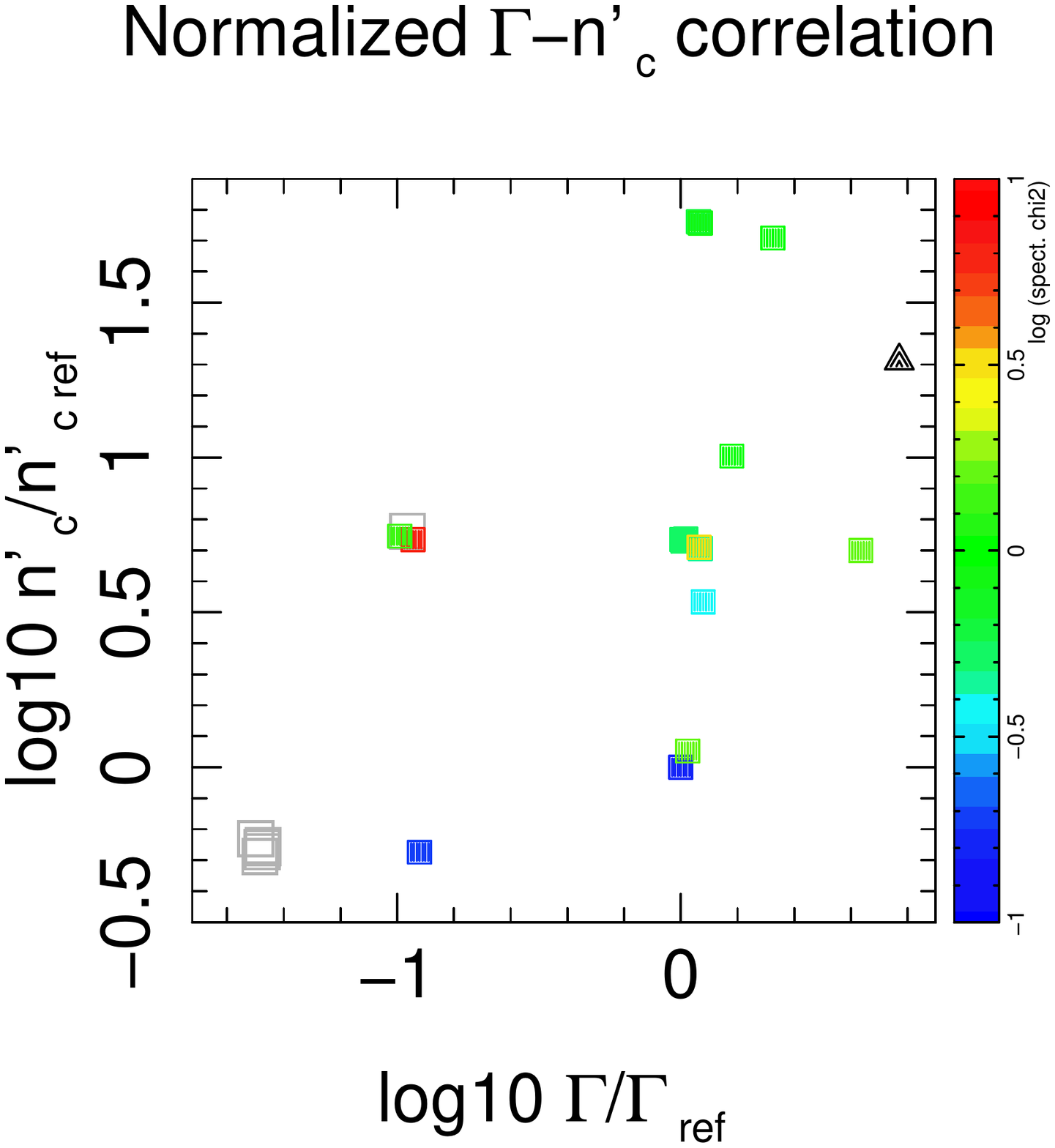} & 
\hspace{-2.5cm}\includegraphics[width=7.5cm]{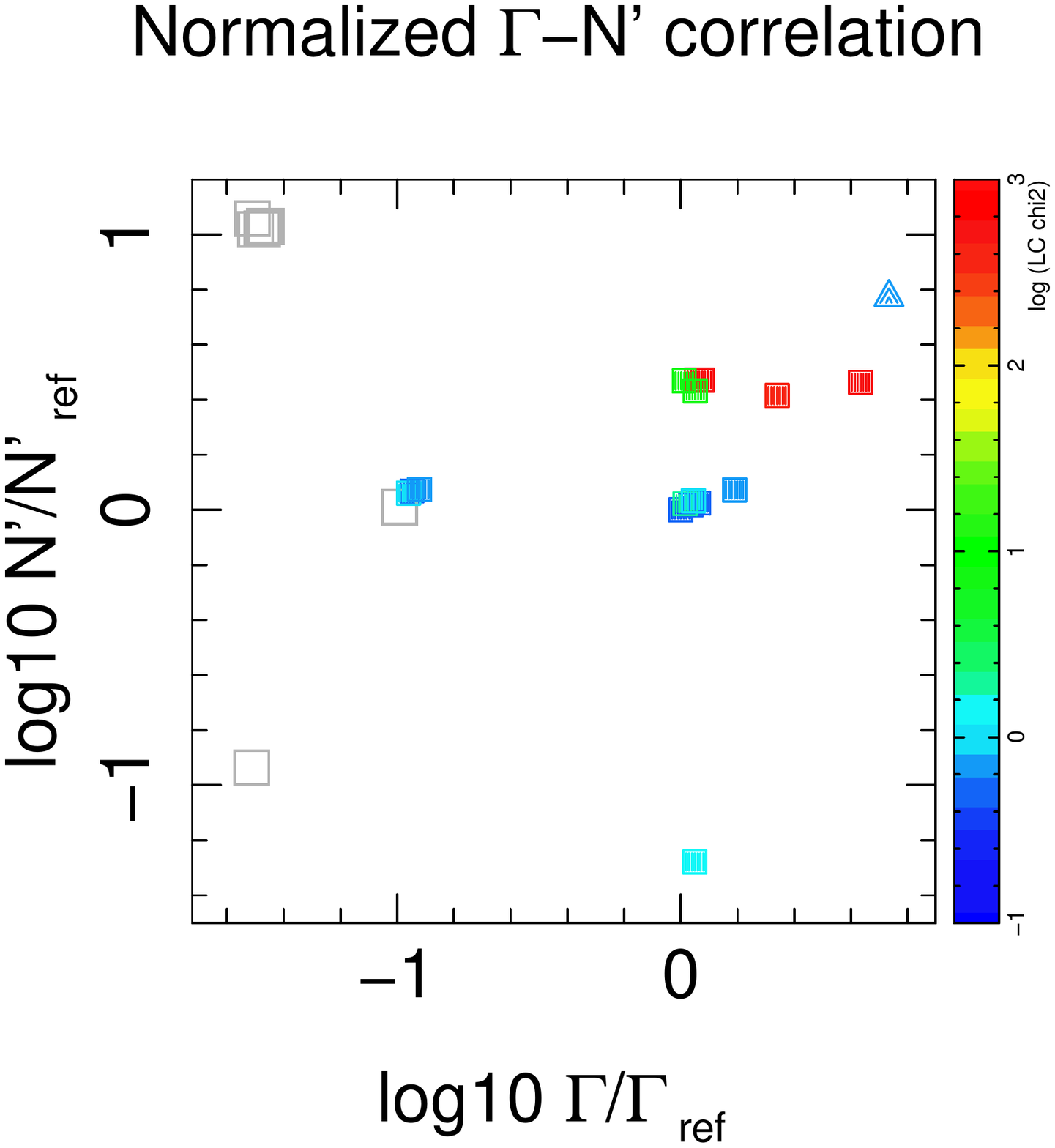} & 
\hspace{-3.cm}\includegraphics[width=5.5cm,angle=90]{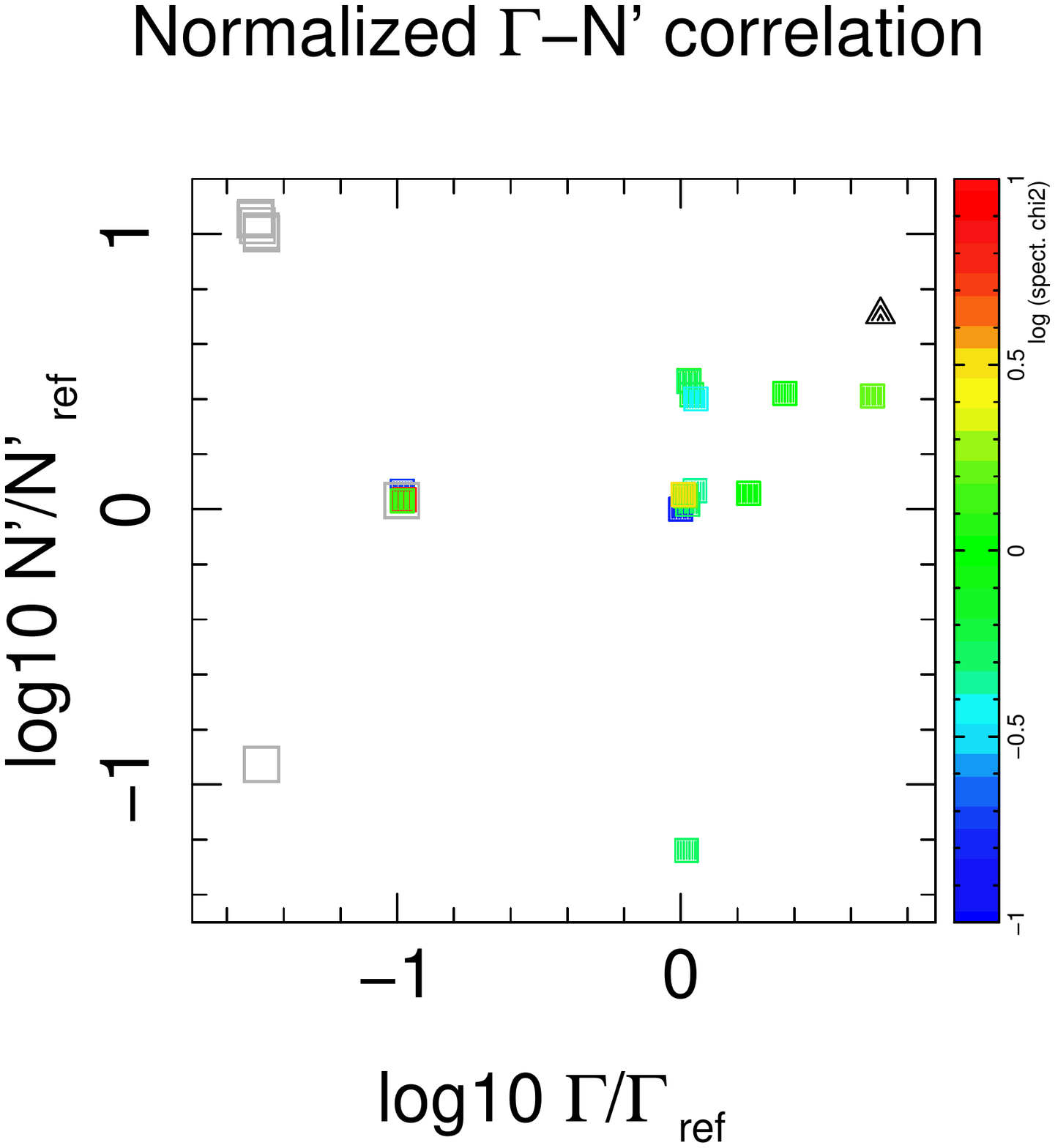} \\
\hspace{-1cm}\includegraphics[width=5.5cm,angle=90]{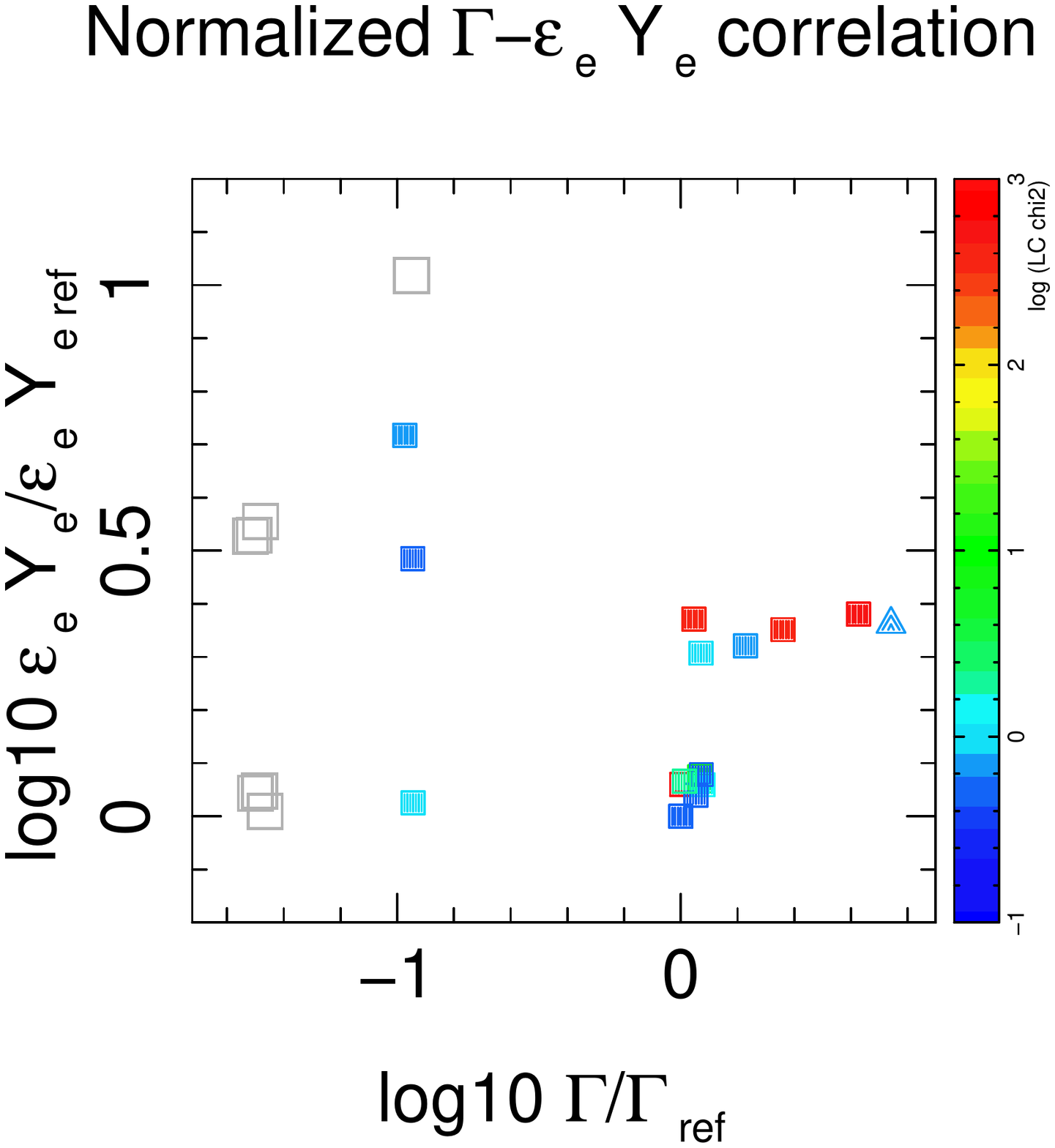} & 
\hspace{-1.75cm}\includegraphics[width=5.5cm,angle=90]{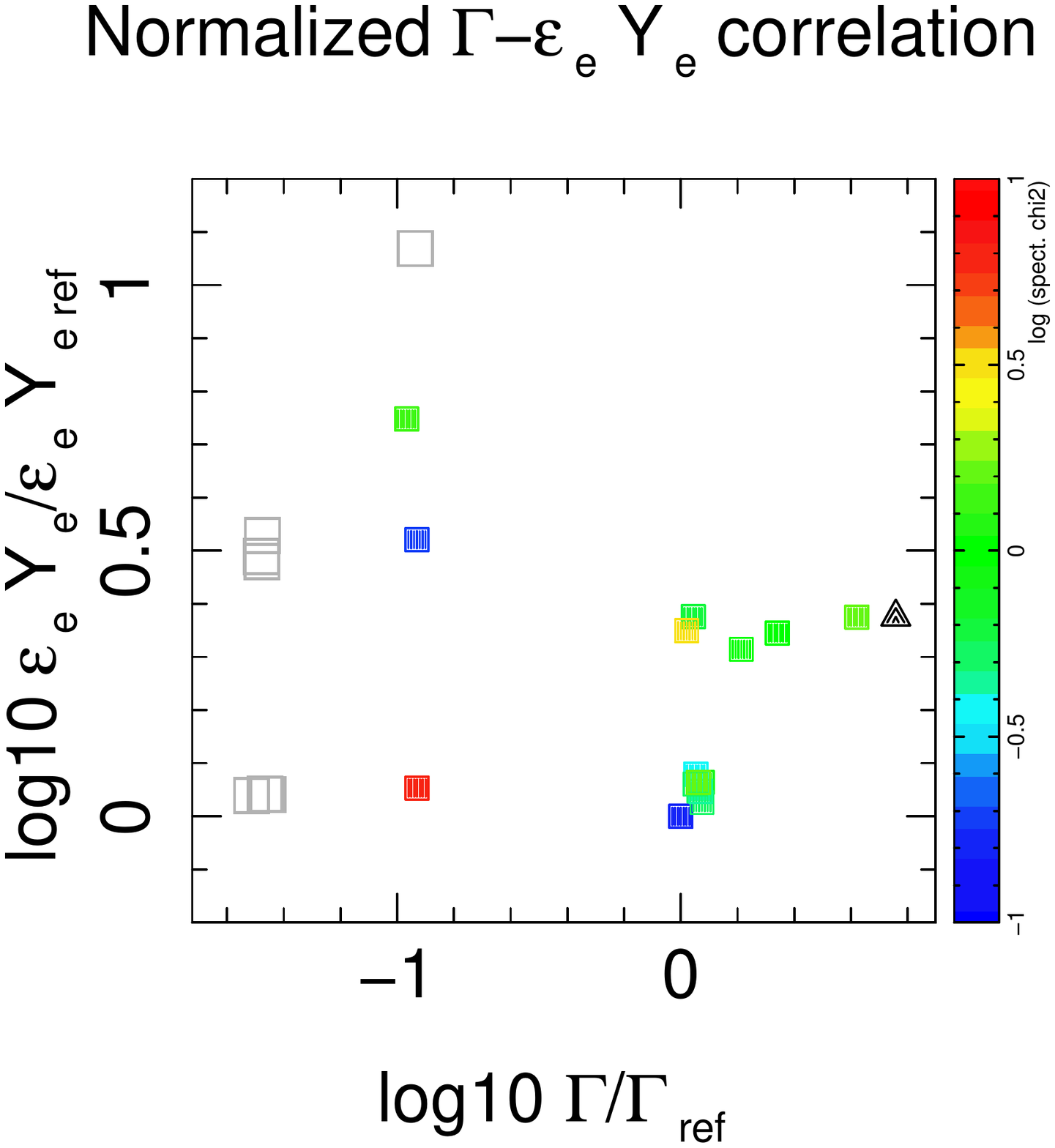} & 
\hspace{-2.5cm}\includegraphics[width=7.5cm]{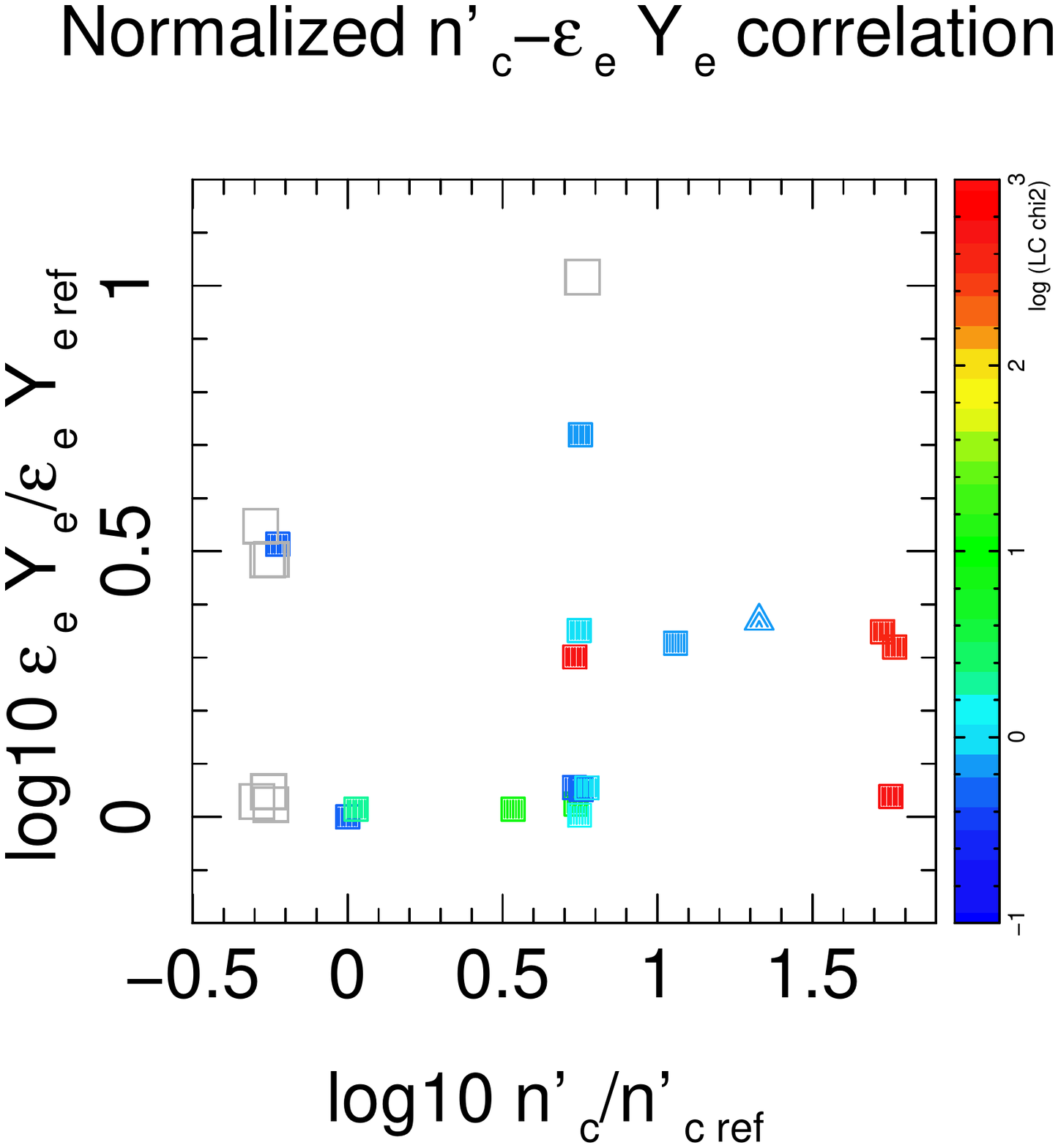} & 
\hspace{-3cm}\includegraphics[width=7.5cm]{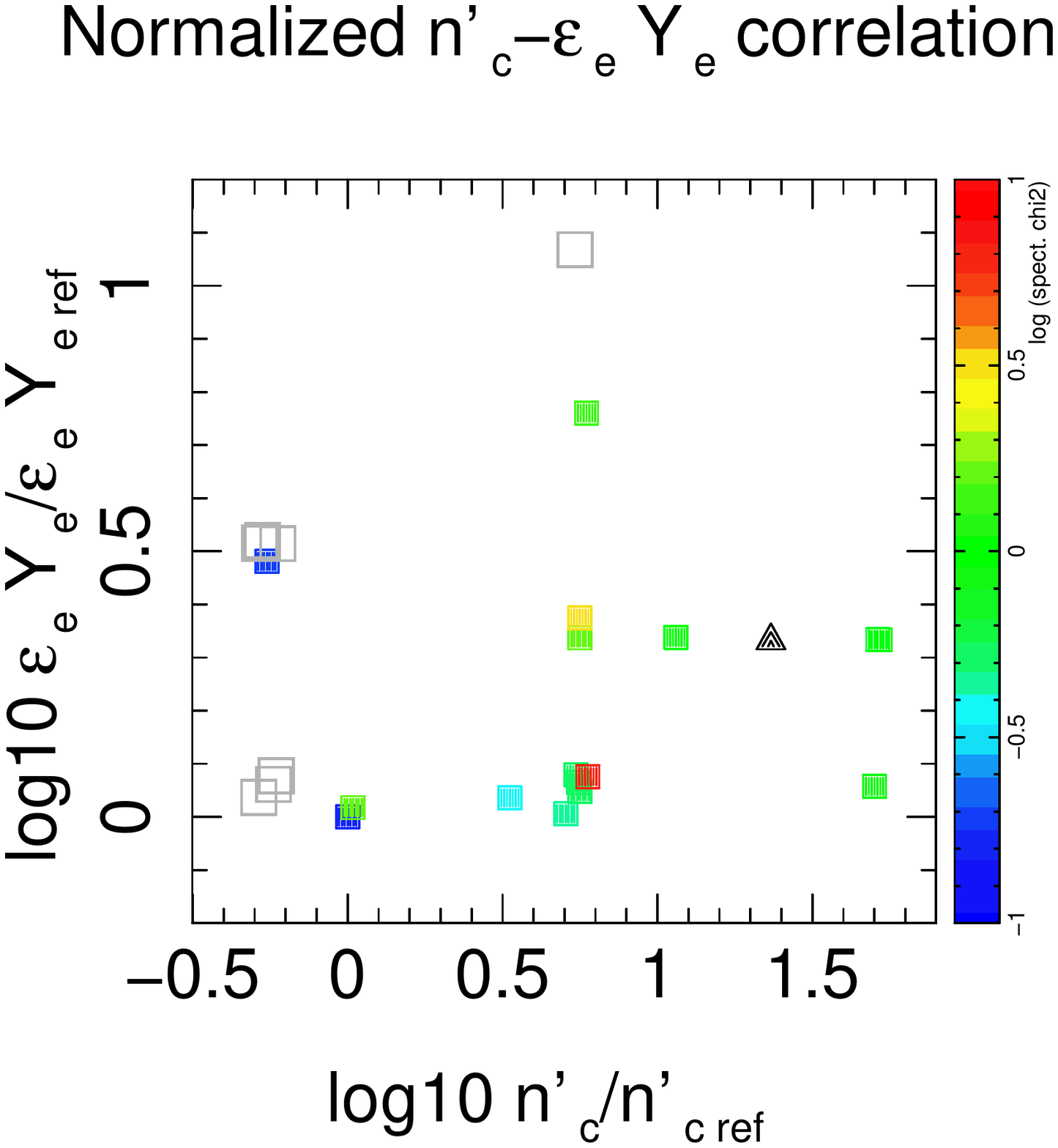} \\
\hspace{-1cm}\includegraphics[width=8cm]{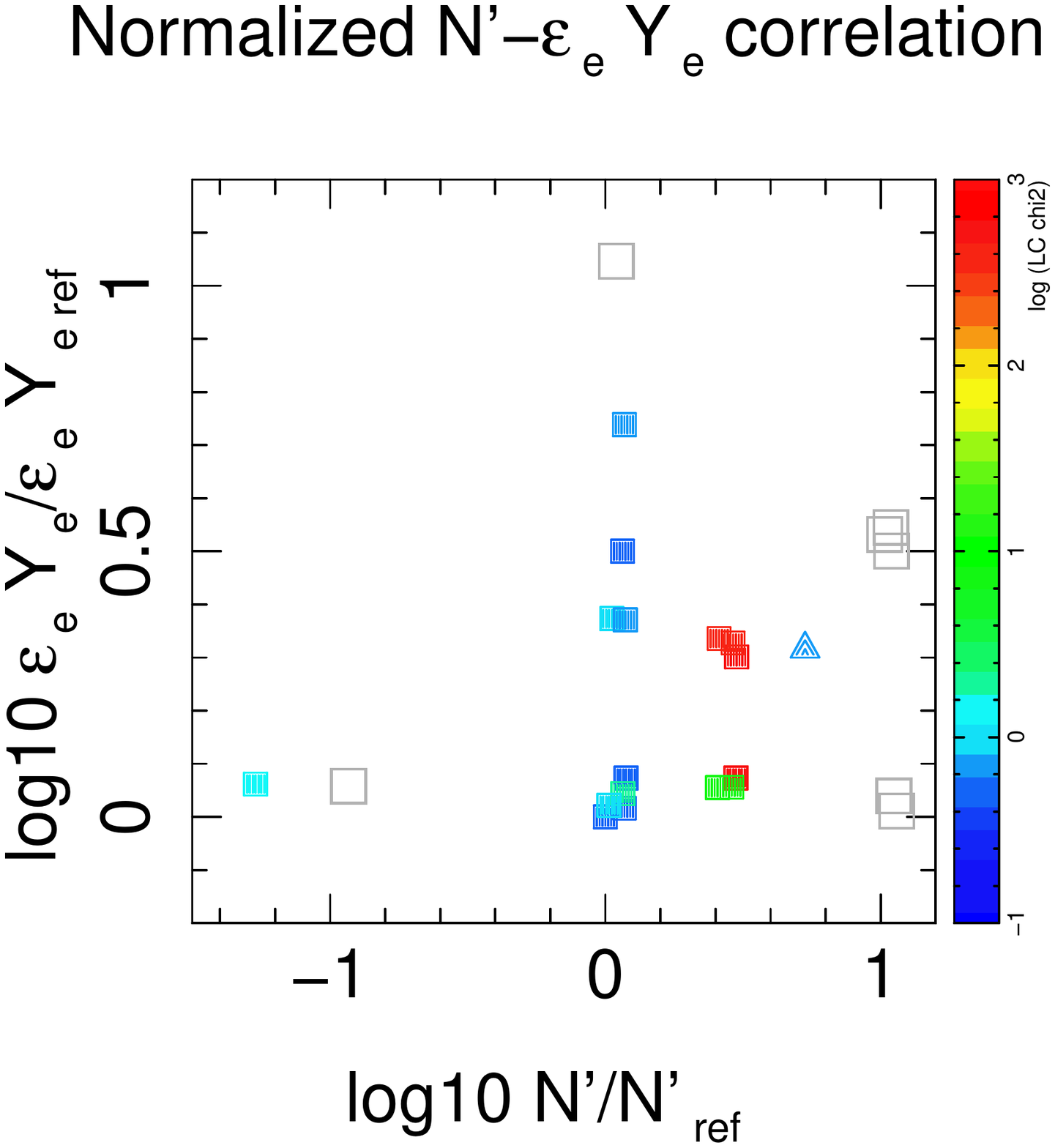} & 
\hspace{-1.75cm}\includegraphics[width=5.5cm,angle=90]{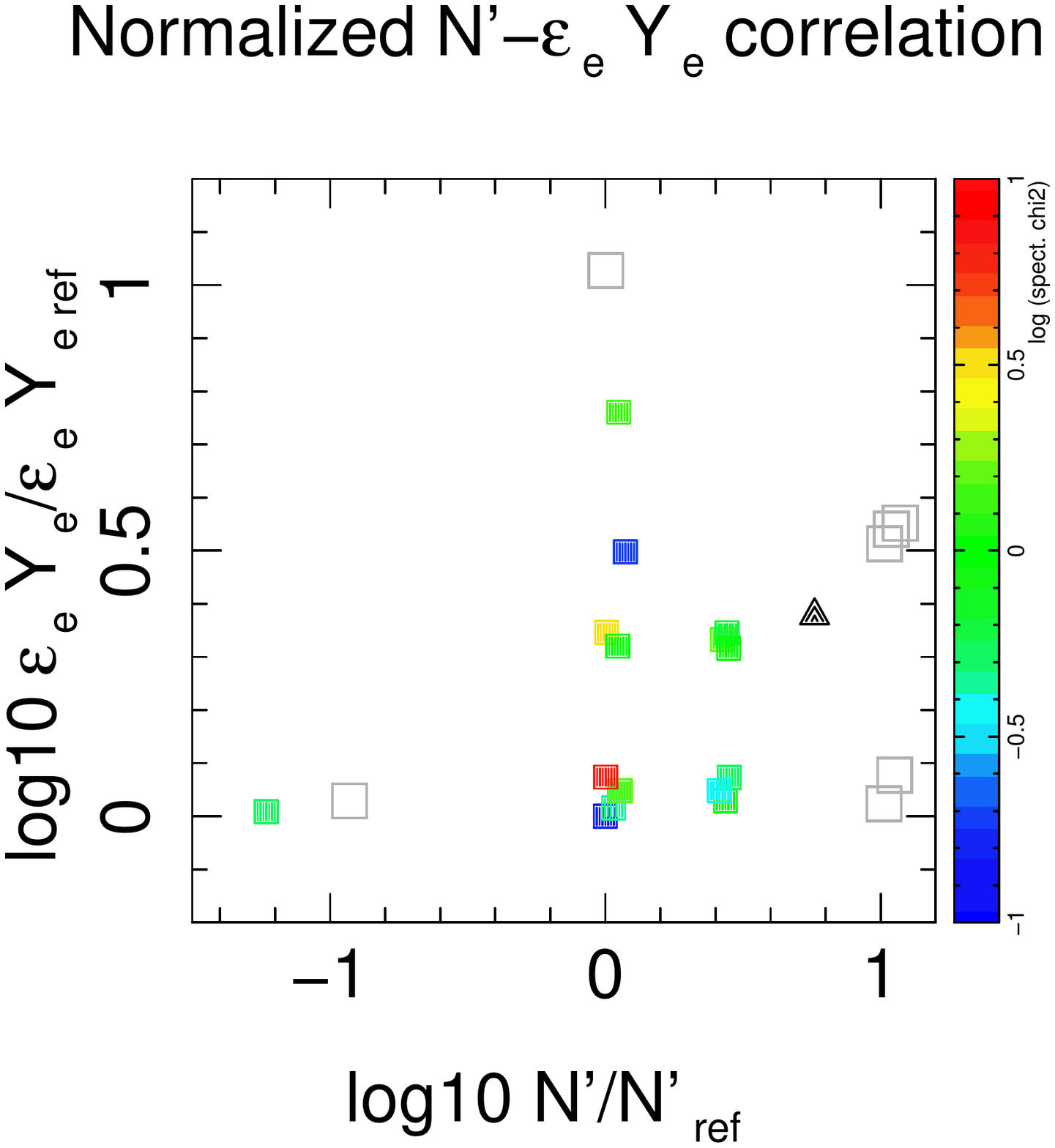} & 
\hspace{-2.5cm}\includegraphics[width=5.5cm,angle=90]{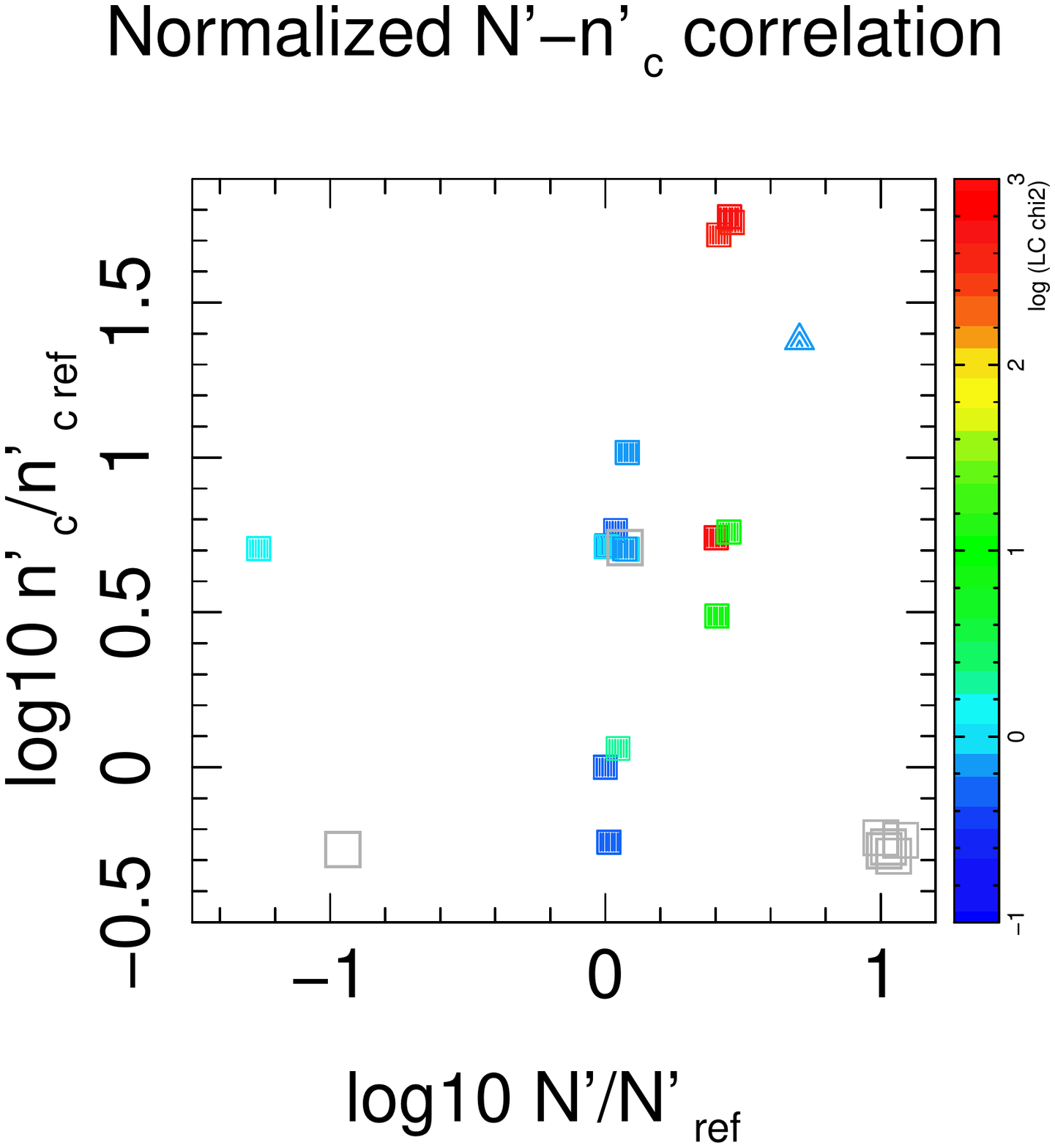} & 
\hspace{-3cm}\includegraphics[width=5.5cm,angle=90]{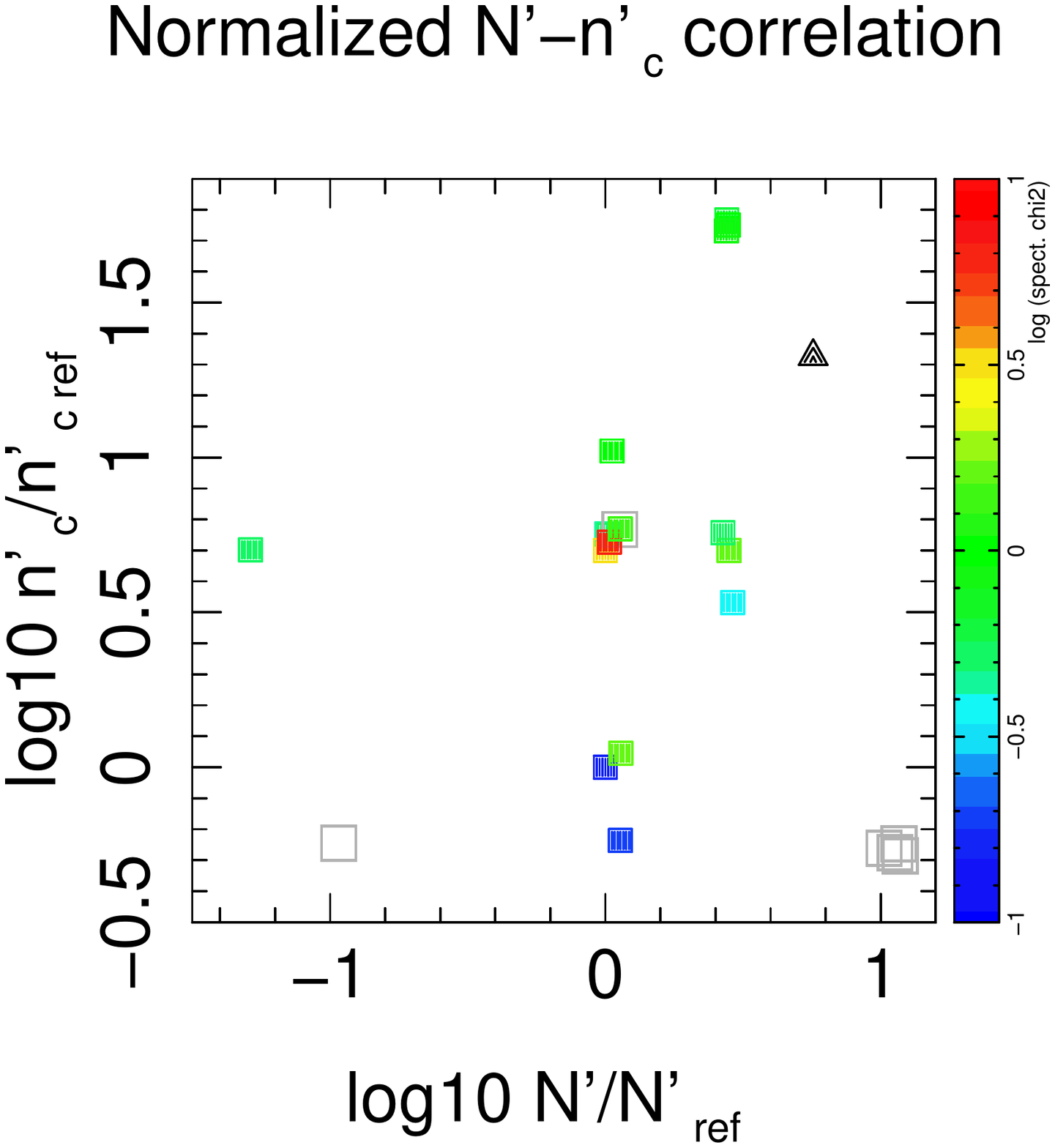}
\end{tabular}
\end {center}
\caption{Color-coded $\chi^2$ per degree of freedom for simulations presented in 
Tables \ref{tab:param} and \ref{tab:notgood} in 2D parameter planes. To make the comparison between 
models easier, all parameters are normalized by the corresponding value in simulation No. 1 in 
Table \ref{tab:param}. From left, in odd columns $\chi^2$ is obtained by fitting broad-band light 
curve to GBM data, and in even columns from fitting the spectrum. Open squares present simulations 
with $\chi^2$ larger than maximum value shown in the color coded scale. 
Triangle symbol presents simulation No. 4 in Table \ref{tab:param} which its properties are similar 
to GRB 130603B. Only data of the first peak is used to estimate $\chi^2$ of this simulation. No broad 
band spectral data for this burst is available and the absence of $\chi^2$ 
is shown by using color black for triangle symbols in the plots. To prevent overlap of symbols they 
are slightly and randomly shifted in both directions. The shift is much smaller than variation of 
parameters among different simulations and very compact clumps of symbols mean the same value for 
two parameters in the corresponding simulations. When one of the plotted parameters is Lorentz factor, 
three groups of simulations with $\Gamma \sim \mathcal{O}(1)$ (cocoon), $\Gamma \sim \mathcal{O}(10)$ 
(off-axis), $\Gamma \sim \mathcal{O}(100)$ (ultra-relativistic) are distinguishable. 
\label{fig:paramcorr}}
\end {figure}

\begin{table}
\begin{center}
\caption{Simulations with altered parameters with respect to those presenting best fits to GRB 170817A data \label{tab:notgood}}
\end{center}
{\scriptsize
\begin{center}
\begin{tabular}{|p{5mm}|p{5mm}p{10mm}p{12mm}p{5mm}p{5mm}p{5mm}p{5mm}p{10mm}p{10mm}p{10mm}|p{10mm}|p{10mm}|p{3cm}|}
\hline
No. & $\Gamma$ & $r_0 (cm)$ & $p~(p_1,~p_2)$& $\gamma_{cut}$ & $\gamma'_0$ & $\epsilon_B$ & 
$Y_e\epsilon_e$ & $N'$ cm$^{-3}$) & $n'_c$ (cm$^{-2}$) & $|B|$ (kG) & $\chi_{lc}^2$ p.d.o.f. & 
$\chi_{spect}^2$ p.d.o.f. & Deficiencies \\
\hline
\multicolumn{14}{|c|}{\bf Relativistic jet models} \\
\hline
1 & 400 & $2 \times 10^{10}$ & 1.5 & 200 & 1.5 & $10^{-4}$ & 0.02 & $5 \times 10^{14}$ & 
$5 \times 10^{25}$ & 1 & 429.9 & 1.49 & Too hard; Too bright \\
2 & 100 & $2 \times 10^{10}$ & 1.5 & 200 & 1.5 & $10^{-4}$ & 0.02 & $5 \times 10^{14}$ & 
$5 \times 10^{26}$ & 1 & 311.9 & 0.58 & Too bright; Positive spectral index at high energies \\
3 & 200 & $2 \times 10^{10}$ & 1.5 & 200 & 1.5 & $10^{-4}$ & 0.02 & $5 \times 10^{14}$ & 
$5 \times 10^{26}$ & 0 & 284.6 & 1.07 & Too hard; Too bright \\
4 & 100 & $2 \times 10^{10}$ & 2.2 & 10 & 1.5 & $10^{-4}$ & 0.01 & $5 \times 10^{14}$ & 
$5 \times 10^{26}$ & 1 & 464.4 & 0.8 & Too soft; Too bright \\
6 & 100 & $2 \times 10^{10}$ & 2.5 & 10 & 1.5 & $10^{-4}$ & 0.01 & $5 \times 10^{14}$ & 
$5 \times 10^{25}$ & 1 & 7.335 & 0.525 & Slightly too hard; Too bright\\
7 & 100 & $2 \times 10^{10}$ & 2.5 & 10 & 1.5 & $10^{-4}$ & 0.01 & $5 \times 10^{14}$ & 
$3 \times 10^{25}$ & 1 & 7.694 & 0.405 & Too bright \\
8 & 100 & $2 \times 10^{10}$ & 2.5 & 10 & 1.5 & $10^{-4}$ & 0.01 & $10^{13}$ & 
$5 \times 10^{25}$ & 1 & 1.382 & 0.568 & Slightly too soft; Too faint \\
9 & 100 & $2 \times 10^{10}$ & 2.5 & 10 & 1.5 & $10^{-4}$ & 0.01 & $2 \times 10^{14}$ & 
$5 \times 10^{25}$ & 1 & 0.642 & 0.439 & Slightly too hard \\
10 & 100 & $2 \times 10^{10}$ & 2.5 & 10 & 1.5 & $10^{-4}$ & 0.01 & $2 \times 10^{14}$ & 
$5 \times 10^{25}$ & 1 ($f=5$ HZ) & 0.644 & 0.524 & Slightly too hard \\
11 & 100 & $2 \times 10^{10}$ & 2.5 & 10 & 1.5 & $10^{-4}$ & 0.01 & $2 \times 10^{14}$ & 
$10^{25}$ & 1 & 1.903 & 1.68 & Peak too soft; Too bright in soft $\gamma$-ray energies\\
12 & 100 & $2 \times 10^{10}$ & 3 & 10 & 1.5 & $10^{-4}$ & 0.02 & $2 \times 10^{14}$ & 
$5 \times 10^{25}$ & 0 & 1.094 & 2.81 & Too hard; Too faint \\
13 & 150 & $2 \times 10^{10}$ & 2.5 & 10 & 1.5 & $10^{-4}$ & 0.02 & $2 \times 10^{14}$ & 
$10^{26}$ & 1 & 0.697 & 1.02 & Too hard\\
\hline
\multicolumn{14}{|c|}{\bf Off-axis models} \\
\hline
15 & 10 & $10^{12}$ & 2.5 & 10 & 1.5 & $10^{-4}$ & 0.1 & $2 \times 10^{14}$ & $5 \times 10^{25}$ & 
0 & $6 \times 10^6$ & - & Too bright; Too long \\
16 & 10 & $2 \times 10^{10}$ & 2.5 & 10 & 1.5 & $10^{-4}$ & 0.01 & $2 \times 10^{14}$ & 
$5 \times 10^{25}$ & 0 & 1.110 & 5.43 & Too soft \\
17 & 10 & $2 \times 10^{10}$ & 2.5 & 10 & 1.5 & $10^{-4}$ & 0.05 & $2 \times 10^{14}$ & 
$5 \times 10^{25}$ & 0 & 0.675 & 1.4 & too faint, too hard \\
\hline
\multicolumn{14}{|c|}{\bf Cocoon models} \\
\hline
18 & 3 & $2 \times 10^{11}$ & 2.5 & 10 & 1.5 & $10^{-6}$ & 0.03 & $2 \times 10^{15}$ & 
$5 \times 10^{24}$ & 26 & $-\dagger$ & $-\dagger$ & Too bright; Too soft; Too long \\
19 & 3 & $2 \times 10^{11}$ & 2.5 & 10 & 1.5 & $10^{-6}$ & 0.03 & $2 \times 10^{15}$ & 
$5 \times 10^{24}$ & 26 & $-\dagger$ & $-\dagger$ & Too soft; Too long \\
20 & 3 & $2 \times 10^{11}$ & 2.5 & 10 & 1.5 & $10^{-6}$ & 0.03 & $2 \times 10^{15}$ & 
$5 \times 10^{24}$ & 2.6 & $-\dagger$ & $-\dagger$ & Too soft; Too long \\
21 & 3 & $2 \times 10^{11}$ & 2.5 & 10 & 1.5 & $10^{-4}$ & 0.01 & $2 \times 10^{15}$ & 
$5 \times 10^{24}$ & 2.6 & - & - & Too bright; Too soft; Too long \\
22 & 3 & $2 \times 10^{11}$ & 2.5 & 10 & 1.5 & $10^{-4}$ & 0.01 & $2 \times 10^{15}$ & 
$5 \times 10^{24}$ & 26 & - & - & Too bright; Too soft; Too long \\
23 & 3 & $2 \times 10^{11}$ & 2.5 & 10 & 1.5 & $10^{-4}$ & 0.01 & $2 \times 10^{13}$ & 
$5 \times 10^{24}$ & 2.6 & - & - & Too soft; Too long \\
\hline
\end{tabular}
\end{center}
}
\begin{description}
\item {$\star$} Values in this table correspond to initial value of parameters in the first regime of each simulation. 
Other parameters and regimes are similar to models given in Table \ref{tab:param} and are not shown here for the sake of clarity.
\item {$-\dagger$} Peak of light curves out of observation time.
\end{description}
\end{table}
To better understand the correlation between parameters of the model, their degeneracies, and how 
they affect the main observables of GRB170817A, namely light curves and spectrum, Fig. \ref{fig:paramcorr} 
shows color-coded $\chi^2$ value of simulations presented in Tables \ref{tab:param} and 
\ref{tab:notgood} in 2D parameter planes for a subset of parameters of the phenomenological model, 
which are related to primary properties of the relativistic jet and vary significantly in our 
simulations. The plots in this figure show that despite small coverage of 2D parameter planes by our 
simulations, for each parameter both high and low values are sampled - except in case of $n'_c$ for 
cocoon models, in which by definition the column density of outflow could not be larger than 
relativistic jets. We notice large parameter degeneracies between models with good light curve fit, 
i.e. $\chi^2 < 1$, which is consistent with similarity of light curves in Fig. \ref{fig:totlc}. 
By contrast, many models with very different spectral $\chi^2$ fall on the same position 
in 2D parameter planes. Such behaviour is present in all the combination of parameters shown in 
Fig. \ref{fig:paramcorr} and means that the spectral behaviour depends on multiple parameters 
and cannot be well presented by a 2D parameter space.

\subsection{Comparison with GRB 130603B} \label{sec:compgrb130603b}
\begin{figure}
\begin{center}
\begin{tabular}{p{6cm}p{6.5cm}p{4.5cm}}
\vspace{-0.5cm}\includegraphics[width=6cm]{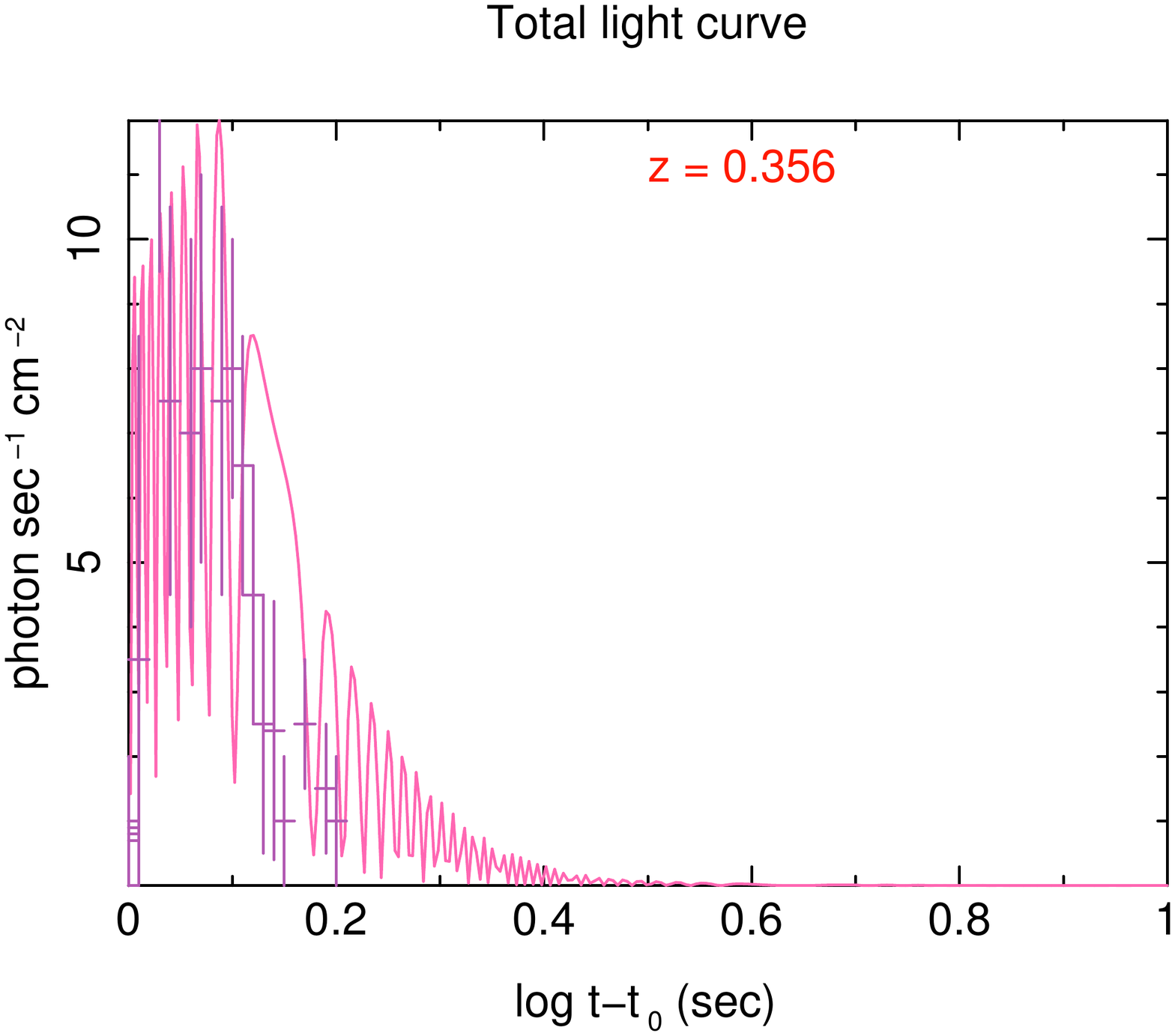} &
\vspace{-0.5cm}\hspace{-0.5cm}\includegraphics[width=6.5cm]{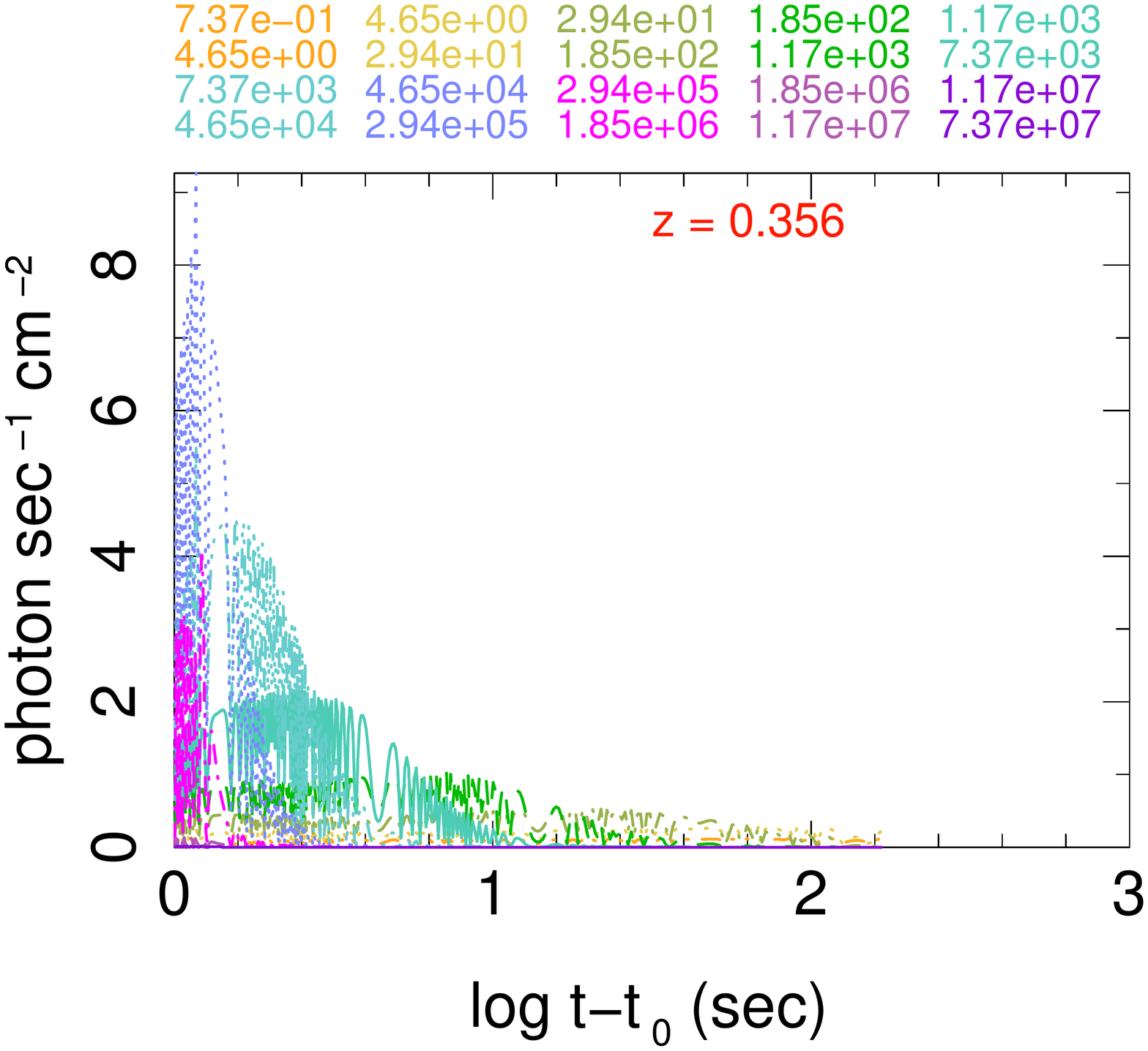} & 
\vspace{-0.5cm}\hspace{-1cm}\includegraphics[width=4.5cm,angle=90]{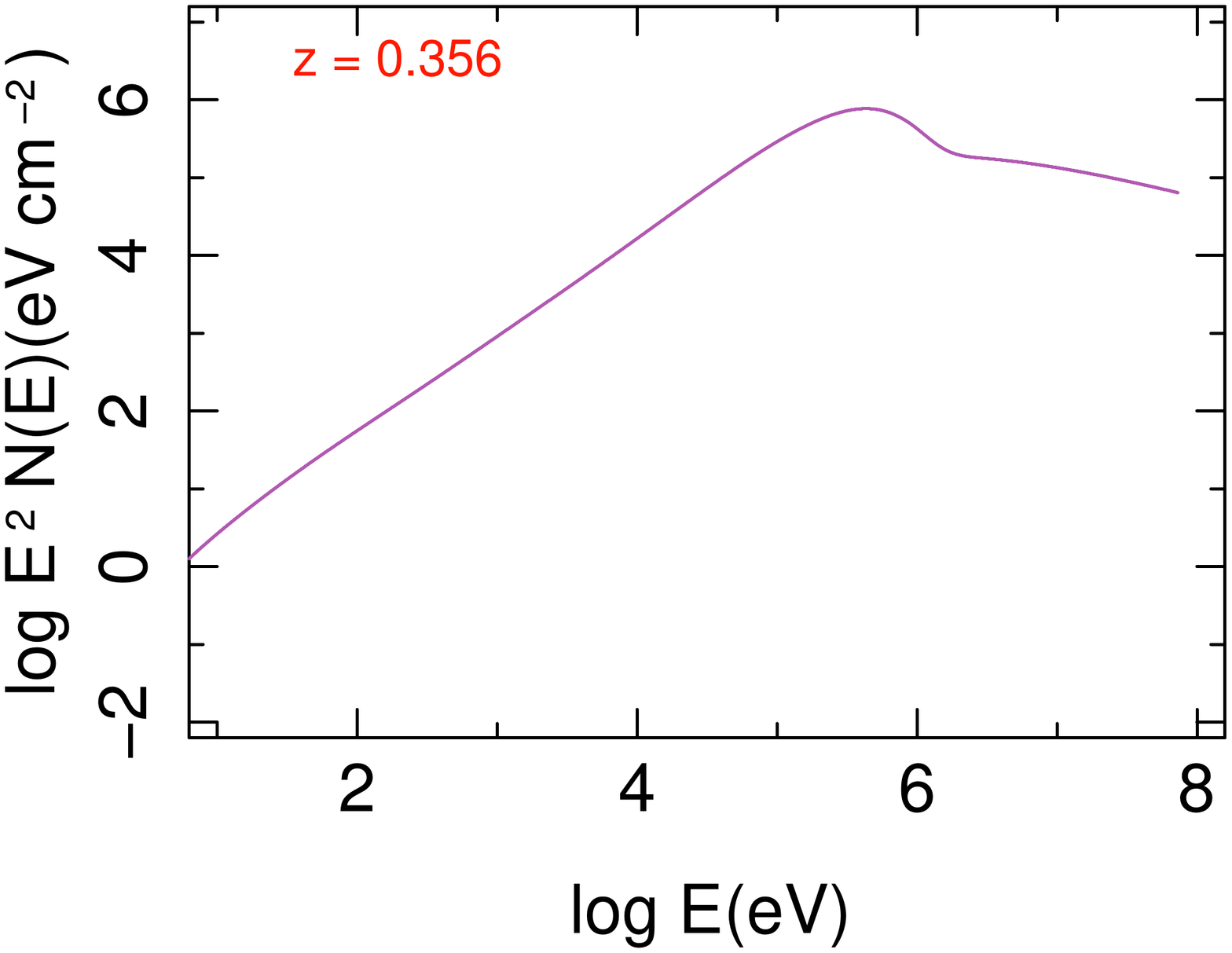}
\end{tabular}
\end{center}
\caption{Broad band (left) and narrow bands (middle) light curves and spectrum (right) of 
model No. 4 in Table \ref{tab:param}, which is a good approximation for the first peak of 
GRB 130603B. Data points are from Swift-BAT observations in 15-350 keV band. The peak 
energy of the spectrum is very close to $E_{peak} = 660 \pm 100$ keV of the short GRB 130603B 
observed by Konus-Wind~\citep{grb130603bkonus}. No broad band spectral data is publicly available 
for this burst to compare with simulated spectrum. \label{fig:grb130603b}}
\end{figure}

For the purpose of comparison the 4$^{th}$ model in Table \ref{tab:param} presents parameters of a 
simulation reproducing properties of the first peak of the bright short burst GRB 130603B, accompanied 
by a kilonova~\citep{grb130603bkilonova,grb130603bkilonova0}\footnote{This simulation 
must be considered only as a simulated burst {\it similar} to GRB 130603B because the parameter 
space was not extensively explored to find the best match. Notably, in contrast to GRB 170817A, no 
attempt was made to simulate the two peaks of this burst separately.}. The simulated light curves 
and spectrum of this model are shown in Fig. \ref{fig:grb130603b}. The difference between 
characteristics of this model and simulations of GRB 170817A is remarkable: the jet extent is 
$\sim 20$ folds (in comparison with model No. 1) or $\sim 40$ folds (in comparison with model 
No. 2) larger; the slow shell is 5 times denser; fraction of kinetic energy transferred to 
electrons weighed by electron yield, that is $\epsilon_e Y_e$ is 2 folds larger than in model 
No. 1; bulk Lorentz factor of ejecta is larger by a factor of 5 (in comparison with model No. 1) 
and by a factor of $50$ (in comparison to model No. 2); external magnetic field is $\sim 30$ times 
stronger than models No. 1 and 2. These differences are easily noticeable in 2D parameter space 
plots in Fig. \ref{fig:paramcorr}. We also notice that in comparison to data, simulation No. 4 in 
Table \ref{tab:param} somehow overestimates X-ray and soft gamma-ray emission. This may be due to 
the opacity of the high density jet for these soft photons and the absence of self-absorption in 
our code.

To see whether degeneracy in the value of Lorentz factor, which we found in simulations of GRB 
170817A, are also present in harder and brighter bursts, we attempted to simulate GRB 130603B with 
a Lorentz factor of $50$. We could not find any model with a flux as high as what was observed for 
this burst, a peak energy of $\sim 900$ keV in the rest-frame of the burst, and 
$Y_e\epsilon_e < 0.1$, which is motivated by PIC 
simulations~\citep{fermiaccspec,fermiaccspec0,fermiaccspec1}. For $\epsilon_e Y_e \sim 0.1$ the 
peak energy is not too far from observed value. However, with an 
electron yield $Y_e \lesssim 0.3$ expected in kilonova ejecta, the fraction of kinetic energy 
transferred to electrons must be $\sim 0.3$, which is too large to be inconsistent with prediction of PIC 
simulations. Therefore, it seems that in what concerns the apparent degeneracy of models with 
$\Gamma \sim 100$ and $\Gamma \sim 10$, GRB 170817 is an exception.

The fluence of GRB 130603B in Konus-wind 20 keV-10 MeV energy band was 
$E_{iso} = (2.1 \pm 0.2) \times 10^{51}$ ergs~\citep{grb130603bkonuseiso}, i.e. $\gtrsim 10^4$ times 
larger than in GRB 170817A. The peak energy of $E_{peak} \approx 900 \pm 100$ keV in the rest frame of this 
burst was $\approx 4$ times higher than the latter. We should also remind that $E_{iso}$ and $E_{peak}$ 
of GRB 130603B were not exceptionally high and present typical values for short bursts, see 
Fig. \ref {fig:allsgrb} and e.g.~\citep{gbmcat2012}. Therefore, we conclude that lower densities of 
shells and smaller jet extend of GRB 170817 with respect to more typical bursts were responsible for 
unusual properties of this GRB.

Finally, because no observation in low energies from trigger time up to few tens of thousands of 
seconds is available, we did not try to simulate afterglows of GRB 170817. Later afterglows are 
expected to be superimposed with emission from slow components such as wind and ejecta from a 
disk, and do not directly present properties of emission generated by external shocks during 
passage of the prompt relativistic jet through circumburst environment. In absence of any early data, 
it would be meaningless and confusing to make conclusions about an unobserved emission only 
based on theoretical assumptions. Nonetheless, modeling of late afterglows is by itself 
interesting, specially because this burst is the first short GRB with long and extended 
observation of its late afterglow. It may provide valuable information about the state of slow 
ejecta and their evolution with time. This a work in progress and will be reported elsewhere.

In conclusion, the interest of using the same phenomenological model for modeling multiple bursts 
is that one can compare their parameters and properties in the same theoretical setup. This allows 
to estimate the effect of variation of physical properties of jet and environment from burst to 
burst, and conclusions should be less affected by theoretical uncertainties than absolute value of 
parameters.

\section{Interpretation of prompt emission simulations} \label{sec:interp}
In Sec. \ref{sec:prompt} we divided candidate models and corresponding simulations of 
GRB 170817A into 3 categories according to their Lorentz factor, namely: 
mildly relativistic cocoon, off-axis view of a structured jet, and on-axis ultra-relativistic jet.
This classification is motivated by short GRB models and suggestions in the literature for the 
origin of this unusually soft and faint 
burst~\citep{gw170817fermi,gw170817swiftnustar,gw170817rprocess,gw170817cocoon}. 
In this section we discuss the plausibility of these hypotheses, based on the results of 
simulations presented in Sec. \ref{sec:gwsimul}.

\subsection {Mildly relativistic cocoon} \label{sec:cocoon}
From results presented in Fig. \ref{fig:notgood} and Table \ref{tab:notgood}, and discussions 
in Sec. \ref{sec:gwsimul}, it is clear that mildly relativistic cocoons with characteristics 
similar to what is suggested in~\citep{gw170817cocoon} cannot reproduce observed properties of 
GRB 170817A prompt emission. All the simulations with $\Gamma \sim 2-3$ and prompt shock at a 
distance of $\mathcal {O}(1) \times 10^{11}$ cm are too soft and have a duration $\gtrsim 10$ sec, 
too long to be consistent with observations of Fermi-GBM and Konus-Wind. However, we remind that 
this conclusion is for synchrotron/self-Compton emission generated by internal shocks in the 
cocoon. There is another version of cocoon model which associates the soft emission to breakout 
of the jet or outflow. This mechanism cannot be directly studied with our simulation code. 
Nonetheless, we in the next subsection we use phenomenological formulation 
of~\citep{grbcocoon,gw170817cocoon} to assess the plausibility of this model as an explanation 
for unusual characteristics of GRB 170817A.

\subsubsection {Cocoon breakout}  \label{sec:cocoonout}
Cocoon breakout model for faint GRBs~\citep{grbcocoon} assumes that the relativistic jet is chocked 
by thick envelop of a collapsing star or dense slow ejecta surrounding a binary merger. The 
intervening material traps radiation generated by shocks and other processes until its expansion 
reduce the opacity and release the trapped photons. Due to multiple scattering during their 
confinement, photons become either fully or partially thermalized and their spectrum become softer. 
In this respect, this model is a low energy analogue of standard fireball mechanism, in which the 
gamma-ray in GRBs is assumed to be due to $e^\pm$ annihilation when a leptonic plasma become optically 
thin. The spectrum of GRBs in fireball model contains a dominant thermal (black body) 
component~\citep{thermal2,thermalmod0,thermal3}.

In what concerns GRB 170817A, its spectrum, specially the first peak, is power-law with exponential 
cutoff~\citep{gw170817fermi}, thus consistent with a synchrotron rather than thermal emission. Indeed, 
observations show that synchrotron emission from relativistic shocks is the main contributor in prompt 
gamma-ray emission of almost all GRBs, although in some bursts addition of a thermal component - from 
photospheric or slower outflows - may improve spectral fit, see e.g.~\citep{grbspectfermi}.

The semi-thermal spectrum of cocoon breakout is presented by an effective temperature $T_{bo}$ and 
a bolometric fluence $E_{bo}$. Duration of emission $t_{bo}$ is a function of latter quantities and 
estimated as~\citep{grbcocoon,gw170817cocoon}:
\be
t_{bo} \sim 1sec (\frac{E_{bo}}{10^{46}~{\text erg}})^{\frac{1}{2}} 
(\frac{T_{bo}}{150~{\text keV}})^{-\frac{9+\sqrt{3}}{4}}  \label{cocoondure}
\ee
Assuming that in GRB 170817A $T_{bo} \sim E_{peak} \sim 230$ keV, with a bolometric fluence of 
$E_{bo} \sim 10^{47}$ ergs, we obtain $t_{bo} \sim 0.8$ sec $\ll 2$ sec duration of this burst. A lower 
$T_{bo}$ can increase duration of emission, but will have problem to explain detection of high energy 
photons up to $\gtrsim 1$ MeV\footnote{Indeed~\citep{gw170817cocoon} use $T_{bo} \sim 150$ keV in 
their analysis to obtain a longer prompt emission. However, they do not present any comparison 
of their model with the Fermi-GBM or Integral-IBIS prompt gamma-ray data.}. For having such a high 
effective temperature, the terminal Lorentz factor before breakout must be 
$\Gamma_f \sim T_{bo} / 50 {\text keV} \sim 5$~\citep{grbcocoon}, which is larger than the value used 
in the literature and in our cocoon simulations. Moreover, using the relation between final Lorentz 
factor and star radius - or in the case of short bursts distance of circum-merger material from 
center, $\Gamma_f \sim 30 M_5^{0.14} R_5^{-0.27}$, where $M_5$ is the mass of material in units of 
$5M_\odot$ and $R_5$ is distance from center in units of $5R_\odot$~\citep{grbcocoon}, we find a 
distance of $\sim 60 R_\odot \sim 4 \times 10^{12}$ cm for $M_5 \sim 0.006 \sim 0.03M_\odot$, which 
corresponds to estimated mass of polar outflow from kilonova 
AT 2017 gfo~\citep{gw170817optkilonovath,gw170817bluekilonova,gw170817optkilonova}. If the material 
is ejected at the time of merger collapse to BH, then there had to be $\lesssim 100$ sec of delay 
between the end of GW and detection of GRB. This is much longer than the observed delay of 
$\sim 1.7$ sec~\citep{gw170817fermi,gw170817integral}. In conclusion, breakout of a mildly 
relativistic cocoon may explain or contribute in the early low energy emissions, but cannot 
explain the observed prompt gamma-ray.

\subsection {Off-axis jet} \label{sec:offaxis}
Table \ref{tab:param} shows that if $\Gamma \sim 10$, other parameters, in particular 
$\epsilon_e Y_e$, can be adjusted to obtain a burst with properties of GRB 170817A prompt emission. 
If this apparent low Lorentz factor is due to an off-axis view of the ultra-relativistic jet, the 
relation between emitted and received power~\citep{emissionbook}:
\be
\frac{\frac{dP_e}{d\omega d\Omega}}{\frac{dP_r}{d\omega d\Omega}} = \frac{1}{\Gamma^2 (1+\beta 
\cos \theta_v)} \label{recpower} 
\ee 
where $\theta_v$ is angle between a far observer and jet axis, shows that an off-axis view of the 
jet alone is not enough to explain the faintness of GRB 170817A. On the other hand, a structured 
jet by definition has lower Lorentz factor at high latitudes and may explain a softer and a few 
orders of magnitude smaller fluence of this burst in comparison to typical short 
bursts, shown in Fig. \ref{fig:allsgrb}. However, in this case one expects significant brightening 
of the afterglow when the dissipation of energy through collision of the jet with circumburst material 
reduces beaming and makes the central region of the jet visible to a far off-axis 
observer~\citep{grboffaxis}. Such a brightening is not observed for GRB 170817A, and as we discussed 
in Sec. \ref{sec:xray} the slight late brightening in X-ray is not unique to this burst and has been 
seen in bright, presumably on-axis, bursts. For instance, in the case of GRB 130603B the brightening 
of the X-ray afterglow was observed at $\sim 10$ days after prompt emission~\citep{grb130603bxray}, 
which is roughly the same epoch as that of GRB 170817A. Therefore, we conclude that the late afterglow 
cannot be uniquely associated to interaction of the relativistic jet with circumburst material and 
its brightening due to the opening of an off-axis jet, as predicted by~\citep{grboffaxis} formulation.
Furthermore, the UV/optical/IR counterpart AT 2017 gfo indicates a dominantly thermal ejecta of mass 
$M_{ejecta} \sim 0.03-0.05 M_\odot$ at $\sim T+0.6~-~T+1.5$ 
days~\citep{gw170817optkilonova,gw170817bluekilonova,gw170817rprocess}. A relativistic jet of mass 
$M_{jet} \sim 10^{-6} M_\odot$~\citep{gw170817cocoon,gw170817swiftnustar} and its 
synchrotron/self-Compton emission cannot explain neither the observed flux nor the features 
seen in the low energy spectrum. Thus, additional ejecta components must be involved in the 
production of these emissions. Indeed, evidence for additional ejecta, such as an expanding thermal 
cocoon, expanding envelop, and/or evaporating disk are found in other GRBs 
too~\citep{grbthermalcocoon,grbtejecta}.

\begin{figure}
\begin{center}
\begin{tabular}{p{7cm}p{7cm}}
\includegraphics[width=7cm]{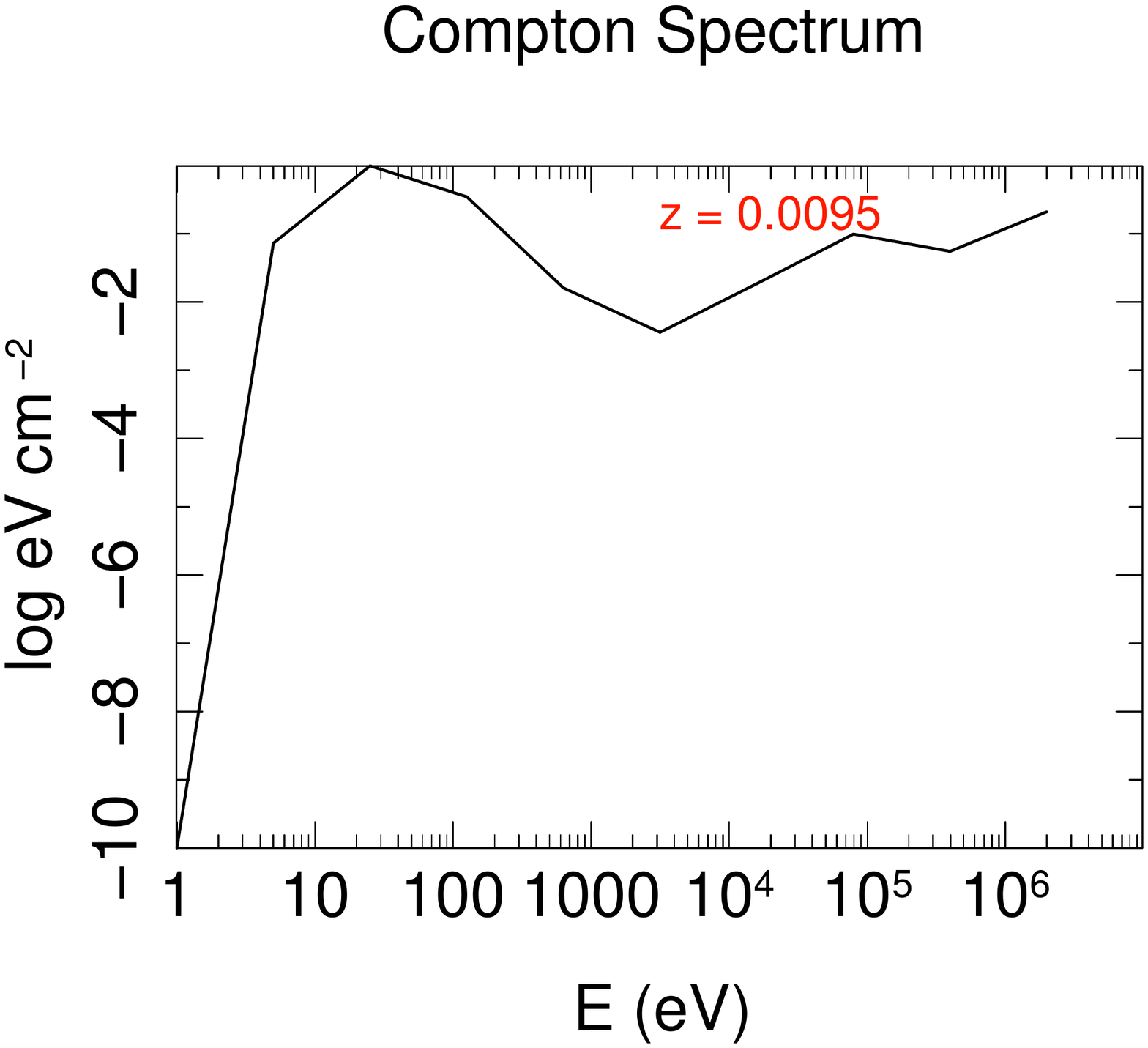} &
\includegraphics[width=7cm]{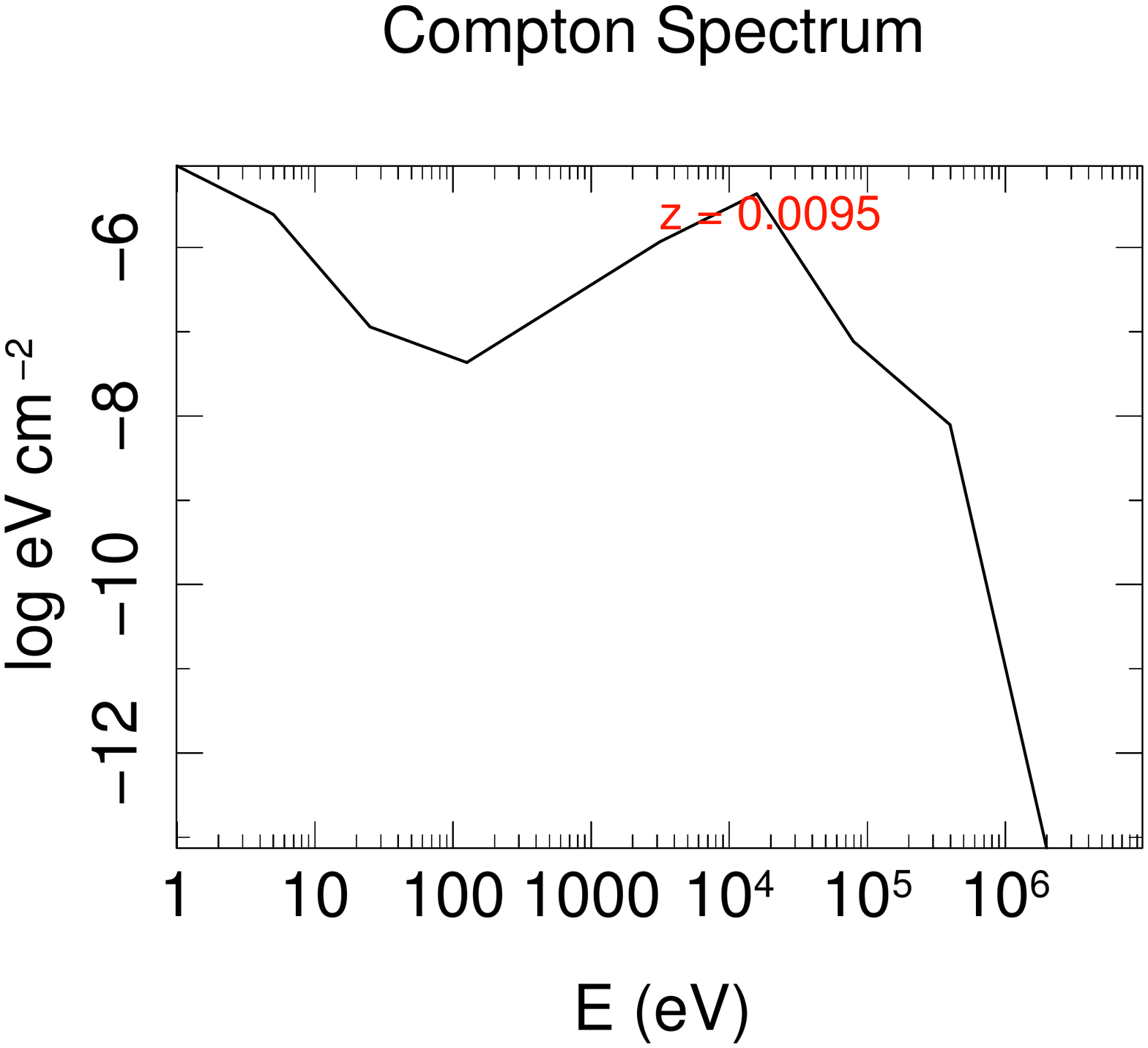}
\end{tabular}
\end{center}
\caption{Spectrum of Compton scattering of photons by fast electrons in simulations No. 1 (left) 
and No. 2 (right). \label{fig:compton}}
\end{figure}

The prompt gamma-ray emission cannot be due to scattered photons either. Fig. \ref{fig:compton} shows 
the spectrum of scatted photons through Compton scattering by fast electrons for simulations 
No. 1 and 2 in Table \ref{tab:param}. The shape of these spectra deviates significantly from observed 
spectrum. Notably, the dominant peak of Compton spectrum is at much lower energies and its amplitude 
is few orders of magnitude less than synchrotron emission for both high and low $\Gamma$ simulations 
No. 1 and 2, respectively. Our simulation code does not include Compton scattering of photons by 
hadrons. However, simulations of this process~\citep{comptonhadron,comptonhadron0} shows that similar 
to leptonic case its effect is dominant at high energy tail i.e. $E \gtrsim 1$ MeV and low energies, 
but not in $10^2-10^3$ keV energy band, which includes the peak energy of most short and long GRBs. 
In addition, no high energy excess, which could be associated to hadronic processes was detected 
in GRB 170817A. Therefore, we conclude that the prompt gamma-ray of this burst couldn't be due to 
scattered of higher energy photons from an off-axis jet.

Alternatively, the GRB forming jet of GW 170817 event might have intrinsically a low Lorentz factor 
of order $\Gamma \sim \mathcal{O}(10)$, rather than $\Gamma \sim \mathcal{O}(100)$ estimated 
for most short GRBs by the phenomenological model and simulations of~\citep{hourigrb,hourigrbmag}, 
see also parameters for GRB 130603B in Table \ref{tab:param}, and estimation of Lorentz factor of 
GRBs in literature~\citep{grblorentz,grblorentz0,grblorentz1,grblorentz2}. Indeed, in 
Sec. \ref {sec:prompt} we reported simulations consistent with the first peak of GRB 170817A with 
both $\Gamma \sim 100$ and $\Gamma \sim 10$. However, Table \ref{tab:param} shows that to 
compensate for low Lorentz factor $\Gamma \sim 10$, the efficiency of energy transfer to electrons 
(more generally charged leptons) must be $\sim 3$ times larger than simulations with 
$\Gamma \sim 100$. On the other hand, as we discussed in Sec. \ref{sec:compgrb130603b}, according 
to our simulations the density of colliding shells in GRB 170817A were more than an order of 
magnitude smaller than e.g. GRB 130603B modeled in the same manner, and from simulated short GRBs 
in~\citep{hourigrbmag} with typical fluence and peak energy, as e.g. one can conclude from 
Fig. \ref{fig:allsgrb}. 

The lower shell density with respect to other short GRBs is necessary to reproduce smaller fluence 
and peak energy of GRB 170817. In this case, transferring a large fraction of kinetic energy of 
baryons to electrons seems even more difficult than when shells are denser. Due to smaller 
cross-section for scattering of particles in a diluted fluid, colliding shells are expected to 
produce less turbulence and weaker induced electromagnetic field in the shock front. Moreover, 
low electron yield of neutron rich BNS ejecta should make the transfer of kinetic energy to 
electrons even harder. Estimation of electron yield $Y_e$ for various components of the ejecta 
of GW 170817 event based on the observation of r-process 
products~\citep{nstarmergeejecta,gw170817rprocessth} are: $Y_e \sim 0.1 - 0.4$ for dynamical 
component, $Y_e \sim 0.3$ for wind, and $Y_e \sim 0.25$ in another wind 
component~\citep{gw170817rprocess}. Considering these yields, and the value of $\epsilon_e Y_e$ 
in low Lorentz factor simulation No. 2, the effective fraction of kinetic energy transferred to 
electrons $\epsilon_e$ had to be $\sim 0.1-0.3$ to generate a prompt gamma-ray emission with 
characteristics of GRB 170817A. However, PIC simulations predict that in relativistic shocks 
$\epsilon_e \lesssim 0.1$~\citep{fermiaccspec,fermiaccspec0,fermiaccspec1}. Therefore, the value 
of $\epsilon_e$ in simulations with low Lorentz factor are at best marginally consistent with 
PIC simulations. On the other hand, simulations of BNS merger 
predict~\citep{nsmergerrprocsimul,nsmergerrprocsimulhres,nsmergerrprocsimul0} that relativistic 
jet is formed from low electron yield dynamical poleward ejecta. In this case, the value of 
$\epsilon_e$ in our simulations with $\Gamma \sim 10$ would be implausibly high. 

One may criticize the above argument because simulations of relativistic shocks with PIC method 
are still very far from being realistic and predictions for $\epsilon_e$ mentioned above are either 
concluded from PIC simulations of $e^\pm$ plasma~\citep{fermiaccspec}, or from simulations with 
smaller mass ratio between opposite charges than in electron-proton 
plasma~\citep{fermiaccspec0,fermiaccspec1,picsimul}. 
Nonetheless, giving the fact that electromagnetic interaction is independent of mass, one expects 
that in a baryon dominated shock the transferred energy to charged leptons make up an even smaller 
fraction of total available energy than in $e^\pm$ case, despite the fact that the total energy 
transferred to electromagnetic fields would be larger for the same relative Lorentz factor of 
colliding shells. Thus, the value of $\epsilon_e$ in short GRB jets may be even less than the 
finding of PIC simulations.

Based on these arguments we conclude that it is unlikely that a structured jet viewed off-axis in 
a region with $\Gamma \sim 10$ can explain the observed properties of prompt gamma-ray of 
GRB 170817A, unless the viewing angle was not too far from on-axis or the jet was not strongly 
structured and Lorentz factor was $\sim 100$, which is only a few folds less than brighter short 
bursts, such as GRB 130603B. 

It is not possible to compare the above conclusion with e.g. analyses of late afterglows 
by~\citep{gw170817lateradio} and~\citep{gw170817latebroad}, because there is no evidence of direct 
relation between the ejecta responsible for these emissions and the prompt ultra-relativistic jet. 
For instance,~\citep{gw170817lateradio} assume a reduction of Lorentz factor from $3.5$ at early 
times to $2.5$ at the time of their observations at $\sim T+100$ days, and model the afterglows by 
a shock with $\epsilon_e = 0.1$ and $\epsilon_B = 0.01$ or $\epsilon_B = 0.003$. However, according 
our arguments about $\epsilon_e$, the value chosen by~\citep{gw170817lateradio} is at upper limit 
of expected range. Moreover, the initial Lorentz factor is suitable for slow outflows and too low 
for the production of the prompt gamma-ray. The values of density, $\epsilon_e$ and $\epsilon_B$ 
reported in~\citep{gw170817latebroad} are consistent with findings of PIC simulations and 
simulations of afterglow using phenomenological model of~\citep{hourigrbmag}, which will be 
reported elsewhere. But they do not provide any information about the state of ejecta/jet 
responsible for the prompt gamma-ray. Notably, they do not provide any value for the Lorentz 
factor at any epoch to compare with our findings.

\subsection {Ultra-relativistic jet} \label{sec:ultrareljet}
Finally, after disfavoring other models, we conclude that the most physically plausible origin 
of GRB 170817A is synchrotron/self-Compton emission from internal shocks in an ultra-relativistic 
jet, which according to the phenomenological model of~\citep{hourigrb,hourigrbmag} its bulk Lorentz 
factor was $\sim 100$ and densities of colliding shells as reported in simulation No. 1 in 
Table \ref{tab:param}. These values are, respectively, few folds and more than one order of 
magnitude less than what is expected for typical short GRBs according to the same model. We notice 
that~\citep{gw170817grbonaxis} have also arrived to a similar conclusion by analysing afterglows 
of GW/GRB 170817. According to simulation No. 1 in Table \ref {tab:param}) such a jet needs 
$\epsilon_e Y_e \sim 0.01$. Considering estimated values for $Y_e$ of various ejecta of GW 170817 
event, we obtain $\epsilon_e \sim 0.03 - 0.1$, which is comfortably in the range of values observed 
in PIC simulations. As mentioned in Sec. \ref{sec:offaxis}, we cannot rule out that a somehow off-axis 
view of a mildly structured jet was responsible for reduced Lorentz factor and densities. However, 
based on arguments given in Sec. \ref{sec:faintness}, it is more plausible that intrinsic properties 
of progenitor neutron stars and dynamics of their merger were responsible for the faintness of 
GRB 170817A. The lack of bright short GRBs at low redshifts is also an additional evidence for this 
conclusion, see Fig. \ref{fig:allsgrb}-c,d and Fig. 28 in~\citep{batcat}, which shows the 
distribution of average flux of Swift-BAT GRBs with known redshift(both short and long) and the 
absence of intrinsically bright bursts at low redshifts.

In the next section we use results from observation of neutron stars, GRMHD simulations 
of BNS merger, and simulations of jet acceleration to assess what might have been different in 
GW 170817 event with respect to progenitor of other short GRBs.

\section {Implication for properties of GW 170817 progenitor} \label{sec:progdy}
Which properties of the progenitor neutron stars of GW 170817 event and their merger may have 
been responsible for lower than usual Lorentz factor of relativistic jet and its lower density, as 
concluded from our modeling of GRB 170817A ? To answer this question we need a full theoretical and 
numerical formulation of neutron star physics and NS-NS merger, including: equation of state and 
interactions of neutron rich material under strong gravity force; magnetic field of progenitor BNS 
and its evolution during merging event; dynamics of merging, specially its latest stages before 
formation of a Hyper Massive Neutron Star (HMNS) or a black hole; evolution of accretion disk; 
evolution of pressurized neutron rich ejecta and its interactions with radiation and neutrino fields; 
and processes involved in acceleration of particles in the ejecta to ultra-relativistic velocities 
and formation of a relativistic jet. 

Such task highly exceeds our analytical and numerical calculation capabilities. In addition, NS-NS 
and NS-BH mergers occur at distances of order of few tens of kilometers, whereas particle 
acceleration occurs in a magnetically loaded outflow along a distance of at least few orders of 
magnitude longer~\citep{grbjetsimul1,grbjetsimul}. These two scales cannot be numerically treated 
with same precision in a same code. Therefore, for relating the results of our modeling of high 
energy electromagnetic emission and other observations of GW/GRB 170817A to properties of 
its progenitor, we have to rely on partial and far from ideal models and simulations, which only 
allow a qualitative assessment of progenitor's characteristics. For this purpose we use mostly, but 
not exclusively, the results of simulations reported in~\citep{nsmergerrprocsimul0} for NS-NS merger 
and those of~\citep{grbjetsimul} for jet acceleration.

\subsection {Equation of State (EoS)} \label{sec:eos}
It is by far the most important characteristic of neutron stars and defines the relation between 
their mass and radius. It also determines other properties such as core and crust densities, 
tidal deformability, which affects ejecta mass, density, and buoyancy during merger, differential 
rotation, maximum mass, magnetic field, and formation of a HMNS and its 
lifetime~\citep{nstarstaterev}. LIGO-Virgo analysis of GW 170817 event disfavors stiff equations 
of state~\citep{gw170817ligo}. Simulations of NS-NS merger in~\citep{nsmergerrprocsimul0} are 
performed for two equations of state: IF-q\footnote{Nomenclature used 
in~\citep{nsmergerrprocsimul0}.} consisting of a single polytropic fluid with polytropic index 
$\Gamma_p = 2$ and polytropic constant $K = 100$; and Hyperon-rich H4 
model~\citep{nstareoshyperobn}. According to classification of NS states in~\citep{nsstateparam}, 
IF-q and H4 are prototypes of soft and stiff equations of state, respectively. We use the results of 
our simulations to assess which of these models is better consistent with GW/GRB 170817A.

In what concerns state dependent properties, which may affect electromagnetic emission from merger, 
for close mass NS progenitors the density of inner part of the accretion disk and poloidal magnetic 
field of the merger are lower in IF-q case than H4~\citep{nsmergerrprocsimul0}. Although currently 
no systematic study of the impact of the equation of state on the properties of polar outflow is 
available, it is known that it is closely related to magnetic field, mass, density, extent of 
accretion disk~\citep{diskjetcorrel}, and accretion rate~\citep{blandfordzenajek}. The smaller 
value of these quantities in IF-q means that it also generates less outflow. Thus, we conclude that 
an equation of state similar to IF-q better represents the state of GW/GRB 170817A progenitor. This 
independent assessment of EoS is consistent with GW observations, which finds that H4 falls just 
on the 90\% exclusion probability curve for both fast and slow rotating progenitor 
BNS~\citep{gw170817ligo}.



\subsection {Strength of magnetic field} \label{sec:mag}
Simulations of~\citep{nsmergerrprocsimul0} show that for equal mass BNS, after the collapse of 
HMNS to black hole if the initial magnetic fields of progenitors are aligned with each other and 
anti-aligned with the rotation axis of the BNS (case DD in nomenclature 
of~\citep{nsmergerrprocsimul0}), the average poloidal magnetic field is about 5 times weaker than 
if both initial fields are aligned with rotation axis (case UU). If the initial fields are 
anti-aligned with each other (case UD), the average poloidal field is even smaller by a few folds. 
Moreover, in DD and UD cases, the average field at $|\theta| \lesssim 20^\circ$, where $\theta$ 
is the angle between magnetic axis and rotation axis of the merger, is a few times weaker than 
UU case. A reduced magnetic field proportionally reduces attainable Lorentz factor for material 
ejected close to polar direction~\citep{grbjetsimul}.

If the progenitor neutron stars of GW/GRB 170817A had dipole magnetic fields which extended out of 
their surface, they should have been surrounded by strongly magnetized atmospheres before their 
merger. In this case, the magnetic interaction during close encounter of the stars might have 
disaligned their fields well before the last stages of inspiral, and at the time of merging they 
were in a state close to UD in the simulations of~\citep{nsmergerrprocsimul0}. 
Moreover, considering the old population of the host galaxy NGC 4993, which have an estimated 
minimum age of $\gtrsim 1$~Gyr~\citep{gw170817ligohost,gw170817ligohost0,gw170817ligohost1}, 
magnetic fields of progenitors could have been as low as $10^8 - 10^9$ G and fast precessing, if 
the progenitors were recycled millisecond pulsars. The field could be even smaller if they had 
evolved in isolation~\citep{nstarrev,nstarrev0}. Such initial field strength is much 
smaller than $|B| \sim 10^{12}-10^{15}$~G used in simulations of BNS merger. Therefore, the magnetic 
field of the short lived HMNS and accretion disk of the final black hole of GW 170817 also could have 
been a few orders of magnitude less than $|B| \sim 10^{15}-10^{16}$~G seen in the 
simulations~\citep{nsmergerrprocsimul,nsmergerrprocsimulhres,nsmergerrprocsimul0,nsmergerrprocsimul1}.

\subsection {Disk/torus, jet, and accretion rate} \label{sec:outflow}
Density and initial Lorentz factor of magnetically collimated polar outflow is expected to depend on 
the Poynting energy carried by the flow. As discussed above, simulations performed with high initial 
magnetic field of $\sim 10^{15}$ G and equal mass progenitors~\citep{nsmergerrprocsimulhres} 
generate a large magnetic field of $\sim 10^{16}$ G for the merger and a relatively large initial 
Lorentz factor of $\Gamma_i \sim 4$ for the outflow. Simulations with smaller initial magnetic field 
of $\sim 10^{12}$ G attain a magnetic field of $\sim 10^{13}$ G on the disk and a polar outflow with 
an axial velocity of $\sim 0.3~c$, where $c$ is the speed of light~\citep{nsmergerrprocsimul0}. 
Thus, we expect that if the initial masses of progenitors of GW 170817 were close to each other and 
their initial magnetic fields similar to those of millisecond pulsars, the magnetic field of the 
merger could be $\lesssim 10^{10}$~G, which is few orders of magnitude less than what is expected 
for younger progenitors. Although the velocity of blue ejecta in GW 179817 event is estimated to 
have been $\sim 0.3~c$~\citep{gw170817rprocess,gw170817bluekilonova,gw170817kilonovaspeed}, which is 
similar to what is obtained in simulations with a merger magnetic field of $\sim 10^{13}$~G, 
a larger disk mass and/or density might have partially energized the outflow. The observed low 
initial velocity of polar outflow and possibility a low magnetic field imply reduced acceleration of 
particles at high altitudes, and thereby a thin relativistic jet with low Lorentz factor, which could 
generate a soft and faint GRB consistent with observations and our estimations for characteristics of 
the ultra-relativistic jet.

In addition to Lorentz factor, the fluence of a GRB depends on the jet extend, i.e. the total 
amount of ejected and accelerated material. According to simulations of~\citep{nsmergerrprocsimul0} 
equal mass NS-NS mergers generate less massive and more diluted disks - by a 
factor of $\sim 100$ in their inner part - than mergers with a mass ratio of $\sim 0.8$. 
In GW/GRB 170817A event progenitor masses were not equal but were close to each other: 
$M_1 \sim 1.36-1.60~M_\odot$ and $M_2 \sim 1.17-1.36~M_\odot$, leading to $M_2/M_1 = 0.855 \pm 0.095$. 
The upper limit of this mass ratio is close to 1. Thus, the merger might have ejected much less 
material than NS binaries with larger mass difference or NS-BH mergers, which based on observations 
of BH-BH merger, are expected to have much larger mass difference. Moreover, relativistic MHD 
simulations of magnetized jet in~\citep{grbjetsimul} show that the reduction of initial total 
kinetic and Poynting energy by a factor of 2 reduces the density of outflow with highest Lorentz 
factor by a factor of 5 or so. Both of these observations are consistent with reduced shell 
densities and extend, and reduced Lorentz factor found in our simulations of GRB 170817A. 

Simulations reported in~\citep{nsmergerrprocsimulout,nsmergerrprocsimul0} show that a poloidal 
coherent magnetic field and an outflow funnel begin to form after the collapse of HMNS to a black 
hole and outflow rate is correlated with accretion rate from 
disk/torus~\citep{blandfordzenajek,diskjetcorrel}. Giving the fact that the outflow had to be 
accelerated gradually at high latitudes, a delay between the end of GW and generation of a 
relativistic jet is expected. It had to be inversely proportional to the strength of poloidal magnetic 
field and the injection velocity. To this acceleration delay one has to add the time delay between 
ejection of density shells and their collision~\citep{hourigrbmag}. Furthermore, if the accretion 
disk was low density and diluted, the accretion rate, and thereby the growth of anisotropies in the 
ejecta might have been slower than in cases with higher magnetic field and faster accretion. These 
delays explain the observed delay between the end of GW 170817 maximum and trigger time of 
GRB 170817A. The origin of the delay is also consistent with arguments we raised to explain low jet 
density and relatively long duration of GRB 170817A. 



\subsection {Effect of initial spin} \label{sec:spin}
Initial spins of progenitor neutron stars have a crucial role in the dynamics of merging process, 
in particular in the amount of ejecta, density and extent of accretion disk/torus, and spin of HMNS 
and black hole. Moreover, they provide information about formation and history of the BNS. 
Gravitational waves from a merger contain information about spins of progenitors and their 
alignment with orbital rotation axis, see e.g. simulations in~\citep{nstarmergespin,nstarmergespingw,nstarmergespingw0}. However, in the case of GW 170817 the weakness of the signal and a glitch in 
LIGO-Livingston data prevented quantitative estimation of progenitors spins. 

Binding energy of NS-NS merger is stronger(weaker) for anti-aligned(aligned) initial spins with 
respect to orbital axis, and leads to shorter(longer) inspiral regime and smaller(larger) ejecta, 
but significantly depends on the mass ratio of progenitors and is smaller in equal 
mergers~\citep{nstarmergespin}. Thus, the direction of differences are similar to those of magnetic 
field discussed in Sec. \ref{sec:mag}. However, spin effect on the amount of ejected 
material is subdominant with respect to other processes and amounts to only few 
percents~\citep{nstarmergespin}. On the other hand, the spin of BNS affects the precision of mass 
determination from gravitational wave observations~\citep{binarynstarspin,gw170817ligoprogen}. 
In the case of GW 170817, variation of normalized angular momentum $a \equiv J/GM^2$, where $J$ is 
the angular momentum, in the range $0.05 < a < 0.89$ results to mass ranges 
$M_1 \sim 1.36-2.26 M_\odot$ and $M_2 \sim 0.86-1.36 M_\odot$ 
for NS progenitors~\citep{gw170817ligoprogen}. However, an error of $\lesssim 50\%$ on the masses of 
progenitors and thereby the ejecta alone cannot explain orders of magnitude faintness of 
GRB 170817A, unless the equation of state changes drastically with mass. Therefore, despite small 
effect of spins on outflow and jet, their anti-alignment with orbital directions is better consistent 
with the weak jet of GRB 170817A.

In conclusion, the observed properties of GW and electromagnetic emissions of GW 170817 event are 
consistent with each others and with our estimation of jet properties in sections \ref{sec:prompt} 
and \ref{sec:interp}.

\section{Implication of progenitors properties for afterglow of GW/GRB 170817 and other short GRBs} \label{sec:implic}
Formation of a GRB is the manifestation of just one component of complicated events which occur during 
merger of binary neutron stars. Therefore, any argument for unusual properties of GRB 170817A must 
be also consistent with low energy afterglows and emissions from other components of the merger 
remnant. Here we verify whether properties of the progenitor neutron stars and their merger discussed 
in the previous section, which may explain the faintness of GRB 170817A prompt gamma-ray, are 
compatible with low energy observations.

\subsection {Evidence from UV/optical/IR/radio counterparts} \label{sec:evid}
From UV/optical/IR/radio observations various conclusions are made in the literature about 
properties of the progenitors and their merger, which are not always consistent with each other 
and with numerical simulations. Here is a summary of conclusions and some of inconsistencies:
\begin{description}
\item{-} The merger made a HMNS which after $\sim 10$ msec or so collapsed to black hole. This is 
a common conclusion in the literature~\citep{gw170817optkilonovath,gw170817rprocess}. The strongest 
evidence is the fact that much larger ejecta - presumably from accretion disk around a black hole - 
is necessary for explaining observed luminosity of the optical counterpart AT 2017 gfo than tidally 
stripped tail of the merger can provide~\citep{nstarmergespin,nstarmergespingw,nstarmergespingw0,kilonovaopacity1,kilonovaopacity0,kilonovaopacity}.
\item{-} According to predictions of theoretical models, AT 2017 gfo was a red kilonova, meaning that 
heavy r-processes occurred in a dense optically thick material ejected from an accretion 
disk/torus~\citep{gw170817optkilonovath}.
\item{-} The early bright blue/UV emission~\citep{gw170817swiftnustar} observed at 
$\lesssim T+1.5$ days is from a Lanthanide-free low density post-merger squeezed polar wind 
consisting of light elements and having a relatively large electron yield of 
$Y_e \sim 0.25-0.3$~\citep{nstarmergeejecta,kilonovaopacity,gw170817optkilonovath,gw170817rprocess}. 
Observation of this component may imply that the viewing angle of observer must have been close 
to polar to be able to detect it. As in NS-NS merger the toroidal field is always much stronger than 
poloidal one~\citep{nsmergerrprocsimul0}, polarization of photons should be mainly parallel to the 
jet axis. The absence of linear polarization even at early times~\citep{gw170817bluekilonovapol} is 
an evidence for scattering of photons in a turbulent funnel rather than direct sideway view of the 
ejecta on the surface of sky.
\item{-} There is not a general consensus about the amount of ejected mass and contribution of 
different components: dynamical tidal tail; poleward outflow, cocoon, and wind; and post-merger 
close to spherical ejecta due to the heating of the accretion disk/torus. Observations can only rule 
out a one-component thermally evolving ejecta~\citep{gw170817optkilonova,gw170817rprocess}.
\item{-} Velocity of ejecta defined as $v_{ej} \equiv E/2M_{ej}$ is estimated to be 
$v_{ej} \sim 0.2-0.3~c$ at $\sim T+1.5$ days. Simulations predicts such a velocity for poleward 
dynamical tail material at 
ejection~\citep{nsmergerrprocsimul,nsmergerrprocsimul0,nsmergerrprocsimul1,nsmergerrprocsimulout}.
\item{-} Based on analysis of optical/IR/radio observations of AT 2017 gfo the ejecta mass 
responsible for the observed r-process rich spectrum is estimated to be as high as $M_{ej} \sim 
0.03 - 0.05 M_\odot$ and its velocity as mentioned 
above~\citep{gw170817rprocess,gw170817optkilonova,gw170817bluekilonova}. However, this is much 
larger than $\sim 0.01 M_\odot$ predicted by simulations for the fast outflow. It is also 
$\sim 3-5$ times larger than tidal ejecta mass estimated for the bright short GRB 130603B, 
which was accompanied by a kilonova event~\citep{grb130603bkilonova,grb130603bkilonova0}.
\item{-} At about $T+100$ days X-ray, optical and radio afterglow do not show any signature 
of a relativistic jet~\citep{gw170817lateradio,gw170817latexary,gw170817latexraystructjet}. 
These observations are consistent with a slow evolving ejecta component and with a weak 
ultra-relativistic jet, which at this late times should be most probably dissipated by interacting 
with circumburst material. 
\end{description}
The tidally stripped dynamical ejecta is expected to be cold and to have high $Y_e$ and 
light-element composition due to interaction with released neutrinos ~\citep{nstarmergespingw0}. 
These predictions are consistent with observations. However, GRMHD simulations of BNS merger 
estimate a mass of $10^-{3} - 10^{-2}~M_\odot$ for dynamical ejecta, irrespective of progenitors mass 
ratio and equations of state~\citep{nstarmergespin,nstarmergespingw,nstarmergespingw0,gw170817optkilonovath}. 
Therefore, a contribution from post-merger ejected material from an accretion disk/torus seems 
necessary to explain the data. Moreover, this additional early ejecta should become optically 
thin and observable in UV and visible bands as early as 
$\sim T+0.6$ days~\citep{gw170817optkilonova,gw170817rprocess}. However, simulations also predict 
that post-merger wind would have a low velocity of $0.02-0.1~c$ and high opacity~\citep{nstarmergespingw,nstarmergespingw0,nstarmergeejecta,kilonovaopacity0,kilonovaopacity}. 

It is shown~\citep{gw170817rprocess} that the optical/IR spectrum during $T+0.6$~days to 
$T+1.5$~days can be reproduced by a 3-component model constructed according to simulations 
of~\citep{kilonovaopacity0,kilonovaopacity}: a $Y_e = 0.1-0.4$ component representing dynamical 
tidal tail ejecta with a velocity of $\sim 0.2~c$; and two components with $Y_e = 0.25$ and 
$Y_e = 0.3$ and a low velocity of $\sim 0.05~c$ representing post-merger ejecta. A scaling of the 
simulated spectra, which was performed for $M_{ej} = 0.01~M_\odot$, is necessary to obtain a correct 
amplitude for the spectrum of At 2017 gfo. However, thermal evolution of this model does not 
reproduce later spectra~\citep{gw170817rprocess}. Moreover, even in the earliest time interval, the 
contribution of slow components - presumably from disk - is subdominant and does not solve the 
problem of too large ejecta mass mentioned above. Therefore, despite overall agreement, current 
predictions of BNS merger simulations poorly fit the AT 2017 gfo data~\citep{gw170817rprocess,gw170817optkilonova,gw170817bluekilonova,gw170817optsss1,gw170817optsss,gw170817optdlt}. 

We should also remind that in~\citep{nsmergerrprocsimul0} a velocity of $\sim 0.3~c$ for post-merger 
magnetically loaded polar outflow is reported only for H4 equation of state. For a softer state the 
velocity is expected to be less. In addition, as we discussed earlier, a weaker magnetic field reduces 
both the amount and velocity of the outflow at ejection time~\citep{diskjetcorrel}. And although 
the ejecta continues to be somehow collimated, it should have a larger opening angle. Because the 
polar wind is accelerated by dissipation of Poynting energy after its ejection~\citep{grbjetsimul}, 
the effect of low ejection velocity may be smeared out by acceleration at high latitudes. At present 
simulations of BNS merger does not include these late processes. Moreover, acceleration of charged 
particles segregates them from neutron rich component and increase the effective $Y_{ej}$ in the fast 
outflow. Energy dissipation of neutrinos may also be involved in increasing the velocity of 
initially slow ejecta from accretion disk~\citep{nstarmergnueff}. However, most GRMHD simulations 
do not include a full treatment of neutrinos and their predictions may be unrealistic.

Another solution for resolving inconsistencies, as we discussed in Sec. \ref {sec:xray} and also 
suggested in~\citep{gw170817optkilonova}, is a contribution from the afterglow of GRB 170817A in the 
observed blue peak during $T+0.6$~days to $T+1.5$~days. In this case, less early ejecta  
would be necessary to explain observations. Indeed, analysis of bolometric light curve 
in~\citep{gw170817bluekilonova} shows that if the bluest data points of the spectrum at 
$\sim T+0,6~-~T+1.5$~days are not included in the fit and a thermalization efficiency 
is added to the model, a smaller ejecta mass of $\sim 0.018 M_\odot$ and a larger opacity - 
a signature of higher atomic mass elements, presumably from accretion disk - fit the data 
better. Thus, the blue part of the early optical spectrum needs another component of the ejecta. 
This analysis is another evidence that the early non-thermal blue emission was at least partially 
due to the afterglow of the relativistic jet, which at times $\gtrsim T+0.5$ days was 
significantly slowed-down by shocks and internal dissipation. Even at $\sim T+110$ days observation 
of optical counterpart shows that it is too bright and blue to be consistent with kilonova emission 
alone and a contribution from GRB 170817A afterglow seems necessary to explain 
observations~\citep{gw170817lateopt}. Unfortunately, in absence of an early observation of optical 
and X-ray afterglow, it is not possible to estimate the contribution of synchrotron emission from 
the GRB afterglow in the observed blue peak.

Late observations of optical/IR~\citep{gw170817lateopt} and 
X-ray~\citep{gw170817latexary,gw170817latexraystructjet} observations of GW 170817 counterparts 
are found to be consistent with off-axis structured jet~\citep{gw170817latexraystructjet} and 
with both the latter model and cocoon models~\citep{gw170817latexary}. Moreover, with the assumption 
of an off-axis viewing angle and structured jet, a broad band analysis of late afterglows rules 
out top hat (uniform) and power-law jet profiles and simplest cocoon model~\citep{gw170817latebroad}, 
but finds that a Gaussian jet profile or a cocoon with energy injection are consistent with data.
However, the models studied in~\citep{gw170817latexary} are only at 2-sigma level consistent with 
the early epoch Chandra data. Additionally, radio observations at roughly the same epoch of X-ray 
and optical/IR observations are not consistent with off-axis view of a structured jet and need a 
quasi-spherical mildly relativistic outflow~\citep{gw170817lateradio}. It may be an evidence for a 
cocoon around a relativistic jet, as predicted by simulations. Alternatively, the brightening of 
radio and higher energy emissions can be due to high velocity tail of neutron rich dynamical 
ejecta~\citep{gw170817latefasttail}.
For instance, continuous heating and evaporation of the outer part of the accretion disk and 
reduction of opacity its due to expansion may explain gradual brightening and the need for energy 
injection found by~\citep{gw170817latebroad}. Collision of outflow with the ISM is another 
possibility. All these models predict the decline of emission, which seems to have begun at 
$\lesssim T+134$ days~\citep{gw170817latedecline}. Although there is not yet enough data to allow 
discrimination between models studied 
in~\citep{gw170817lateradio,gw170817latebroad,gw170817latedecline}, some of off-axis models, such as 
those simulated in~\citep{gw170817latexraystructjet} with a viewing angle $\theta \gtrsim 16^\circ$ 
are ruled out. Moreover, models which best fit late observations in X-ray and radio according 
to~\citep{gw170817latexraystructjet} (see their Figs. 2, 3, and 4) are ruled out, because they 
predict the brightening of afterglow for at least few hundreds of days, which 
is inconsistent with the decline observed in X-ray and optical by~\citep{gw170817latedecline}.

In any case, as we argued in Sec. \ref{sec:offaxis}, the ejecta component(s) responsible for the late 
afterglow may have little direct relation with the relativistic jet that generated prompt gamma-ray, 
except probably some contribution from external shock of the relativistic jet. However, external 
shocks depend on surrounding material and do not directly present the state of ultra-relativistic 
jet to be compared with our conclusions.

\subsection{A qualitative picture of GW/GRB/kilonova 170817}
Finally, our interpretation of data and simulations of GW/GRB/kilonova 170817 can be summarized in 
the following qualitative picture:

\begin{description}
\item{-} Progenitors were old and cool neutron stars with close masses, i.e. $M_2 / M_1 \lesssim 1$;
\item{-} They had soft equations of state and small initial magnetic fields of $\lesssim 10^9$ G. 
Their fields were anti-aligned with respect to orbital rotation axis and each other.
\item{-} For dynamical or historical reasons, such as encounter with similar mass objects, their 
spins before final inspiral were anti-aligned.
\item{-} The merger produced a HMNS with a moderate magnetic field of $\lesssim 10^{10}$ G. 
This value is in the lowest limit of what is obtained in GRMHD simulations. 
\item{-} The HMNS eventually collapsed to a black hole and created a moderately magnetized 
disk/torus and a low density, low magnetized and mildly collimated polar outflow.
\item{-} A total amount of $\sim 0.03-0.05~M_\odot$ material, including $10^{-3} - 10^{-2}~M_\odot$ of 
tidally stripped pre-merger and a post-merger wind were ejected to high latitudes. They were 
subsequently collimated and accelerated by transfer of Poynting to kinetic energy. The same 
process increased electron yield by segregation of charged particles.
\item{-} A small mass fraction of the polar ejecta was accelerated to ultra-relativistic velocities 
and made a weak GRB. The reason for low Lorentz factor, density, and extent of this component was the 
weakness of the magnetic field. 
\item{-} Either due to the weakness of the relativistic jet, which soon after internal shocks had 
a break and lost its collimation, or due to the lack of sufficient circumburst material, the 
afterglows of the GRB in X-ray and lower energy bands were very faint at $\gtrsim T+0.6$ days, 
and only detectable as a non-thermal addition in UV/blue emission of cocoon/wind. 
\item{-} The late X-ray brightening is most probably independent of unusual weakness of GRB 170817A 
and is generated by interaction of a slower component of ejecta with the ISM or other surrounding 
material. The remnant of the relativistic jet may have some contribution in these emissions, 
specially at earlier times.
\end{description}



At present faint GRBs similar to GRB 170817A are detectable by high energy satellites only if they 
occur in the local Universe. Therefore, the small accessible volume significantly suppresses the rate 
of such events. Indeed, since the launch of the Swift satellite until present only 7 confirmed short 
bursts without early X-ray counterpart\footnote{By early afterglow we mean from $\gtrsim T+100$ sec 
up to $\sim T+\mathcal{O}(1) \times 10^4$ sec. Usually if no X-ray afterglow is found in this 
interval, no further detection attempt would be made. GRB 170817A was an exception due to its 
association with a GW event.} were observed\footnote{According to our search in the on-line Swift-BAT 
database \href{https://swift.gsfc.nasa.gov/archive/grb\_table/}{https://swift.gsfc.nasa.gov/archive/grb\_table/}.}. 
In addition, association of these transients to BNS merger is not certain and some 
of them may be giant flares from SGRs in nearby galaxies. Only long duration follow-up of future 
early time X-ray faint or dark GRBs with or without associated GW can prove or refute hypotheses 
raised here to explain the unusual characteristics of GRB 170817A.

\subsection{Progenitors of bright short GRBs} \label{sec:othergrb}
Due to observational bias most of GRBs with known redshift or their host galaxies are bright. 
If our explanation of reasons behind the weakness of GRB 170817A are correct, BNS mergers at 
higher redshifts must be on average intrinsically brighter, because younger neutron stars have
stronger magnetic fields. Their spin-orbit orientation is apriori independent of redshift,  
but it should somehow depend on age, formation history, and environment of the BNS. Older binaries 
on average had more opportunity to interact with other celestial bodies. For instance, neutron 
stars in the dense environment of globular clusters have higher chance of forming BNS and being 
gravitationally disrupted, which may change their spin-orbit orientation. 

Despite small number of short GRBs with known redshift, the impact of BNS aging on the outcome of 
the merger gradually become discernible in the data. Indeed, in Fig. \ref{fig:allsgrb}-c,d 
there is a clear trend of increasing total fluence and average flux with redshift. Although the 
absence of faint bursts at higher redshifts can be interpreted as an observational bias, the lack 
of bright bursts at lower redshifts is not explainable and their rarity does not seem sufficient 
to explain clear stratification of fluence and average flux with redshift.

Although no BH-NS merger is so far detected, they remain a plausible origin for energetic short GRBs,  
because the larger mass difference of the pair may produce larger ejecta and magnetic 
field~\citep{nstarbhmergsimul,nstarbhmergsimul0,nstarbhmergsimul1,nstarbhmergsimul2}. Only with 
detection of more gravitational wave events with electromagnetic counterparts it would be possible 
to verify this hypothesis.

\section{Outline} \label{sec:outline}
GW/GRB 170817A event gave us the opportunity to discover the nature of short GRBs, which for 
decades their origin was a subject of speculation and no direct evidence or proof of hypotheses 
about their sources was in hand. 

In this work through simulation of prompt emission of GRB 170817A we showed that despite its outlier 
characteristics, it was generated by the same physical processes involved in the production of more 
ordinary and typical short GRBs. Based on the results of 3D GRMHD from the literature and published 
analysis of multi-wavelength observations of this event we argued that the faintness of GRB 170817A 
was caused by old age, coolness, and reduced magnetization of its progenitor neutron stars. These 
intrinsic factors were probably helped by environmental effects and history of gravitational 
disturbances, which had induced an anti-aligned spin-orbit orientation.

Current observation facilities, specially gravitational wave detectors, are sensitive to events 
similar to GW/GRB 170817 only if they occur at redshifts $\lesssim 0.1$. Moreover, despite the 
ability of the Swift satellite to detect the counterpart of GRBs in X-ray, UV and white bands from 
on average $\gtrsim 70$ sec onward, since its launch in November 2004 only a very small fraction of 
short GRBs have had long duration follow up, i.e. for more than few days, either by the Swift 
instruments, or by other ground and space based telescopes. For this reason, the state of our 
knowledge about late behaviour of their afterglow and 
associated physical processes is incomplete. Nonetheless, there is hope that the huge scientific 
outcome achieved from intense observation of GW/GRB 170817A/AT 2017 gfo, which was a first in its 
kind, would encourage more intense and long duration follow up of short GRBs, even without any 
associated GW. Such observations would help verify some of hypotheses suggested in this work about 
the progenitors of short GRB/kilonova events. For instance, whether the late brightening of their 
afterglow is a common behaviour, and whether there is any systematic correlation between age and 
star formation history of their host galaxy and properties of NS-NS and NS-BH mergers.

\appendix
\section{Evolution models of active region} \label{app:modes}
In the phenomenological model of~\citep{hourigrb} the evolution of $\Delta r'(r')$ cannot be 
determined from first principles. For this reason we consider the following phenomenological 
models:
\bea
&& \Delta r' = \Delta r'_0 \biggl (\frac {\gamma'_0 \beta'}{\beta'_0 \gamma'} 
\biggr )^{\tau}\Theta (r'-r'_0) \quad \text {dynamical model, Model = 0} \label {drdyn} \\
&& \Delta r' = \Delta r'_{\infty} \bigg [1-\biggl (\frac{r'}{r'_0} \biggr )^
{-\delta}\biggr ] \Theta (r'-r'_0) \quad \text {Steady state model, Model = 1} \label {drquasi} \\
&& \Delta r' = \Delta r'_0 \biggl (\frac{r'}{r'_0} \biggr )^{-\delta} 
\Theta (r'-r'_0) \quad \text {Power-law model, Model = 2} \label {drquasiend} \\
&& \Delta r' = \Delta r_{\infty} \bigg [1- \exp (- \frac{\delta(r'-r'_0)}{r'_0}) \biggr ] 
\Theta (r-r'_0) \quad \text {Exponential model, Model = 3} \label {expon} \\
&& \Delta r' = \Delta r'_0 \exp \biggl (-\delta\frac{r'}{r'_0} \biggr )
\Theta (r'-r'_0) \quad \text {Exponential decay model, Model = 4} \label {expodecay}
\eea
The initial width $\Delta r'(r'_0)$ in Model = 1 \& 3 is zero. Therefore, they are suitable for 
description of initial formation of an active region in internal or external shocks. Other models 
are suitable for describing more moderate growth or decline of the active region. In Table 
\ref{tab:param} the column $mod.$ indicates which evolution rule is used in a simulation regime - as 
defined in the foot notes of this table - using model number given in \ref{drdyn}-\ref{expodecay}.

\paragraph*{Acknowledgment} The author thanks Phil Evans for providing her with Swift-XRT data 
before their publication.

\end{document}